\documentclass[10pt,journal]{IEEEtran}

\usepackage[utf8]{inputenc}
\usepackage[T1]{fontenc}
\usepackage[nocompress]{cite}

\usepackage{graphicx}
\usepackage{microtype}
\usepackage{ragged2e}
\usepackage{amsmath,amssymb}
\usepackage{booktabs}
\usepackage{multirow}
\usepackage{array}
\usepackage{url}

\usepackage[table]{xcolor}
\usepackage[
    colorlinks=true,
    citecolor=blue,
    linkcolor=blue,
    urlcolor=black
]{hyperref}

\usepackage[capitalize]{cleveref}

\usepackage{enumitem}
\usepackage{tikz}
\usepackage{forest}
\usepackage{xurl}
\usepackage{tabularx}
\usepackage{placeins}

\definecolor{RowGray}{RGB}{245,245,245}

\newcolumntype{L}[1]{>{\RaggedRight\arraybackslash}p{#1}}
\newcolumntype{C}[1]{>{\Centering\arraybackslash}p{#1}}
\newcolumntype{Y}{>{\RaggedRight\arraybackslash}X}

\useforestlibrary{edges}
  \definecolor{hidden-draw}{RGB}{110,110,110}
  \definecolor{hidden-pink}{RGB}{233,170,183}
  
\crefname{section}{Section}{Sections}
\crefname{table}{Table}{Tables}
\crefname{figure}{Fig.}{Figs.}

\newcommand{\etal}{\emph{et al.}}

\begin{document}
\bstctlcite{IEEEexample:BSTcontrol}

\title{From Acoustic Sensing to Physical Actions: A Comprehensive Survey of Security Risks in Edge-Deployed Audio LLM-Based Agents}

\author{
Binhong Wang,
Haidong~Kang,~\IEEEmembership{Member,~IEEE},
Yanfeng~Zhang,~\IEEEmembership{Member,~IEEE},
Hong-Ning~Dai,~\IEEEmembership{Senior~Member,~IEEE},
Guo~Yu,~\IEEEmembership{Member,~IEEE},
and~Shangce~Gao,~\IEEEmembership{Senior~Member,~IEEE}%
\thanks{
Binhong Wang, Haidong Kang, and Yanfeng Zhang are with
Northeastern University, China
(e-mail: wangbinhong@mails.neu.edu.cn; kanghaidong@qhd.neu.edu.cn;
zhangyf@mail.neu.edu.cn).
}%
\thanks{
Hong-Ning Dai is with the Department of Computer Science,
Hong Kong Baptist University, Hong Kong
(e-maill: hndai@ieee.org).
}%
\thanks{
Guo Yu is with the Institute of Intelligent Manufacturing,
Nanjing Tech University, Nanjing, China
(e-mail: gysearch@163.com).
}%
\thanks{
Shangce Gao is with the Faculty of Engineering,
University of Toyama, Toyama, Japan
(e-mail: gaosc@eng.u-toyama.ac.jp).
}%
\thanks{
Corresponding author: Haidong Kang
(e-mail: kanghaidong@qhd.neu.edu.cn).
}
}

\maketitle
\begin{abstract}
Audio Large Language Models (Audio LLMs) are embedded in agents for smart homes, vehicles, wearables, and embodied systems, where acoustic inputs can propagate from perception into persistent state and consequential execution. This setting extends security analysis beyond model behavior to the system conditions that determine how acoustic inputs can influence agent behavior and external actions. Existing surveys examine acoustic security, LLM-based agents, and edge deployment in isolation, leaving their security interactions in deployed Audio LLM-based agents insufficiently characterized. This survey introduces an Acoustic Sensing-to-Action (S2A) system model that traces these security implications along the path from acoustic sensing to external execution, while accounting for edge–cloud runtime and lifecycle transformation. We use the model to synthesize attacks, defenses, and evaluation methods across the acoustic, agent, and edge-system literature. The synthesis shows that security gaps can emerge within components and at the transitions between stages, where protections developed for one stage do not necessarily provide end-to-end assurance in the deployed system. These findings motivate security evaluation that follows acoustic influence from sensing to action across persistent state, system boundaries, and changing deployment conditions in real-world agents.
\end{abstract}


\section{Introduction}
\label{sec:introduction}
\subsection{Motivation}

Large Language Models (LLMs) are increasingly integrated into agents that
maintain interaction state and use external tools to execute multi-step
tasks. Growing demand for intelligent services is driving the deployment of such agents in real-world applications. DESAMO targets embedded smart-home assistance~\cite{choi2025desamodeviceelderfriendlysmart}, SoundHound reports on-chip voice Artificial Intelligence (AI) for in-vehicle operation~\cite{soundhound_onchip}, and LLM-enabled robotic systems connect model decisions to physical actuators~\cite{ravichandran2026safetyguardrailsllmenabledrobots}. Operation
in these settings requires an agent to interpret environmental information
before using it to guide subsequent decisions. The perceptual interface is
therefore part of the deployed security surface.

Among the modalities available for such perception, audio provides a direct
channel between an agent and its users or surrounding environment. Speech carries linguistic content together with speaker and paralinguistic
information, and acoustic input can expose environmental attributes beyond the transcript~\cite{he-etal-2026-audioprivacy}. Audio Large Language Models (Audio LLMs) allow acoustic information to
participate directly in subsequent agent state, reasoning, or external action.
CarMem demonstrates persistent voice-assistant memory across
sessions~\cite{kirmayr-etal-2025-carmem}, while Audio2Tool evaluates
speech-driven tool invocation~\cite{pahwa2026audio2toolspeakcallact}. These capabilities extend the security surface beyond acoustic perception to persistent state and external execution.

\begin{figure}[!t]
    \centering
    \includegraphics[
        width=0.55\textwidth,
        keepaspectratio
    ]{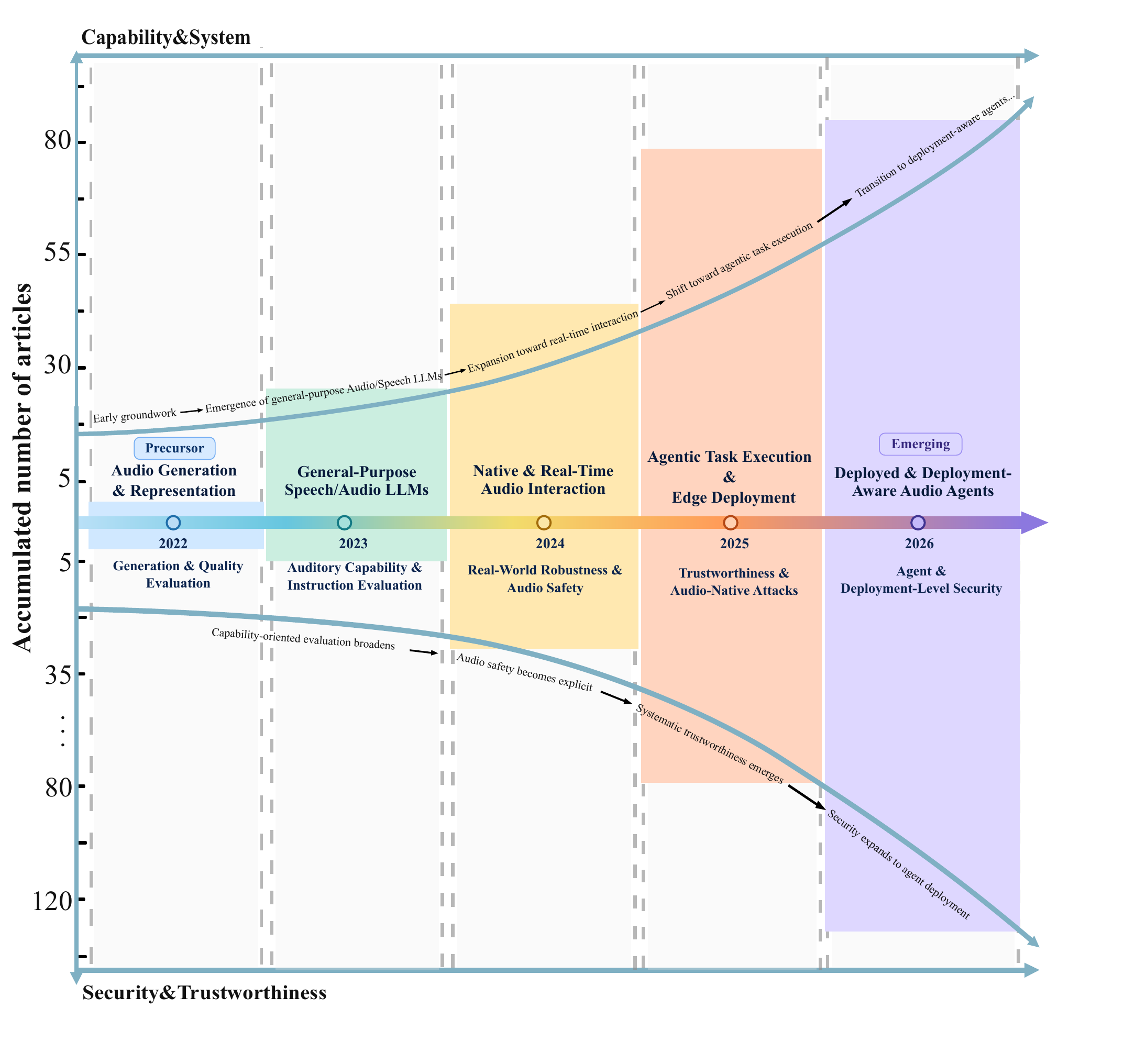}
    \caption{Evolution of the capabilities and security of Audio and Speech LLMs
    toward real-time, agentic, and deployment-aware systems.}
    \label{fig:audio_llm_evolution}
\end{figure}

As shown in Fig.~\ref{fig:audio_llm_evolution}, studies on Audio and Speech LLMs
have progressed from general audio and speech modeling toward real-time
interaction, agentic task execution, and deployment-aware operation. Studies on security
and trustworthiness have also expanded from model robustness and
audio safety toward agent- and deployment-level risks. As these models become
components of deployed agents, the scope of security analysis therefore extends beyond model behavior to the surrounding agent system.

\begin{figure*}[!t]
    \centering
    \includegraphics[
        width=0.96\textwidth,
        keepaspectratio
    ]{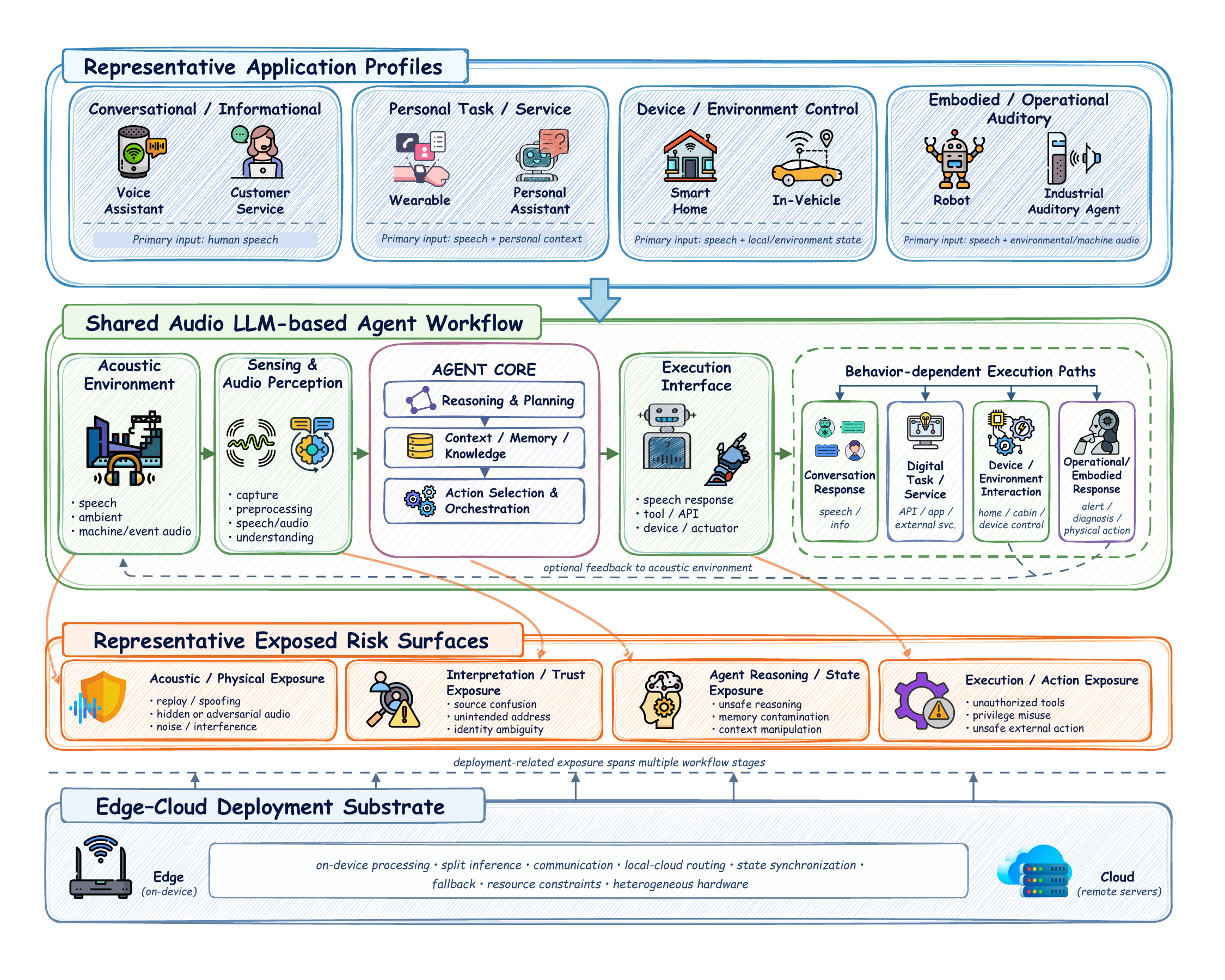}
    \caption{Representative application settings and the shared
    perception-to-action workflow of edge-deployed Audio LLM-based agents.
    Individual systems may omit, combine, or repartition the illustrated
    functions.}
    \label{fig:edge_audio_llm_pipeline_exposure}
\end{figure*}

Fig.~\ref{fig:edge_audio_llm_pipeline_exposure} illustrates four
representative application profiles and a shared perception-to-action workflow
for Audio LLM-based agents. Their interaction flows differ in persistent
context, delegated authority, environmental state, and execution consequences,
motivating a common system abstraction with profile-dependent security
priorities. Section~\ref{subsec:application_profiles} formalizes this comparison.

These workflows introduce security concerns that are not captured by model-level evaluation. Acoustic observations may originate from unintended or
adversarial sources, and successful perception does not establish whether
the information should be trusted or authorized. SH-Bench~\cite{zhan-etal-2026-protecting} and VoxPrivacy~\cite{wang2026voxprivacybenchmarkevaluatinginteractional} document failures
associated with shared-speaker interaction and information flow. Aegis~\cite{li2026aegisgovernanceintegritysecurity} further evaluates security
failures in voice-agent pipelines that reach tools and external actions.

Prior surveys cover complementary aspects of the scope considered here.
Voice-assistant security surveys systematize attacks and defenses involving
the acoustic channel, speaker verification, speech recognition, and device
components~\cite{li2023securityprivacyproblemsvoice}. Surveys of Audio and Speech LLMs instead emphasize model architectures, capabilities, evaluation,
and trustworthiness~\cite{yang2025holisticlalm,
luo2026trustworthylalmsurvey}. These studies establish the modality-specific foundations of the field, but they do not primarily analyze security across persistent agent state, external execution, and distributed edge runtime. Recent surveys increasingly address agent-level settings. Rahimov et al.\ review safety and robustness of Audio LLM-based agents across failure modes, agent architectures, deployment domains, and mitigation paradigms, including persistent-memory and tool-use risks~\cite{rahimov2026audiollmagentsurvey}.
General surveys on the security of LLM-based agents provide broader
systems-oriented analyses of prompt injection, persistent state, delegated authority, tool use, and external actions~\cite{ling2026securellmagents,
xu2026llmagentsecurityduality}. Long-term memory security has also emerged as a dedicated topic spanning memory write, retrieval, execution, propagation, and rollback~\cite{lin2026agentmemorysecuritysurvey}. In parallel, edge-LLM surveys focus on model preparation, runtime optimization, and on-device execution~\cite{zheng2024reviewedgellms}, while recent work on agentic edge intelligence emphasizes communication-constrained coordination among distributed agents~\cite{dib2026agenticedgeintelligencesurvey}. Table~\ref{tab:survey_comparison} summarizes the resulting division of
coverage. Existing modality-centric, agent-centric, and edge-centric surveys
address important components, but interactions among acoustic trust, agent state and execution, and deployment conditions remain distributed across separate lines of work rather than integrated within a common system model.

\begin{table}[t]
\caption{Scope comparison with representative surveys.}
\label{tab:survey_comparison}
\centering
\footnotesize
\setlength{\tabcolsep}{2.3pt}
\renewcommand{\arraystretch}{1.06}

\begin{tabularx}{\columnwidth}{
    @{}
    L{0.405\columnwidth}
    *{6}{C{0.066\columnwidth}}
    @{}
}
\toprule
\textbf{Survey}
& \textbf{L}
& \textbf{A}
& \textbf{P}
& \textbf{T}
& \textbf{E}
& \textbf{D} \\
\midrule

\rowcolor{gray!10}
Li et al.~\cite{li2023securityprivacyproblemsvoice}
& --
& $\checkmark$
& --
& $\triangle$
& $\triangle$
& -- \\

\rowcolor{gray!4}
Yang et al.~\cite{yang2025holisticlalm}
& $\checkmark$
& $\triangle$
& --
& --
& --
& -- \\

\rowcolor{gray!10}
Luo et al.~\cite{luo2026trustworthylalmsurvey}
& $\checkmark$
& $\checkmark$
& --
& $\triangle$
& --
& -- \\

\rowcolor{gray!4}
Rahimov et al.~\cite{rahimov2026audiollmagentsurvey}
& $\checkmark$
& $\checkmark$
& $\triangle$
& $\checkmark$
& $\triangle$
& -- \\

\rowcolor{gray!10}
Ling et al.~\cite{ling2026securellmagents}
& --
& --
& $\checkmark$
& $\checkmark$
& $\triangle$
& -- \\

\rowcolor{gray!4}
Lin et al.~\cite{lin2026agentmemorysecuritysurvey}
& --
& --
& $\checkmark$
& $\triangle$
& --
& -- \\

\rowcolor{gray!10}
Zheng et al.~\cite{zheng2024reviewedgellms}
& --
& --
& --
& --
& $\checkmark$
& $\checkmark$ \\

\rowcolor{gray!4}
Dib~\cite{dib2026agenticedgeintelligencesurvey}
& --
& --
& $\triangle$
& $\triangle$
& $\checkmark$
& $\triangle$ \\

\midrule
\rowcolor{gray!15}
\textbf{This survey}
& $\checkmark$
& $\checkmark$
& $\checkmark$
& $\checkmark$
& $\checkmark$
& $\checkmark$ \\

\bottomrule
\end{tabularx}

\vspace{0.7mm}
\begin{minipage}{\columnwidth}
\footnotesize
\textbf{L} = Focus on Audio and Speech LLMs;
\textbf{A} = Acoustic and Physical-Channel Security;
\textbf{P} = Persistent Agent State;
\textbf{T} = Tool and Action Security;
\textbf{E} = Edge--Cloud Runtime and Distributed Execution;
\textbf{D} = Deployment and Lifecycle Transformation.
$\checkmark$ indicates dedicated coverage,
$\triangle$ partial coverage, and
-- that the topic is not a primary focus.
\end{minipage}
\end{table}

To fill this gap, this survey develops an Acoustic Sensing-to-Action (S2A)
system model that integrates acoustic trust, agent state and execution, and
edge-deployment conditions within a common security analysis.

\subsection{Contributions}
\label{subsec:scope_contributions}

The main contributions are as follows.

\begin{itemize}[leftmargin=1.5em]

    \item We develop the S2A system model for edge-deployed Audio LLM-based
    agents. The model connects the perception-to-action workflow with Edge--Cloud Runtime and Lifecycle
    Transformation, captures the security-relevant characteristics of the
    deployed-agent setting, and distinguishes a common system structure from profile-dependent deployment conditions.

    \item We develop a paired attack and defense taxonomy comprising six
    analysis areas, namely \emph{Acoustic Sensing}, \emph{Provenance and Authority},
    \emph{State and Reasoning}, \emph{Tool and Action Execution},
    \emph{Edge--Cloud Runtime}, and \emph{Lifecycle Transformation}. Within
    this framework, we integrate evidence from acoustic and voice security,
    Audio and Speech LLMs, the security of LLM-based agents, and edge systems, and compare representative mechanisms, demonstrated capabilities or defensive strengths, and limitations.

    \item We examine evaluation resources and open problems from the
    perspective of the final deployed agent configuration. The resulting
    analysis identifies unresolved issues in instruction provenance and
    authority, persistent multi-user state, policy-preserving edge--cloud
    transitions and fallback, action-level containment and recovery, and
    security beyond speech-centric agents.

\end{itemize}

\subsection{Organization of This Survey}

The remainder of this survey is organized as follows.
\Cref{sec:background} defines the operational scope of edge-deployed
Audio LLM-based agents, compares representative application profiles,
develops the system model, reviews architectural, deployment, and lifecycle
realizations, and introduces the six-area security analysis framework.
\Cref{sec:attacks} and \cref{sec:defenses} organize attacks and defenses
according to this framework and compare their mechanisms, capabilities or
strengths, and limitations.
\Cref{sec:openissues} examines current evaluation gaps and unresolved
security problems in deployed Audio LLM-based agents, and
\cref{sec:conclusion} concludes the survey.
Fig.~\ref{fig:overview_of_edge_audio}  provides an overview of the survey organization.

\tikzstyle{my-box}=[
    rectangle,
    draw=hidden-draw,
    rounded corners,
    text opacity=1,
    minimum height=1.5em,
    minimum width=5em,
    inner sep=2pt,
    align=center,
    fill opacity=.5,
    line width=0.8pt,
]

\tikzstyle{leaf}=[my-box,
    minimum height=1.5em,
    fill=hidden-pink!80,
    text=black,
    align=center,
    font=\fontsize{8.25}{9.25}\selectfont,
    inner xsep=4pt,
    inner ysep=4pt,
    line width=0.8pt,
]

\tikzstyle{sectionbox}=[my-box,
    text=black,
    text width=10em,
    align=center,
    font=\normalsize,
    inner xsep=3pt,
    inner ysep=4pt,
    line width=0.8pt,
]

\tikzstyle{examples}=[my-box,
    minimum height=1.5em,
    fill=white,
    text=black,
    align=left,
    font=\fontsize{7.2}{8.20}\selectfont,
    inner xsep=5pt,
    inner ysep=4pt,
    line width=0.7pt,
]

\newcommand{\surveyLeaf}[1]{\parbox[c]{7.35cm}{\centering #1}}
\newcommand{\surveyExamples}[1]{\parbox[t]{11.85cm}{\raggedright #1}}

\begin{figure*}[!th]
    \centering
    \resizebox{0.98\textwidth}{!}{%
        \begin{forest}
            forked edges,
            for tree={
                grow=east,
                reversed=true,
                anchor=base west,
                parent anchor=east,
                child anchor=west,
                base=center,
                rectangle,
                draw=hidden-draw,
                rounded corners,
                align=left,
                minimum width=4em,
                edge+={darkgray, line width=1pt},
                s sep=2pt,
                l sep=5pt,
                inner xsep=2pt,
                inner ysep=3pt,
                line width=0.8pt,
                font=\normalsize,
                ver/.style={
                    rotate=90,
                    child anchor=north,
                    parent anchor=south,
                    anchor=center
                },
            },
            [
                {Security of Edge-Deployed \\ Audio LLM-Based Agents}, ver
                [
                    {Background\\ (\S~\ref{sec:background})}, sectionbox
                    [
                        {\surveyLeaf{Operational Definition and Scope}}, leaf
                        [
                            {\surveyExamples{%
                            SALMONN~\cite{tang2024salmonngenerichearingabilities};
                            Qwen-Audio~\cite{chu2023qwenaudioadvancinguniversalaudio};
                            AudioPaLM~\cite{rubenstein2023audiopalmlargelanguagemodel};
                            Moshi~\cite{defossez2024moshispeechtextfoundationmodel}
                            }}, examples
                        ]
                    ]
                    [
                        {\surveyLeaf{S2A System Model and Security-Relevant Characteristics}}, leaf
                    ]
                    [
                        {\surveyLeaf{Architectural, Deployment, and Lifecycle Realizations}}, leaf
                        [
                            {\surveyExamples{%
                            DESAMO~\cite{choi2025desamodeviceelderfriendlysmart};
                            ASTA~\cite{torkamani2025adaptiveedgecloudinferencespeechtoaction};
                            CoFi-Agent~\cite{zhang2026bridgingperceptiongaplightweight};
                            CarMem~\cite{kirmayr-etal-2025-carmem}.\\[1pt]
                            SELMA~\cite{wagner2025selma};
                            Tiny-Align~\cite{qin2025tinyalignbridgingautomaticspeech};
                            Gemma 3n~\cite{gemma3n};
                            LFM2.5-Audio~\cite{lfm25_audio}
                            }}, examples
                        ]
                    ]
                    [
                        {\surveyLeaf{Six-Area Security Analysis Framework}}, leaf
                    ]
                ]
                [
                    {Threats and Attacks \\ (\S~\ref{sec:attacks})}, sectionbox
                    [
                        {\surveyLeaf{Acoustic Sensing Attacks}}, leaf
                        [
                            {\surveyExamples{%
                            AdvWave~\cite{kang2025advwave};
                            DolphinAttack~\cite{Zhang_2017};
                            Sirens' Whisper~\cite{ling2026sirenswhisper};
                            AudioHijack~\cite{chen2026hijackinglargeaudiolanguagemodels};
                            HearSay~\cite{wang-etal-2026-hearsay};
                            AudioPrivacy~\cite{he-etal-2026-audioprivacy}
                            }}, examples
                        ]
                    ]
                    [
                        {\surveyLeaf{Provenance and Authority Attacks}}, leaf
                        [
                            {\surveyExamples{%
                            ASVspoof~\cite{yamagishi2021asvspoof};
                            SH-Bench~\cite{zhan-etal-2026-protecting};
                            VoxPrivacy~\cite{wang2026voxprivacybenchmarkevaluatinginteractional};
                            Indirect Audio Instructions~\cite{bagdasaryan2023abusingimagessoundsindirect};
                            Audio Injection Evaluation~\cite{hou-etal-2025-evaluating}
                            }}, examples
                        ]
                    ]
                    [
                        {\surveyLeaf{State and Reasoning Attacks}}, leaf
                        [
                            {\surveyExamples{%
                            VoiceJailbreak~\cite{shen2024voicejailbreakattacksgpt4o};
                            StyleBreak~\cite{li2026stylebreak};
                            JAMA~\cite{krishnan2026optimizingmultimodaljailbreaksspoken};
                            AgentPoison~\cite{chen2024agentpoison};
                            MINJA~\cite{dong2025minja}
                            }}, examples
                        ]
                    ]
                    [
                        {\surveyLeaf{Tool and Action Execution Attacks}}, leaf
                        [
                            {\surveyExamples{%
                            ToolHijacker~\cite{shi2026toolhijacker};
                            InjecAgent~\cite{zhan-etal-2024-injecagent};
                            Mind the GAP~\cite{cartagena2026mindgaptextsafety};
                            Evo-Attacker~\cite{yan-etal-2026-evo};
                            Aegis~\cite{li2026aegisgovernanceintegritysecurity}
                            }}, examples
                        ]
                    ]
                    [
                        {\surveyLeaf{Edge--Cloud Runtime Attacks}}, leaf
                        [
                            {\surveyExamples{%
                            PCAT~\cite{285505};
                            LLM Internal-State Inversion~\cite{309752};
                            ModelSpy~\cite{xiao2026modelspy};
                            NaturalSloth~\cite{chen-etal-2026-naturalsloth};
                            Systems-Level Edge-Agent Attacks~\cite{zhan2026systemslevelattacksurfaceedge}
                            }}, examples
                        ]
                    ]
                    [
                        {\surveyLeaf{Lifecycle Transformation Attacks}}, leaf
                        [
                            {\surveyExamples{%
                            Hidden in the Noise~\cite{lin2025hiddennoiseunveilingbackdoors};
                            Component-Level Backdoors~\cite{fortier2026backdoorslivecomponentlevelanalysis};
                            LoRATK~\cite{liu-etal-2025-loratk};
                            Backdoored Tool Use~\cite{zhang-pei-2026-llm};
                            Quantization Backdoors~\cite{ma2024quantizationbackdoors}
                            }}, examples
                        ]
                    ]
                ]
                [
                    {Defenses and Safeguards \\ (\S~\ref{sec:defenses})}, sectionbox
                    [
                        {\surveyLeaf{Acoustic Sensing Defenses}}, leaf
                        [
                            {\surveyExamples{%
                            WaveGuard~\cite{272238};
                            AudioPure~\cite{wu2023defendingadversarialaudiodiffusion};
                            EarArray~\cite{DBLP:conf/ndss/ZhangJLQX21};
                            MicGuard~\cite{299545};
                            Stream-Voice-Anon~\cite{kuzmin2026streamvoiceanonenhancingutilityrealtime}
                            }}, examples
                        ]
                    ]
                    [
                        {\surveyLeaf{Provenance and Authority Defenses}}, leaf
                        [
                            {\surveyExamples{%
                            BPFT / SH-Bench~\cite{zhan-etal-2026-protecting};
                            Adaptive DDSD~\cite{chi2025adaptiveddsd};
                            StruQ~\cite{307694};
                            Task Shield~\cite{jia-etal-2025-task};
                            SafeSpeech~\cite{307846}
                            }}, examples
                        ]
                    ]
                    [
                        {\surveyLeaf{State and Reasoning Defenses}}, leaf
                        [
                            {\surveyExamples{%
                            AudioGuard~\cite{kang2026audioguardcomprehensiveaudiosafety};
                            SPIRIT~\cite{djanibekov-etal-2025-spirit};
                            A-MemGuard~\cite{wei2025amemguard};
                            CognitiveGuard~\cite{qian-2026-visual};
                            SENTINEL~\cite{karamchandani2026forgedreasoning};
                            PPMF~\cite{xu2026memoryprovenancelaundering}
                            }}, examples
                        ]
                    ]
                    [
                        {\surveyLeaf{Tool and Action Execution Defenses}}, leaf
                        [
                            {\surveyExamples{%
                            CaMeL~\cite{debenedetti2025defeatingpromptinjectionsdesign};
                            IPIGuard~\cite{an-etal-2025-ipiguard};
                            AgentSpec~\cite{wang2025agentspeccustomizableruntimeenforcement};
                            VIGIL~\cite{lin-etal-2026-vigil};
                            MiniScope~\cite{zhu2025miniscopeprivilegeframeworkauthorizing};
                            RoboGuard~\cite{ravichandran2026safetyguardrailsllmenabledrobots}
                            }}, examples
                        ]
                    ]
                    [
                        {\surveyLeaf{Edge--Cloud Runtime Defenses}}, leaf
                        [
                            {\surveyExamples{%
                            PriPert~\cite{fan2026doesserverseeunderstanding};
                            DEL~\cite{gu2026differentiallyprivatecommunicationefficient};
                            AgenTEE~\cite{abdollahi2026agentee};
                            PD$^{3}$F~\cite{zhang-etal-2025-pd3f};
                            AegisMCP~\cite{zhan2025aegismcponlinegraphintrusion}
                            }}, examples
                        ]
                    ]
                    [
                        {\surveyLeaf{Lifecycle Transformation Defenses}}, leaf
                        [
                            {\surveyExamples{%
                            STEP~\cite{wang2026stepdetectingaudiobackdoor};
                            GN-FT~\cite{zhou2025gradientnormbasedfinetuningbackdoor};
                            PEFTGuard~\cite{Sun_2025};
                            Safe LoRA~\cite{hsu2025safelorasilverlining};
                            EFRAP~\cite{Li_2024_CVPR};
                            QuantGuard~\cite{zheng2026breakingroundingtrapsecuring}
                            }}, examples
                        ]
                    ]
                ]
                [
                    {Open Issues \\ (\S~\ref{sec:openissues})}, sectionbox
                    [
                        {\surveyLeaf{Evaluation Gaps}}, leaf
                        [
                            {\surveyExamples{%
                            JALMBench~\cite{peng2026jalmbenchbenchmarkingjailbreakvulnerabilities};
                            AudioTrust~\cite{li2026audiotrustbenchmarkingmultifacetedtrustworthiness};
                            PersistBench~\cite{pulipaka2026persistbenchlongtermmemoriesforgotten};
                            GAP~\cite{cartagena2026mindgaptextsafety};
                            Aegis~\cite{li2026aegisgovernanceintegritysecurity};
                            ACIArena~\cite{an-etal-2026-aciarena}
                            }}, examples
                        ]
                    ]
                    [
                        {\surveyLeaf{Instruction Provenance and Authority}}, leaf
                        [
                            {\surveyExamples{%
                            SH-Bench~\cite{zhan-etal-2026-protecting};
                            Adaptive DDSD~\cite{chi2025adaptiveddsd};
                            Task Shield~\cite{jia-etal-2025-task}
                            }}, examples
                        ]
                    ]
                    [
                        {\surveyLeaf{Persistent Multi-User State and Memory Provenance}}, leaf
                        [
                            {\surveyExamples{%
                            CarMem~\cite{kirmayr-etal-2025-carmem};
                            PersistBench~\cite{pulipaka2026persistbenchlongtermmemoriesforgotten};
                            SENTINEL~\cite{karamchandani2026forgedreasoning};
                            PPMF~\cite{xu2026memoryprovenancelaundering}
                            }}, examples
                        ]
                    ]
                    [
                        {\surveyLeaf{Deployment Consistency and Safe Fallback}}, leaf
                        [
                            {\surveyExamples{%
                            ASTA~\cite{torkamani2025adaptiveedgecloudinferencespeechtoaction};
                            CoFi-Agent~\cite{zhang2026bridgingperceptiongaplightweight};
                            Systems-Level Edge-Agent Analysis~\cite{zhan2026systemslevelattacksurfaceedge}
                            }}, examples
                        ]
                    ]
                    [
                        {\surveyLeaf{Action-Level Containment, Verification, and Recovery}}, leaf
                        [
                            {\surveyExamples{%
                            AgentSpec~\cite{wang2025agentspeccustomizableruntimeenforcement};
                            VIGIL~\cite{lin-etal-2026-vigil};
                            Proof-of-Execution Memory~\cite{rahman2026proofexecutionmemory}
                            }}, examples
                        ]
                    ]
                    [
                        {\surveyLeaf{Security Beyond Speech-Centric Agents}}, leaf
                    ]
                ]
            ]
        \end{forest}
    }
    \caption{Overview and literature roadmap of the survey. The attack and
    defense branches follow the six S2A security analysis areas; white nodes
    list representative works discussed in the corresponding sections.}
    \label{fig:overview_of_edge_audio}
\end{figure*}
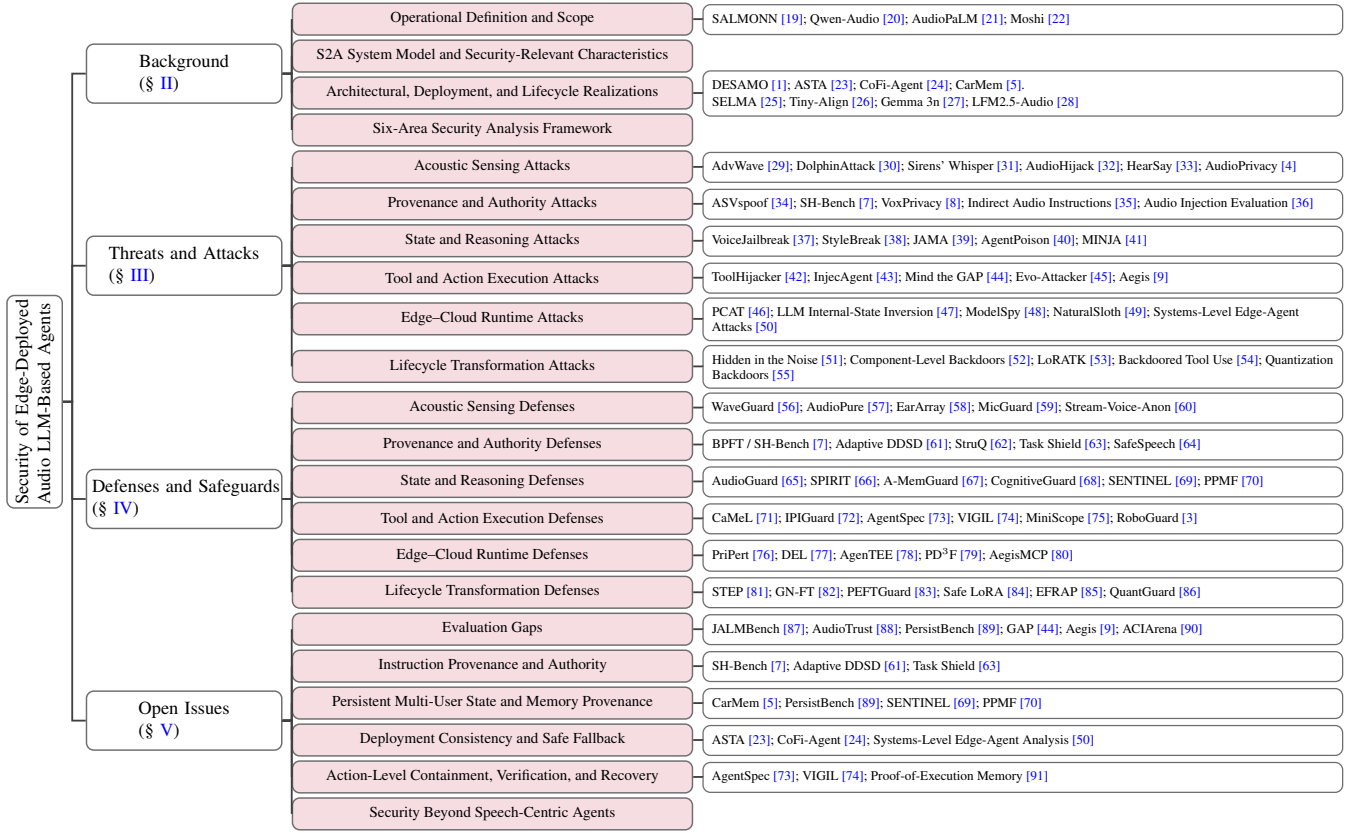

\section{Background and System Model}
\label{sec:background}
To support the security analysis that follows, this section establishes
the technical scope and system abstraction of edge-deployed Audio
LLM-based agents. We first define the surveyed system, its deployment scope, and representative application profiles. We then develop a general system model and examine its architectural, deployment, and lifecycle realizations. Finally, we derive the security analysis framework used to organize the attack and defense literature.

\subsection{Operational Definition and Scope}
\label{subsec:scope_definition}

In this survey, \emph{audio} refers to the acoustic modality processed by
Audio and Speech LLMs and includes both speech and non-speech signals
captured from the physical environment. Speech carries linguistic content together
with speaker identity, paralinguistic cues, and other information not
contained in the transcript~\cite{he-etal-2026-audioprivacy}. The broader audio domain also includes environmental and other non-speech
acoustic events, as represented by general audio-language systems such as
SALMONN and Qwen-Audio~\cite{tang2024salmonngenerichearingabilities,
chu2023qwenaudioadvancinguniversalaudio}. \emph{Audio LLM-based agents} are
systems in which Audio and Speech LLMs process acoustic inputs, and the resulting information can influence subsequent agent
states, reasoning, or external actions. These models may serve as core
reasoning models or as perceptual components coupled to other controllers. Our scope therefore centers on edge-deployed Audio LLM-based
agents in which audio-derived information participates in a broader agent
workflow and an edge device directly interacts with users or the physical
environment.

\subsection{Application Profiles and Deployment Differences}
\label{subsec:application_profiles}

The application profiles in Fig.~\ref{fig:edge_audio_llm_pipeline_exposure}
differ in their interaction flows and security priorities but share a common
perception-to-action structure. Table~\ref{tab:application_profiles} compares
four functional profiles that recur in current systems and benchmarks. A
deployed assistant may combine multiple profiles, and the security emphasis
changes with persistent state, tool authority, environment state, and physical
consequences.

\begin{table*}[!t]
\caption{Representative application profiles and scenario-specific security implications.}
\label{tab:application_profiles}
\centering
\scriptsize
\setlength{\tabcolsep}{3.2pt}
\renewcommand{\arraystretch}{1.12}
\begin{tabularx}{\textwidth}{
    @{}
    L{0.135\textwidth}
    L{0.205\textwidth}
    L{0.18\textwidth}
    L{0.245\textwidth}
    Y
    @{}
}
\toprule
\textbf{Application profile} &
\textbf{Dominant input and context} &
\textbf{Typical execution pattern} &
\textbf{Representative evidence} &
\textbf{Security emphasis} \\
\midrule

\rowcolor{gray!10}
Conversational / Informational &
Human speech, multi-turn dialogue, paralinguistic cues &
Speech or text response with limited external actuation &
URO-Bench~\cite{yan-etal-2025-uro} &
Source and intended-address attribution, interactional privacy, and content safety \\

\rowcolor{gray!4}
Personal Task / Service &
Speech combined with personal context, memory, and task history &
Personalized response, API access, or tool invocation &
CarMem~\cite{kirmayr-etal-2025-carmem}; Audio2Tool~\cite{pahwa2026audio2toolspeakcallact} &
State ownership and provenance, delegated permissions, and tool authorization \\

\rowcolor{gray!10}
Device / Environment Control &
Speech plus device state, environmental variables, implicit intent, and temporal dependencies &
Device actuation, automation, and scheduled multi-step workflows &
DESAMO~\cite{choi2025desamodeviceelderfriendlysmart}; Audio2Tool~\cite{pahwa2026audio2toolspeakcallact}; SimuHome~\cite{seo2026simuhome}; HomeGuardian~\cite{zou2026homeguardian} &
Identity and intended address, state verification, context-conditioned authorization, and physical consequences \\

\rowcolor{gray!4}
Embodied / Operational &
Speech together with environmental or machine audio and other physical context &
Diagnosis, robotic action, operational response, or physical control &
RoboGuard~\cite{ravichandran2026safetyguardrailsllmenabledrobots}; industrial Audio LLM~\cite{cui2026industrial_audio_llm} &
Sensing integrity, non-speech provenance, runtime availability, and action containment or recovery \\

\bottomrule
\end{tabularx}
\end{table*}

The comparison separates deployment-specific conditions from the shared
perception-to-action path. Application profiles determine which functions
dominate and what consequences follow from a failure, while the common path motivates the system abstraction introduced next. SimuHome, HomeGuardian, RoboGuard, and the industrial Audio LLM study are used here as scenario-specific evidence rather than classified as complete Audio LLM-based agent systems.

\subsection{System Model and Security-Relevant Characteristics}
\label{subsec:system_model}

\begin{figure*}[!t]
    \centering
    \includegraphics[
        width=0.97\textwidth,
        keepaspectratio
    ]{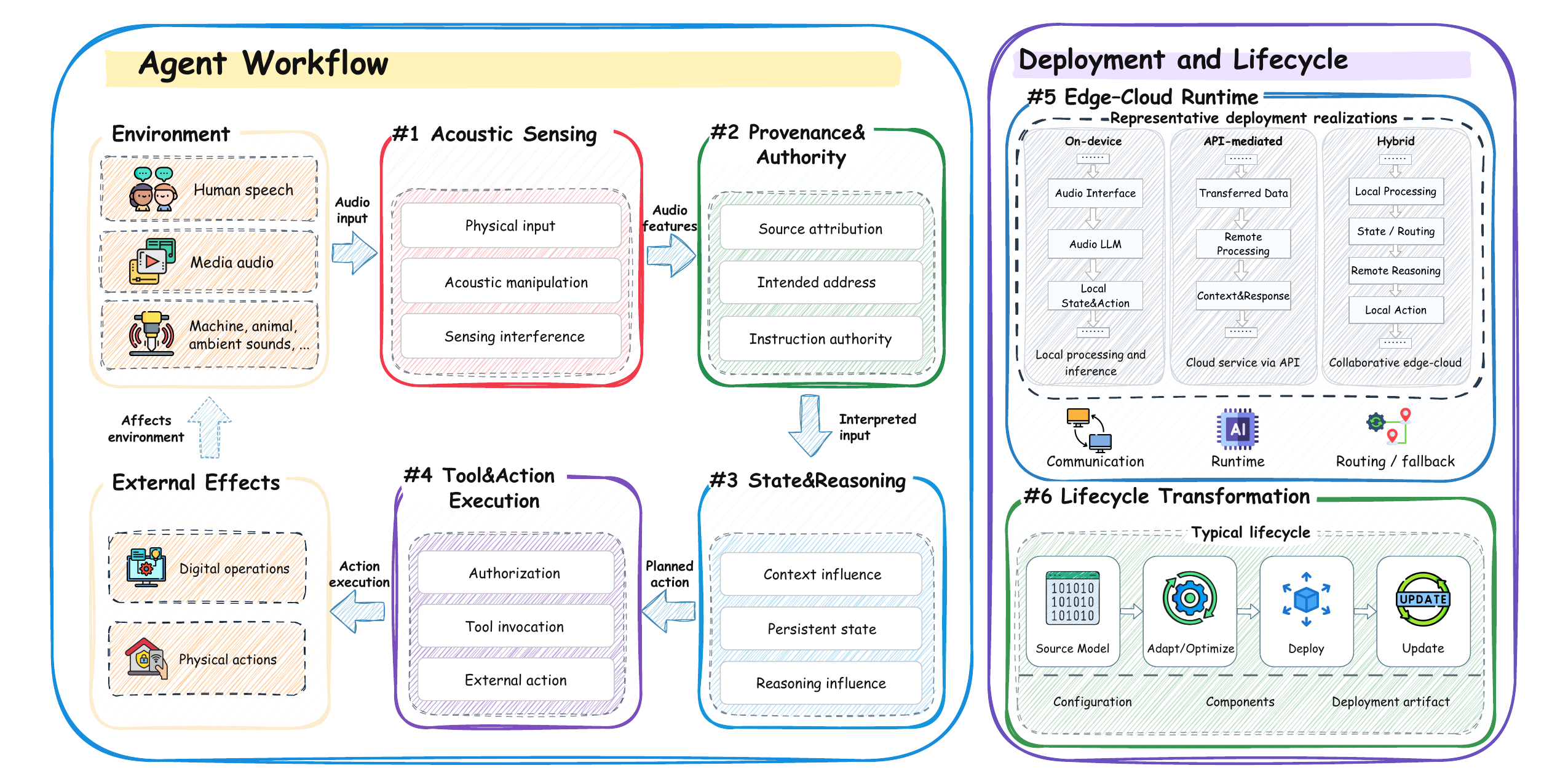}
    \caption{Acoustic Sensing-to-Action (S2A) system model and security
    analysis areas for edge-deployed Audio LLM-based agents. }
    \label{fig:agent_system_model}
\end{figure*}

To establish a common basis for security analysis, we define an Acoustic
Sensing-to-Action (S2A) system model as a system abstraction that traces
acoustic information from environmental input to external effects. In this
model, \emph{Acoustic Sensing} covers the capture and perceptual processing of
acoustic input. \emph{Provenance and Authority} covers source attribution,
intended address, and instruction authority, where instruction authority
refers to whether attributed content is permitted to influence subsequent
agent decisions or actions. \emph{State and Reasoning} covers the integration,
persistence, and reasoning influence of audio-derived information, while
\emph{Tool and Action Execution} covers authorization, tool invocation, and
external actions. \emph{Edge--Cloud Runtime} denotes the distributed
communication, execution, and routing conditions across local and remote
components, whereas \emph{Lifecycle Transformation} denotes changes from
source models and components to deployable artifacts and later adapted
configurations. As shown in Fig.~\ref{fig:agent_system_model}, these six areas
together describe how acoustic information is processed and acted upon in
deployed systems.

Four characteristics distinguish this deployed-agent setting from
model-centric interaction. \emph{Shared acoustic exposure} places the agent
in a physical environment where acoustic observations are not confined to
a controlled input source. \emph{Persistent agent state} allows
audio-derived information to remain available beyond the current
interaction and affect later reasoning. \emph{External execution} connects
agent decisions to digital operations or physical effects.
\emph{Distributed edge deployment} places agent functions across multiple
components and introduces runtime dependencies among them. Table~\ref{tab:security_characteristics} contrasts these characteristics
with model-centric settings and summarizes their security implications.

\begin{table}[t]
\caption{Security-relevant characteristics of edge-deployed Audio
LLM-based agents.}
\label{tab:security_characteristics}
\centering
\footnotesize
\setlength{\tabcolsep}{3pt}
\renewcommand{\arraystretch}{1.12}

\begin{tabularx}{\columnwidth}{
    @{}
    L{0.27\columnwidth}
    L{0.22\columnwidth}
    Y
    @{}
}
\toprule
\textbf{Characteristic}
&
\textbf{Model-centric setting}
&
\textbf{Security implication}
\\
\midrule

\rowcolor{gray!10}
\textbf{Shared acoustic exposure}
&
Controlled input
&
Perception does not establish source provenance, intended address, or
instruction authority.
\\

\rowcolor{gray!4}
\textbf{Persistent agent state}
&
Transient context
&
Misattributed or malicious information may persist and affect later
reasoning or other interactions.
\\

\rowcolor{gray!10}
\textbf{External execution}
&
Response generation
&
Security failures can extend from model behavior to unauthorized digital
operations or physical effects.
\\

\rowcolor{gray!4}
\textbf{Distributed edge deployment}
&
Fixed execution environment
&
Security depends on cross-component communication and on the behavior of
the final deployed configuration.
\\

\bottomrule
\end{tabularx}
\end{table}

\subsection{Architectural, Deployment, and Lifecycle Realizations}
\label{subsec:system_realizations}

The S2A model admits multiple architectural and deployment realizations.
We examine three dimensions: functional placement, Audio--LLM interfaces, and Lifecycle Transformation.

\paragraph{Functional Placement}

Edge-deployed Audio LLM-based agents span both local and distributed
execution. DESAMO performs Audio LLM inference, acoustic-event detection, and device interaction on an embedded platform~\cite{choi2025desamodeviceelderfriendlysmart}. ASTA combines local Automatic Speech Recognition (ASR)
and lightweight LLM inference with adaptive edge--cloud routing~\cite{torkamani2025adaptiveedgecloudinferencespeechtoaction}. CoFi-Agent
places an Audio LLM and tools on the edge and invokes remote reasoning for
computationally demanding cases~\cite{zhang2026bridgingperceptiongaplightweight}. These systems illustrate three representative placement patterns, including predominantly local execution,
partitioned edge--cloud execution, and runtime-dependent routing.

Table~\ref{tab:edge_audio_llm_implementations} summarizes representative
systems together with models and frameworks that enable edge deployment.
Functional placement determines which agent functions and intermediate
information cross local--remote boundaries.

\begin{table}[t]
\caption{Representative systems for edge-deployed Audio LLM-based agents and
deployment enablers.}
\label{tab:edge_audio_llm_implementations}
\centering
\footnotesize
\setlength{\tabcolsep}{2.4pt}
\renewcommand{\arraystretch}{1.12}

\begin{tabularx}{\columnwidth}{
    @{}
    L{0.215\columnwidth}
    Y
    C{0.12\columnwidth}
    L{0.17\columnwidth}
    @{}
}
\toprule
\textbf{Work}
& \textbf{Edge-Side Role}
& \textbf{Mode}
& \textbf{Status}
\\
\midrule

\rowcolor{gray!15}
\multicolumn{4}{@{}l@{}}{\textit{System implementations and prototypes}}
\\

\rowcolor{gray!10}
DESAMO~\cite{choi2025desamodeviceelderfriendlysmart}
& Audio LLM inference, event detection, and device action
& On-device
& Device eval.
\\

\rowcolor{gray!4}
ASTA~\cite{torkamani2025adaptiveedgecloudinferencespeechtoaction}
& ASR, lightweight LLM, adaptive routing, and action execution
& Hybrid
& Device eval.
\\

\rowcolor{gray!10}
SoundHound Chat AI~\cite{soundhound_onchip}
& In-vehicle LLM and on-chip voice interaction
& On-device
& Industrial demo
\\

\rowcolor{gray!4}
CoFi-Agent~\cite{zhang2026bridgingperceptiongaplightweight}
& Local Audio LLM and tools with remote reasoning
& Hybrid
& Framework eval.
\\

\midrule

\rowcolor{gray!15}
\multicolumn{4}{@{}l@{}}{\textit{Edge models and integration enablers}}
\\

\rowcolor{gray!10}
Tiny-Align~\cite{qin2025tinyalignbridgingautomaticspeech}
& Audio encoder--LLM alignment and device-side adaptation
& On-device
& Framework eval.
\\

\rowcolor{gray!4}
Gemma 3n~\cite{gemma3n}
& Audio-capable multimodal model and hardware-tailored loading
& On-device
& Open release
\\

\rowcolor{gray!10}
LFM2.5-Audio-1.5B~\cite{lfm25_audio}
& Native audio/text input and audio/text output
& On-device
& Open release
\\

\bottomrule
\end{tabularx}

\vspace{1mm}
\begin{minipage}{\columnwidth}
\scriptsize
\textit{Status} summarizes the publicly reported implementation,
evaluation, demonstration, or release form. Component descriptions refer to reported
edge-resident functions and do not imply that the complete agent workflow
executes locally.
\end{minipage}

\end{table}

\paragraph{Agent State and Interaction Memory}

Agentic operation also depends on how information is retained across turns and sessions. CarMem instantiates a long-term memory subsystem for an in-car LLM voice assistant that extracts, maintains, and retrieves user preferences across multi-session interactions~\cite{kirmayr-etal-2025-carmem}. Persistent state is therefore distinct from the current acoustic observation and transient prompt context. Stored
information can be selected and reintroduced into later context as the
interaction continues. The state store, retrieval path, and association
between stored information and a user or session are consequently part of
the agent system even when the memory mechanism is implemented outside the Audio LLM.

\paragraph{Audio--LLM Interfaces}

Audio--LLM coupling determines the representation through which acoustic information enters language-model processing. Speech-agent front ends may also perform
interaction-gating functions before higher-level reasoning. SELMA jointly
supports voice-trigger detection, device-directed speech detection, and ASR
within a speech-enabled language model~\cite{wagner2025selma}, illustrating that whether speech belongs to the current interaction can be
determined during acoustic processing rather than inferred only after transcription. We distinguish three interface
types for coupling acoustic information to language-model computation.

Text-mediated systems first convert acoustic input into an explicit textual
representation before language-model processing, as in cascaded
ASR--LLM pipelines such as ASTA~\cite{torkamani2025adaptiveedgecloudinferencespeechtoaction}.
Latent-bridging architectures pass continuous encoder representations to
the language model through learned projection or adaptor modules.
Representative systems include BLSP~\cite{wang2024blspbootstrappinglanguagespeechpretraining},
LLaSM~\cite{shu2023llasmlargelanguagespeech},
SALMONN~\cite{tang2024salmonngenerichearingabilities}, and
Qwen-Audio~\cite{chu2023qwenaudioadvancinguniversalaudio}.
Token-native architectures instead represent speech or audio with discrete
units that participate in language-model sequence processing. SpeechGPT~\cite{zhang2023speechgptempoweringlargelanguage},
AudioPaLM~\cite{rubenstein2023audiopalmlargelanguagemodel}, and
Moshi~\cite{defossez2024moshispeechtextfoundationmodel} provide
representative realizations.

\paragraph{Lifecycle Transformation}

Deployment can transform the source model before it becomes an executable
edge artifact. Deployment preparation may include model compression,
reduced numerical precision, and hardware-oriented optimization to satisfy
device constraints. Audio-capable models exhibit a similar deployment trend. Gemma 3n incorporates mechanisms intended for
resource-constrained execution~\cite{gemma3n}, while
LFM2.5-Audio-1.5B provides a compact native audio model designed for local
deployment~\cite{lfm25_audio}. The deployed artifact may therefore differ
from the source model in numerical representation, model organization, or
runtime form.

The configuration can also change through adaptation after the initial
deployment preparation. Tiny-Align performs resource-efficient
cross-modal alignment and personalization on edge hardware~\cite{qin2025tinyalignbridgingautomaticspeech}, illustrating how local
adaptation and personalization can continue to modify the deployed
configuration. These changes constitute the \emph{Lifecycle Transformation}
defined in Section~\ref{subsec:system_model}.

\subsection{Security Analysis Framework}
\label{subsec:security_framework}

Building on the S2A model in Fig.~\ref{fig:agent_system_model}, we organize
security analysis into the six areas introduced above. \emph{Acoustic
Sensing}, \emph{Provenance and Authority}, \emph{State and Reasoning}, and
\emph{Tool and Action Execution} follow the workflow from acoustic input to
external action, whereas \emph{Edge--Cloud Runtime} and \emph{Lifecycle
Transformation} capture cross-cutting deployment conditions. Attacks are
organized according to the area in which their primary mechanism acts, and
defenses according to where protection is primarily enforced.

\section{Threats and Attacks}
\label{sec:attacks}
Following the security analysis framework in
Section~\ref{subsec:security_framework}, we organize attacks according to
the area in which their primary mechanism acts. Because end-to-end security
studies of Audio LLM-based agents remain limited, we also include
component-level attacks when their mechanisms directly affect a
security-relevant S2A stage, while distinguishing this evidence from complete
agent evaluations. Earlier voice-assistant attacks on inaudible commands and
physical channels provide representative examples~\cite{Zhang_2017,yan2020surfingattack,Yakura_2019}.

\begin{table*}[t]
\centering
\caption{Taxonomy and comparison of attacks on edge-deployed Audio LLM-based agents.}
\label{tab:attack_comparison}
\fontsize{7.0}{7.45}\selectfont
\setlength{\tabcolsep}{2.35pt}
\renewcommand{\arraystretch}{0.96}
\begin{tabularx}{\textwidth}{
    @{}
    L{0.235\textwidth}
    L{0.245\textwidth}
    L{0.225\textwidth}
    Y
    @{}
}
\toprule
\textbf{Attack Type} &
\textbf{Key Mechanism} &
\textbf{Security Impact} &
\textbf{Main Limitation} \\
\midrule

\rowcolor{gray!15}
\multicolumn{4}{@{}l}{\textbf{Acoustic Sensing}} \\
\rowcolor{gray!4}
Adversarial waveform manipulation~\cite{kang2025advwave,wang2026grmutilityawarejailbreakattacks,
fang2026sparsetokenssufficejailbreaking,roh2026codecrobustattacksaudiollms}
& Perturbations optimized in waveform, frequency, token, or codec-latent spaces.
& Targeted behavior or jailbreaks; some variants survive lossy compression.
& Often model- or access-dependent; physical transfer remains inconsistent. \\
\rowcolor{gray!10}
Covert and sensor-mediated injection~\cite{Zhang_2017,yan2020surfingattack,ling2026sirenswhisper}
& Commands encoded in inaudible, structural, optical, or other transduced carriers.
& Covert prompts or commands outside ordinary audible addressed speech.
& Depends on transducer and device characteristics, calibration, and delivery conditions. \\
\rowcolor{gray!4}
Propagation-aware physical manipulation~\cite{Yakura_2019,chen2026hijackinglargeaudiolanguagemodels,
sadasivan-etal-2026-attackers}
& Attack construction models playback, room propagation, reverberation, and recapture.
& Maintains malicious influence over the air and can hijack downstream behavior or actions.
& Sensitive to room, device, distance, and mobility conditions. \\
\rowcolor{gray!10}
Acoustic attribute inference and privacy exposure~\cite{wang-etal-2026-hearsay,he-etal-2026-audioprivacy}
& Infers identity or environmental attributes from voiceprints and non-linguistic cues.
& Reveals sensitive information beyond transcript content.
& Mostly evaluated as profiling or leakage, not propagation into later state or execution. \\

\rowcolor{gray!15}
\multicolumn{4}{@{}l}{\textbf{Provenance and Authority}} \\
\rowcolor{gray!4}
Voice identity spoofing and impersonation~\cite{yamagishi2021asvspoof}
& Replay, synthesis, or voice conversion reproduces identity-bearing characteristics.
& Undermines voice authentication and user-specific access assumptions.
& Identity spoofing alone does not establish instruction or action authority; downstream consequences remain underexplored. \\
\rowcolor{gray!10}
Intended-speaker and cross-user confusion~\cite{zhan-etal-2026-protecting,
wang2026voxprivacybenchmarkevaluatinginteractional}
& Exploits ambiguity over which audible speaker belongs to the active interaction.
& Bystander speech may be processed, disclosed, or bound to the wrong interaction.
& Benchmarks emphasize selective hearing and disclosure more than downstream authorization or action. \\
\rowcolor{gray!4}
Instruction-authority confusion and indirect audio injection~\cite{bagdasaryan2023abusingimagessoundsindirect,
hou-etal-2025-evaluating}
& Injects instructions through audio or analyzed content without preserving source role.
& Untrusted content can steer responses or dialogue as if authoritative.
& Binding among source, session, permission, and execution authority remains weakly evaluated. \\

\rowcolor{gray!15}
\multicolumn{4}{@{}l}{\textbf{State and Reasoning}} \\
\rowcolor{gray!4}
Contextual and paralinguistic manipulation~\cite{shen2024voicejailbreakattacksgpt4o,
roh2025multilingual,li2026stylebreak}
& Changes framing, language, accent, prosody, or style while preserving the harmful objective.
& Weakens refusal and exposes sensitivity to linguistic or paralinguistic variation.
& Mostly immediate-response evaluation; persistent-state and execution effects are rarely tested. \\
\rowcolor{gray!10}
Representation-level and cross-modal manipulation~\cite{geng-etal-2025-mathsf,ma2025audiojailbreakattacksexposing,
krishnan2026optimizingmultimodaljailbreaksspoken,
wang2026acousticinterferencenewparadigm}
& Manipulates latent audio, discrete speech tokens, or audio--text interactions.
& Bypasses transcript-centric safeguards through non-textual representations.
& Often requires internal, tokenization, or optimization access; evaluation is mainly generation-level. \\
\rowcolor{gray!4}
Persistent-state and memory exploitation~\cite{chen2024agentpoison,dong2025minja,
wang2025agentmemoryprivacy,pulipaka2026hiddenmemory}
& Poisons, retrieves, or extracts persistent records reused across interactions.
& Steers later reasoning or actions across sessions or exposes retained private data.
& Evaluation of Audio LLM-based agents is limited; success depends on memory-write, retrieval, and retention policies. \\

\rowcolor{gray!15}
\multicolumn{4}{@{}l}{\textbf{Tool and Action Execution}} \\
\rowcolor{gray!4}
Tool selection and invocation manipulation~\cite{shi2026toolhijacker,zhan-etal-2024-injecagent,
ye-etal-2024-toolsword}
& Manipulates tool descriptions, metadata, or untrusted tool-returned content.
& Redirects tool selection, invocation, or later tool-mediated behavior.
& Mostly text-agent evaluation; acoustic provenance and speaker authority are rarely incorporated. \\
\rowcolor{gray!10}
Authorization and execution-policy failures~\cite{yuan-etal-2024-r-judge,cartagena2026mindgaptextsafety,
li2026aegisgovernanceintegritysecurity}
& Exploits mismatches between model safety judgments and external-operation permissions.
& Produces prohibited or privileged actions despite apparently safe language behavior.
& Few studies jointly bind source identity, instruction authority, and action permission. \\
\rowcolor{gray!4}
Multi-step and cross-tool escalation~\cite{NEURIPS2024_97091a51,yan-etal-2026-evo,
zhang-pei-2026-llm,li2026aegisgovernanceintegritysecurity}
& Propagates malicious influence through tool feedback, state, and successive actions.
& Enables privilege accumulation, action redirection, or persistent-state exfiltration.
& Speech-originated long-horizon attack chains remain sparsely evaluated. \\

\rowcolor{gray!15}
\multicolumn{4}{@{}l}{\textbf{Edge--Cloud Runtime}} \\
\rowcolor{gray!4}
Intermediate-state leakage and split inference~\cite{285505,309752,fan2026doesserverseeunderstanding}
& Targets activations or intermediate states transmitted during split inference.
& Reconstructs input information or infers client-side attributes.
& Audio-specific leakage, including speaker and environmental attributes, remains poorly characterized. \\
\rowcolor{gray!10}
Local runtime compromise~\cite{298278,309838,xiao2026modelspy,294635,255272}
& Exploits local artifacts, side channels, shared resources, or low-level faults.
& Leaks model or internal information or corrupts inference below the model interface.
& Platform and access assumptions vary; heterogeneous runtimes for Audio LLMs remain little studied. \\
\rowcolor{gray!4}
Resource-exhaustion attacks~\cite{haque23_interspeech,
shumailov2021spongeexamplesenergylatencyattacks,
zhang-etal-2025-crabs,chen-etal-2026-naturalsloth}
& Crafts audio or requests that increase decoding, generation, latency, memory, or energy.
& Degrades availability without requiring semantic failure.
& Full sensing--reasoning--execution workloads remain largely untested. \\
\rowcolor{gray!10}
Communication provenance and fallback manipulation~\cite{zhan2026systemslevelattacksurfaceedge}
& Targets routing, failover, coordination, or provenance across local and remote paths.
& Exploits provenance gaps, fallback windows, or policy differences across execution paths.
& Audio-specific state and authority transitions remain largely untested. \\

\rowcolor{gray!15}
\multicolumn{4}{@{}l}{\textbf{Lifecycle Transformation}} \\
\rowcolor{gray!4}
Acoustic backdoors and training-time poisoning~\cite{294504,10096332,lin2025hiddennoiseunveilingbackdoors}
& Associates acoustic, environmental, or paralinguistic triggers with malicious behavior during training or adaptation.
& Preserves benign behavior until a trigger activates targeted failure.
& Physical and complete-agent validation remains limited. \\
\rowcolor{gray!10}
Compromised components and adaptation artifacts~\cite{fortier2026backdoorslivecomponentlevelanalysis,
hanif-etal-2025-trojanwave,liu-etal-2025-loratk,
zhang-pei-2026-llm}
& Compromises reusable components, prompts, parameter-efficient updates, or fine-tuned agent models.
& Survives integration or adaptation and can activate memory- or tool-mediated behavior.
& Coverage varies by artifact; audio-specific PEFT attacks and integrated consequences are less established. \\
\rowcolor{gray!4}
Deployment-transformation attacks~\cite{chen2025roundingguidedbackdoorinjectiondeep,
ma2024quantizationbackdoors}
& Exploits rounding or precision conversion during quantization.
& Creates low-precision backdoor behavior absent or dormant in the inspected full-precision model.
& Validation on transformed artifacts of Audio LLMs and full-agent integration remains limited. \\

\bottomrule
\end{tabularx}
\end{table*}

\subsection{Acoustic Sensing Attacks}
\label{subsec:perception}

\begin{figure}[t]
    \centering
    \includegraphics[width=\columnwidth]{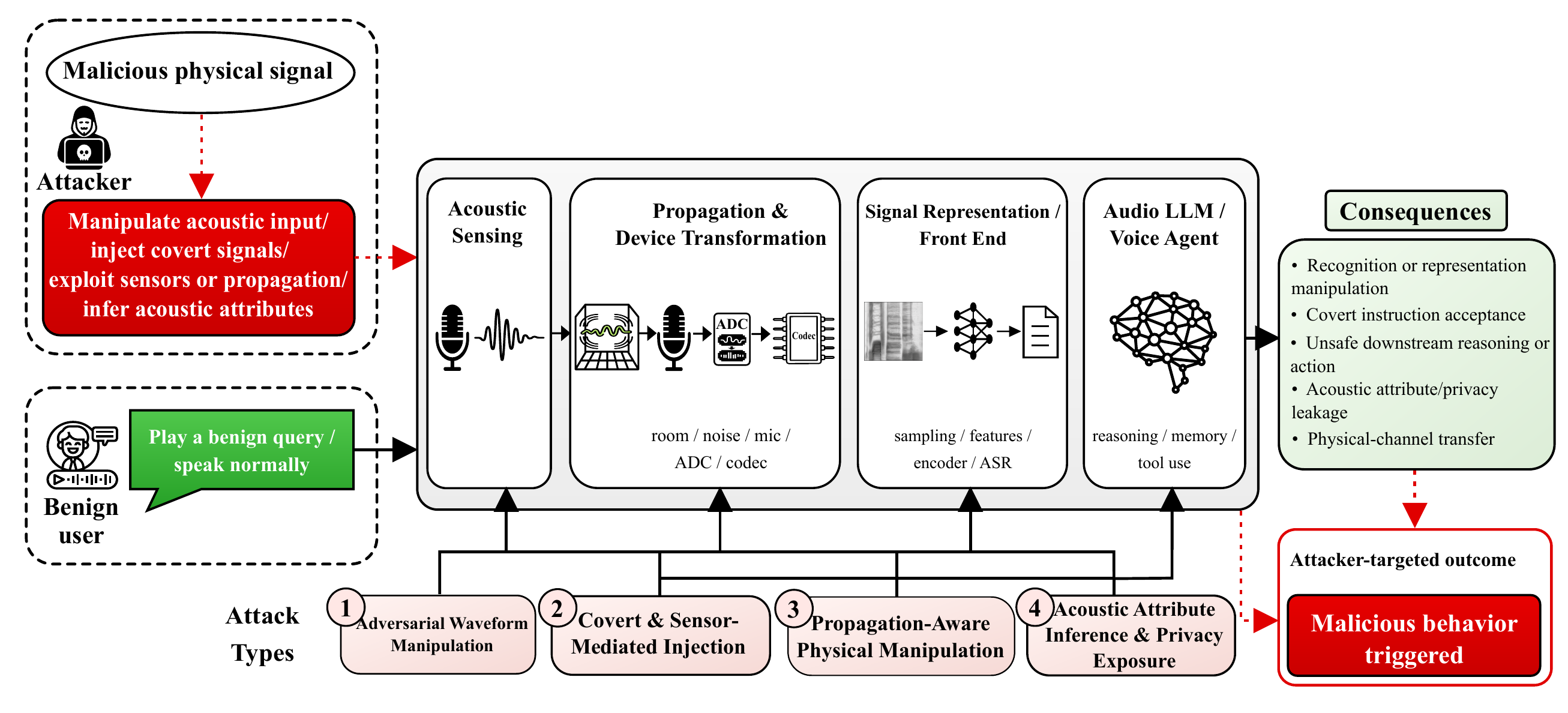}
    \caption{Overview of Acoustic Sensing Attacks.}
    \label{fig:attack_acoustic_sensing}
\end{figure}

At the acoustic boundary, attacks can manipulate the waveform presented
to the system, exploit the physical transduction path of the sensor, or
extract information exposed by the observation itself. Fig.~\ref{fig:attack_acoustic_sensing}
summarizes these entry paths into downstream agent processing.

\paragraph{Adversarial waveform manipulation}

Adversarial audio modifies the received waveform so that machine processing
diverges from the benign acoustic content. Early attacks on automatic speech
recognition demonstrated targeted transcription manipulation and improved
perceptual stealth and robustness through psychoacoustic masking and
simulated acoustic transformations~\cite{carlini2018audioadversarialexamplestargeted,pmlr-v97-qin19a}.
Recent studies extend the attack target from transcription to the behavior of Audio and Speech LLMs, using reusable or representation-aware perturbations that can be
embedded in otherwise benign acoustic content~\cite{kang2025advwave,ma-etal-2025-universal,
ziv2025breakingaudiolargelanguage,
dingeto2026goodsoundsadversarialjailbreaking,
wang2026grmutilityawarejailbreakattacks,
fang2026sparsetokenssufficejailbreaking}.
Such manipulations alter the acoustic representations consumed by
downstream language-model processing.

Their practical significance depends on whether the manipulation survives
transformations encountered before inference. Codec-robust attacks show
that adversarial effects can persist through lossy audio compression~\cite{roh2026codecrobustattacksaudiollms}, whereas physical effectiveness
depends more strongly on the delivery channel and device configuration.

\paragraph{Covert and sensor-mediated injection}

Covert commands can be distorted or embedded in other audio while remaining
actionable to speech-recognition systems~\cite{197215,yuan2018commandersongsystematicapproachpractical}.
Sensor-mediated attacks instead exploit physical transduction through
ultrasonic, structural, or optical carriers~\cite{Zhang_2017,yan2020surfingattack,255276,287266,299709,
ling2026sirenswhisper}. In both cases, the sensing path admits
machine-interpretable input outside ordinary addressed speech.

\paragraph{Propagation-aware physical manipulation}

Physical attacks must remain effective after playback, propagation, and
recapture. Early methods incorporated room and device transformations into
attack construction~\cite{Yakura_2019,chen2020metamorph,10368060}, while
recent studies on Audio LLMs evaluate channel-aware injection and over-the-air
manipulation~\cite{chen2026hijackinglargeaudiolanguagemodels,
sadasivan-etal-2026-attackers}. Their evaluations vary substantially across
physical settings, making deployment-level transferability difficult to
compare.

\paragraph{Acoustic attribute inference and privacy exposure}

Acoustic observations can expose sensitive attributes beyond linguistic
content. HearSay studies privacy leakage from acoustic voiceprints, and
AudioPrivacy extends such profiling to broader vocal and environmental
signals~\cite{wang-etal-2026-hearsay,he-etal-2026-audioprivacy}.

Recent studies on Audio LLMs cover digital manipulation, acoustic privacy
exposure, and selected over-the-air attacks. Cross-device and
cross-environment transferability remains inconsistently evaluated.

\subsection{Provenance and Authority Attacks}
\label{subsec:provenance_authority}

Once acoustic content has been perceived, security depends on how the
system attributes its source and assigns interaction authority. Determining
who produced an utterance, whether it was addressed to the agent, and what
permissions apply together governs whether that input should influence the
interaction. We express this dependency without assuming a particular
recognition architecture. For acoustic observation $x$ in interaction
context $c$, let $G$ infer the source and intended addressee, and let $P$
denote the authorization policy for a requested operation $u$:
\begin{equation}
\begin{aligned}
(\hat{s},\hat{d}) &= G(x,c), \\
\hat{a} &= P(\hat{s},\hat{d},u,c), \qquad
 a^{\star}=P(s^{\star},d^{\star},u,c).
\end{aligned}
\label{eq:provenance_authority}
\end{equation}
Here $(s^{\star},d^{\star})$ denotes the true source and addressee. A
security failure occurs when $\hat{a}=1$ while $a^{\star}=0$. Voice
spoofing, cross-user confusion, and indirect instruction injection differ
in which part of this chain they corrupt, but each can cause unauthorized
content to acquire effective authority.

\paragraph{Voice identity spoofing and impersonation}

Inter-speaker differences in vocal anatomy and habitual speaking patterns
produce speaker-discriminative acoustic characteristics that can support
biometric speaker verification. Replay, speech synthesis, and voice conversion
can reproduce these characteristics and challenge systems that rely on
voice-based authentication. The ASVspoof challenge series evaluates replayed and generated speech
against automatic speaker-verification systems~\cite{yamagishi2021asvspoof}.

In personalized edge deployments, speaker identity may gate user-specific
context and privileged services. Successful impersonation can therefore
expose another user's state or invoke actions under that user's privileges.
Identity attribution determines who is speaking, whereas instruction
authority determines what that identity may control in the current
interaction.

\paragraph{Intended-speaker and cross-user confusion}

Shared acoustic environments introduce ambiguity about which audible speech
belongs to the active interaction. Multiple speakers may be captured by the
same device even though only one is addressing the agent. SH-Bench evaluates
whether Audio LLMs can selectively attend to an intended speaker while
protecting bystander speech~\cite{zhan-etal-2026-protecting}. VoxPrivacy
examines interactional privacy across speakers and reports failures in
conditional and proactive disclosure decisions~\cite{wang2026voxprivacybenchmarkevaluatinginteractional}.

Misassigning bystander or overlapping speech to the active interaction can
cause unrelated content to be incorporated into the active request and alter
downstream agent behavior.

\paragraph{Instruction-authority confusion and indirect audio injection}

Perceived acoustic content must be assigned an interaction role before it
can guide agent behavior. Indirect instruction injection embeds commands in
multimodal content presented for analysis~\cite{bagdasaryan2023abusingimagessoundsindirect}. Earlier
voice-assistant studies also show that remotely delivered speech can
trigger services or devices when presence and access-control assumptions
are weak~\cite{lei2019insecurityhomedigitalvoice}. Hou~\etal~further
evaluate several forms of audio injection and hijacking, finding that
susceptibility varies across models and injection positions~\cite{hou-etal-2025-evaluating}.

Continuous listening increases this exposure because ambient content can
enter the active interaction unless its provenance and role are preserved.

Existing studies treat identity spoofing, intended-speaker selection, and
indirect instruction injection largely as separate problems. Joint
evaluation of source attribution, intended address, instruction authority,
and downstream action in Audio LLM-based agents remains limited.

\subsection{State and Reasoning Attacks}
\label{subsec:context_reasoning}

After acoustic information enters the agent context, attacks may target its
interpretation, integration, or retention during subsequent processing.

\paragraph{Contextual and paralinguistic manipulation}

The behavior of Audio and Speech LLMs can change with both conversational
framing and speech delivery. VoiceJailbreak and \textit{Now You Hear Me}
show that narrative structure can weaken refusal behavior even when the
underlying harmful objective remains unchanged~\cite{shen2024voicejailbreakattacksgpt4o,yu-etal-2026-now}.
Multi-AudioJail and StyleBreak further demonstrate sensitivity to
variations in language, accent, prosody, and speaking style~\cite{roh2025multilingual,li2026stylebreak}. These results indicate that
safety behavior depends not only on propositional content but also on how
that content is contextualized and delivered.

\paragraph{Representation-level and cross-modal manipulation}

Other attacks operate on the representations through which acoustic
information participates in model computation. Recent work manipulates
non-textual audio representations or discrete speech tokens to induce
jailbreak behavior~\cite{geng-etal-2025-mathsf,ma2025audiojailbreakattacksexposing}.
Related studies exploit interactions across audio and text, including
speech--audio composition, joint multimodal optimization, and latent
acoustic interference~\cite{yang2026speechaudiocompositionalattacksmultimodal,
krishnan2026optimizingmultimodaljailbreaksspoken,
wang2026acousticinterferencenewparadigm}. These studies indicate that security-relevant influence can arise from
representations that are not visible in the linguistic transcript. Internal speech representations can
also retain speaker-identifying information, while membership inference
can expose associations between vocal identity and linguistic content~\cite{kuzmin2026privacypreservingendtoendfullduplexspeech,
dong2026membershipinferenceattackslarge}.

\paragraph{Persistent-state and memory exploitation}

Persistent memory converts transient observations into reusable state and
therefore creates attack paths that extend beyond the interaction in which
the information was first observed. AgentPoison demonstrates that poisoning
long-term memory or retrieval knowledge can activate attacker-selected
behavior when malicious records are later retrieved~\cite{chen2024agentpoison}. MINJA further shows that malicious records can
be inserted through query-only interaction, without direct read or write
access to the memory store, and can subsequently induce attacker-directed
reasoning and actions~\cite{dong2025minja}. Recent studies identify multiple
memory-write channels and demonstrate delayed or single-interaction poisoning
that persists across later retrievals~\cite{dash2026untrustedinputtrustedmemory,
pulipaka2026hiddenmemory,tian2026injecmem}.

Persistent state also creates confidentiality exposure. MEXTRA demonstrates
black-box extraction of private information retained in agent memory~\cite{wang2025agentmemoryprivacy}. BackdoorAgent further shows that triggers
introduced at the memory stage can persist through intermediate states and
propagate into later planning and tool-use stages~\cite{feng-etal-2026-backdooragent}. These results establish persistent
memory as an agent-specific security surface with both integrity and privacy
failure modes.

Memory poisoning and extraction have been studied extensively in general
LLM-based agents, but corresponding attacks have rarely been evaluated in Audio
LLM-based agents. The effects of speaker attribution, shared-device interaction, and
audio-derived state on memory admission, retrieval, and cross-session
propagation remain insufficiently evaluated.

\subsection{Tool and Action Execution Attacks}
\label{subsec:agentic}

\begin{figure}[t]
    \centering
    \includegraphics[width=\columnwidth]{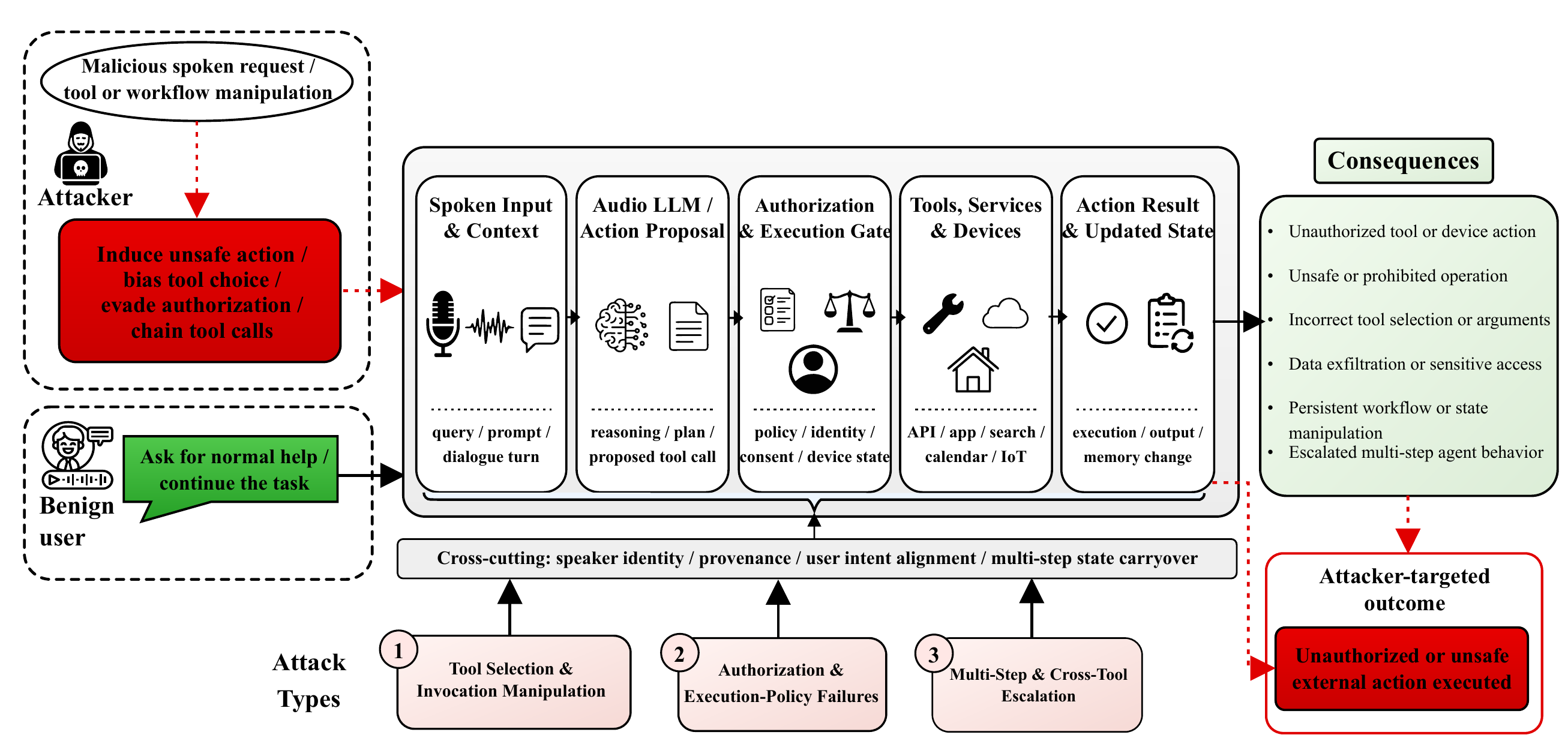}
    \caption{Overview of Tool and Action Execution Attacks.}
    \label{fig:attack_tool_execution}
\end{figure}

External execution creates a distinct attack surface because agent decisions
must be translated into concrete operations. This interface is already
present in speech-driven applications. Audio2Tool evaluates spoken tool use
across smart-car, smart-home, and wearable domains, including direct,
multi-intent, implicit, corrective, and multi-turn requests~\cite{pahwa2026audio2toolspeakcallact}. Weaknesses may therefore arise in tool selection,
execution authorization, or the coordination of actions across multi-step
workflows. Fig.~\ref{fig:attack_tool_execution} summarizes representative
paths from accepted requests to external operations.

\paragraph{Tool selection and invocation manipulation}

Tool-enabled agents can be redirected by manipulating the information used
to select or invoke external capabilities. ToolHijacker poisons tool
descriptions to bias retrieval toward attacker-controlled tools~\cite{shi2026toolhijacker}. InjecAgent shows that malicious instructions
returned through external tool content can redirect subsequent agent
behavior, while ToolSword identifies related failures across tool-learning
and execution workflows~\cite{zhan-etal-2024-injecagent,ye-etal-2024-toolsword}. These attacks
exploit the agent's dependence on tool metadata and tool-returned content
during execution.

\paragraph{Authorization and execution-policy failures}

Correct tool selection still requires the proposed action to satisfy
execution-time permissions and policy constraints. R-Judge reports limited
ability of LLMs to identify behavioral risks from agent histories~\cite{yuan-etal-2024-r-judge}, and \textit{Mind the GAP} shows cases in
which a model verbally refuses a harmful request while still producing a
prohibited tool call~\cite{cartagena2026mindgaptextsafety}. Voice-agent
evaluation further exposes privilege- and governance-related failures during
external execution~\cite{li2026aegisgovernanceintegritysecurity}. These
results separate language-level safety judgments from the authority actually
granted when an operation is executed.

\paragraph{Multi-step and cross-tool escalation}

Execution risk can accumulate as tool outputs, persistent state, and
completed actions alter the context available to later steps. AgentDojo
evaluates indirect prompt injection through untrusted tool outputs, while
cross-tool attacks show that information or malicious influence can
propagate between otherwise separate capabilities~\cite{NEURIPS2024_97091a51,li2025lesdissonancescrosstoolharvesting}.
Evo-Attacker extends this threat model to long-horizon workflows by using
attack memory and reinforcement learning to adapt tool-level interventions
over multiple execution steps~\cite{yan-etal-2026-evo}.

BackdoorAgent extends this pattern across planning, memory, and tool-use
stages~\cite{feng-etal-2026-backdooragent}, while backdoored agents have
been shown to retrieve stored user context and exfiltrate it through disguised
tool calls~\cite{zhang-pei-2026-llm}. Voice-agent evaluations such as Aegis
show that extended execution can expose privilege and resource-related
failures~\cite{li2026aegisgovernanceintegritysecurity}. The security outcome
of an agent workflow can therefore depend on the composition and history of
multiple individually plausible actions.

Tool injection, long-horizon manipulation, and cross-stage propagation have
received substantial attention in general agent-security studies. Speech-originated
attack chains that terminate in externally consequential actions remain
insufficiently evaluated. An unresolved issue is whether acoustic provenance and user authority are
preserved as influence propagates through multi-step execution.

\subsection{Edge--Cloud Runtime Attacks}
\label{subsec:edge_runtime}

Distributing agent functions across local devices and remote services
introduces runtime exposures that do not arise in a purely local pipeline.
These include exchanged intermediate representations, locally accessible
execution state, and coordination between distributed components.

\paragraph{Intermediate-state leakage and split inference}

Partitioned inference can expose representations that retain information
about the original input even when raw data remains local. Studies of split
learning and LLM inference show that transmitted states can support input
reconstruction or reveal client-side information~\cite{285505,309752,fan2026doesserverseeunderstanding}. For Audio LLMs,
encoder or connector representations may additionally preserve speaker and
other acoustic attributes, making the communication interface a potential
privacy boundary.

\paragraph{Local runtime compromise}

On-device execution exposes model artifacts and runtime behavior to
adversaries with access to the local platform. Existing work demonstrates
model extraction and information leakage through software and physical
side channels~\cite{298278,309838,xiao2026modelspy}. Runtime integrity can
also be altered below the model interface through fault injection or
targeted corruption of model parameters~\cite{294635,255272}. These mechanisms become relevant whenever components of Audio LLMs
execute on devices outside a fully trusted computing
environment.

\paragraph{Resource-exhaustion attacks}

Resource constraints create another runtime exposure. Crafted speech can
increase ASR decoding cost, while adversarial requests against language
models can induce excessive computation or prolonged generation~\cite{haque23_interspeech,
shumailov2021spongeexamplesenergylatencyattacks,
zhang-etal-2025-crabs,chen-etal-2026-naturalsloth}. Such attacks are
particularly relevant to edge execution, where sustained latency or energy
consumption can degrade availability even without changing the model's
semantic output.

\paragraph{Communication provenance and fallback manipulation}

Hybrid systems also depend on routing and coordination between local and
remote components. Systems-level analysis of edge agents identifies
provenance, coordination, and fallback as security-sensitive parts of the
runtime~\cite{zhan2026systemslevelattacksurfaceedge}. Weak provenance can
associate exchanged state with the wrong user or component, while fallback
may redirect a request to an execution path governed by different security
policies.

Current studies focus mainly on split learning, on-device machine learning,
and general LLM-based agents. Evidence on these runtime mechanisms in deployed Audio LLM-based agents
remains limited.

\subsection{Lifecycle Transformation Attacks}
\label{subsec:lifecycle}

Not all security failures originate during online operation. Model
preparation, adaptation, integration, and deployment transformation can
alter security behavior before the resulting artifact enters normal use.
A compact abstraction treats deployment as a transformation $\mathcal{T}$
of source parameters $\theta_0$. Let $\xi$ collect benign deployment
choices and let $\Delta$ denote an attacker-controlled change to training
data, components, adaptation artifacts, or transformation rules. The
resulting security objective can be written as
\begin{equation}
\begin{aligned}
\theta_d^{\Delta} &= \mathcal{T}(\theta_0;\xi,\Delta),\\
\max_{\Delta}\quad & \mathcal{R}(\theta_d^{\Delta}) \\
\text{s.t.}\quad & \mathcal{U}(\theta_d^{\Delta})
\ge \mathcal{U}(\theta_d^{0})-\varepsilon,
\end{aligned}
\label{eq:lifecycle_attack}
\end{equation}
where $\mathcal{R}$ measures security risk, $\mathcal{U}$ denotes benign
utility, and $\theta_d^{0}=\mathcal{T}(\theta_0;\xi,0)$. This formulation
captures the common objective of changing deployed security behavior while
leaving ordinary utility sufficiently intact to avoid detection by routine
performance evaluation.

\paragraph{Acoustic backdoors and training-time poisoning}

Backdoor attacks associate an attacker-selected acoustic pattern with
malicious behavior during training or adaptation. Existing studies show
that triggers can be embedded in environmental and paralinguistic
characteristics rather than explicit command content~\cite{294504,10096332,schoof2024emobackdoorattacksspeaker,
cai2022vsvcbackdoorattackkeyword}. Studies on Audio LLMs further demonstrate
backdoors that affect safety alignment through latent acoustic patterns~\cite{lin2025hiddennoiseunveilingbackdoors}. These mechanisms allow
apparently normal model behavior to coexist with trigger-dependent failures
after deployment.

\paragraph{Compromised components and adaptation artifacts}

Modular pipelines for Audio and Speech LLMs create additional entry points
through reusable components and lightweight adaptation artifacts.
Component-level studies show that compromised speech encoders can remain
security-relevant after integration and downstream adaptation~\cite{fortier2026backdoorslivecomponentlevelanalysis,291116}. Learned
prompts and parameter-efficient updates can introduce related triggered
behavior without modifying the complete model~\cite{hanif-etal-2025-trojanwave,liu-etal-2025-loratk}.

Agent-level studies further show that the consequence of upstream
compromise may emerge only after integration with persistent state and tools.
Backdoored fine-tuned agents can invoke memory-access tools and exfiltrate
stored context through disguised tool calls after a semantic trigger is
activated~\cite{zhang-pei-2026-llm}. Security evaluation of an adaptation
artifact therefore depends on the agent workflow in which the resulting
model is deployed.

\paragraph{Deployment-transformation attacks}

Model conversion can also change security behavior after training.
Quantization attacks exploit rounding or precision changes to introduce or
activate behavior that is absent from the original full-precision model~\cite{chen2025roundingguidedbackdoorinjectiondeep,
ma2024quantizationbackdoors}.

Studies on Audio and Speech LLMs in this area primarily address acoustic
backdoors, compromised speech components, and prompt-level adaptation.
Studies of general LLM-based agents demonstrate that backdoors can propagate into persistent
state and tool-mediated execution. Security effects introduced specifically
by deployment transformations, including quantization and hardware-oriented conversion, remain less studied in complete Audio LLM-based agents.

\section{Defenses and Safeguards}
\label{sec:defenses}
Following the security analysis framework in
Section~\ref{subsec:security_framework}, we organize defenses according to
the system area in which protection is primarily enforced. Because complete
agent-level evaluations remain limited, component-level defenses are included
when they protect a security-relevant S2A stage; they are treated as
stage-specific evidence rather than validated end-to-end protection for Audio
LLM-based agents.

\begin{table*}[t]
\centering
\caption{Taxonomy and comparison of defenses for edge-deployed Audio LLM-based agents.}
\label{tab:defense_comparison}
\fontsize{7.0}{7.45}\selectfont
\setlength{\tabcolsep}{2.35pt}
\renewcommand{\arraystretch}{0.96}
\begin{tabularx}{\textwidth}{
    @{}
    L{0.235\textwidth}
    L{0.245\textwidth}
    L{0.225\textwidth}
    Y
    @{}
}
\toprule
\textbf{Defense Type} &
\textbf{Key Mechanism} &
\textbf{Main Strength} &
\textbf{Main Limitation} \\
\midrule

\rowcolor{gray!15}
\multicolumn{4}{@{}l}{\textbf{Acoustic Sensing}} \\
\rowcolor{gray!4}
Adversarial screening and input purification~\cite{272238,wu2023defendingadversarialaudiodiffusion,chen24_interspeech}
& Transformation-based screening or signal reconstruction before inference.
& Model-external protection across multiple perturbation families without retraining Audio LLMs.
& Mostly ASR or classification; purification may remove useful cues, and adaptive codec-robust attacks remain challenging. \\
\rowcolor{gray!10}
Physical-channel and sensor protection~\cite{DBLP:conf/ndss/ZhangJLQX21,285423,299545}
& Uses array attenuation, time--frequency statistics, or carrier artifacts to detect injection.
& Blocks or detects malicious input before reasoning using physical-channel properties.
& Device, sensor, and attack-carrier dependent; transfer to unseen mechanisms is uncertain. \\
\rowcolor{gray!4}
Acoustic privacy transformation~\cite{kuzmin2026streamvoiceanonenhancingutilityrealtime,soh2025securespeech}
& Anonymizes speaker identity and/or transforms sensitive linguistic content.
& Reduces leakage near the sensing boundary and supports streaming protection.
& Privacy--utility--latency trade-offs; may remove speaker or affective cues needed downstream. \\

\rowcolor{gray!15}
\multicolumn{4}{@{}l}{\textbf{Provenance and Authority}} \\
\rowcolor{gray!4}
Selective hearing and intended-address detection~\cite{zhan-etal-2026-protecting,chi2025adaptiveddsd}
& Separates intended speech from bystanders, background conversation, or non-directed speech.
& Reduces unintended admission in shared acoustic environments.
& Does not authenticate identity or establish action authority. \\
\rowcolor{gray!10}
Instruction provenance and trusted-context separation~\cite{307694,jia-etal-2025-task}
& Separates trusted instructions from untrusted data and checks task alignment.
& Constrains indirect injection while preserving a task-level authority boundary.
& Often text-agent based; acoustic source and intended-address metadata may be lost before enforcement. \\
\rowcolor{gray!4}
Voice authenticity and impersonation protection~\cite{307846,307838,11430623}
& Uses anti-cloning, provenance marking, or physical liveness checks.
& Reduces acceptance of replayed, synthesized, or converted speech as legitimate.
& Authenticity does not establish instruction authority; methods may require enrollment, hardware, or channel assumptions. \\

\rowcolor{gray!15}
\multicolumn{4}{@{}l}{\textbf{State and Reasoning}} \\
\rowcolor{gray!4}
Audio-native safety filtering and representation control~\cite{kang2026audioguardcomprehensiveaudiosafety,
yang2026speechaudiocompositionalattacksmultimodal,
djanibekov-etal-2025-spirit,
lin2026sarsteersafeguardinglargeaudiolanguage}
& Uses guard models or activation-level steering before or during generation.
& Covers speech, non-speech harms, compositional attacks, and inference-time refusal control.
& Adds latency or requires model-state access; transfer and over-refusal remain concerns. \\
\rowcolor{gray!10}
Safety hardening during model adaptation~\cite{alexos25_interspeech,yang-etal-2025-reshaping}
& Uses adversarial or safety-oriented fine-tuning during adaptation.
& Embeds robustness in the target model without requiring an external guard at inference.
& Coverage dependent; robustness may reduce benign utility and miss unseen acoustic conditions. \\
\rowcolor{gray!4}
Persistent-state and memory protection~\cite{ICLR2025_8c234d9c,wei2025amemguard,qian-2026-visual,
karamchandani2026forgedreasoning,xu2026memoryprovenancelaundering}
& Sanitizes, validates, or preserves provenance for memory admission, reasoning history, and reuse.
& Checks low-trust state before reuse and can constrain memory-triggered high-risk actions.
& Architecture dependent; trusted metadata must survive summarization, consolidation, and cross-session reuse. \\

\rowcolor{gray!15}
\multicolumn{4}{@{}l}{\textbf{Tool and Action Execution}} \\
\rowcolor{gray!4}
Control and tool-flow constraints~\cite{debenedetti2025defeatingpromptinjectionsdesign,
an-etal-2025-ipiguard,jia-etal-2025-task,
doshi2026verifiablysafetooluse}
& Derives trusted data/control flow, pre-plans dependencies, or checks task contribution.
& Limits untrusted content from redefining execution and reduces reliance on model refusal.
& Assumes a correct trusted task or plan; rigid structures may constrain legitimate dynamic workflows. \\
\rowcolor{gray!10}
Runtime authorization and mediation~\cite{wang2025agentspeccustomizableruntimeenforcement,
lin-etal-2026-vigil,xu2026memoryprovenancelaundering}
& Applies execution-time policy, verify-before-commit, or memory-aware authority checks.
& Separates reasoning from commitment and blocks low-trust state from gaining high-risk authority.
& Depends on accurate policy, intent, and provenance; the mediator becomes a critical enforcement point. \\
\rowcolor{gray!4}
Least privilege and action constraints~\cite{zhu2025miniscopeprivilegeframeworkauthorizing,
ji2026seagent,ravichandran2026safetyguardrailsllmenabledrobots}
& Enforces minimal tool permissions, MAC/ABAC policies, or formal actuator constraints.
& Bounds consequences and blocks privilege escalation despite unsafe proposals.
& Requires reliable tool semantics and policy models; acoustic authority is rarely incorporated. \\

\rowcolor{gray!15}
\multicolumn{4}{@{}l}{\textbf{Edge--Cloud Runtime}} \\
\rowcolor{gray!4}
Privacy-preserving collaborative inference~\cite{fan2026doesserverseeunderstanding,
gu2026differentiallyprivatecommunicationefficient}
& Perturbs, projects, or privatizes intermediate states transmitted in split inference.
& Reduces remote reconstruction while retaining distributed inference.
& Privacy, utility, and communication trade off; acoustic attributes may remain exposed. \\
\rowcolor{gray!10}
Secure local execution~\cite{Mo_2020,10179382,abdollahi2026agentee}
& Uses TEEs or attested confidential VMs to isolate models, runtime state, and applications.
& Provides hardware-backed confidentiality and integrity for local agent components.
& Requires confidential-computing hardware and trusted interfaces; audio buffers and state remain underexplored. \\
\rowcolor{gray!4}
Resource and availability control~\cite{zhang-etal-2025-pd3f}
& Applies resource-aware scheduling, generation budgets, or adaptive termination.
& Bounds excessive computation and availability loss without semantic attack detection.
& May terminate legitimate long tasks; resource accounting for complete Audio LLM-based agents remains incomplete. \\
\rowcolor{gray!10}
Communication integrity and coordination~\cite{zhan2025aegismcponlinegraphintrusion,abdollahi2026agentee}
& Monitors cross-component activity or uses attested, verifiable communication channels.
& Protects integrity and provenance across distributed agent components.
& Dynamic routing, fallback, and synchronization complicate preservation of acoustic authority metadata. \\

\rowcolor{gray!15}
\multicolumn{4}{@{}l}{\textbf{Lifecycle Transformation}} \\
\rowcolor{gray!4}
Backdoor detection and remediation~\cite{wang2026stepdetectingaudiobackdoor,
zhou2025gradientnormbasedfinetuningbackdoor,
fortier2026backdoorslivecomponentlevelanalysis}
& Uses stability or gradient profiling and component-level backdoor analysis.
& Supports pre-deployment detection and remediation, including black-box audio cases.
& Attack- and artifact-dependent; absence of a detected trigger does not establish full-agent safety. \\
\rowcolor{gray!10}
Artifact inspection and provenance~\cite{Sun_2025,wang2024mitigatingdownstreammodelrisks}
& Inspects adapter weights and records lineage across models and components.
& Supports pre-integration screening and supply-chain auditability.
& Provenance does not prove behavioral safety; detector generalization depends on artifact and attack distributions. \\
\rowcolor{gray!4}
Adaptation and transformation validation~\cite{hsu2025safelorasilverlining,
liu2025targetedvaccinesafetyalignment,pmlr-v267-lu25g,
Li_2024_CVPR,zheng2026breakingroundingtrapsecuring}
& Constrains fine-tuning updates and modifies quantization rounding.
& Targets lifecycle stages that can invalidate source-model safety while preserving utility.
& Transformation-specific; protection must be revalidated on the final artifact and hardware path. \\

\bottomrule
\end{tabularx}
\end{table*}

\subsection{Acoustic Sensing Defenses}
\label{subsec:defense_input}

Protection begins at the acoustic boundary, where received audio and the
sensing path can be examined or transformed before subsequent agent
processing.

\paragraph{Adversarial screening and input purification}

Transformation-based screening detects suspicious inputs from changes in
model behavior under controlled audio transformations. WaveGuard applies
this principle to adversarial ASR inputs, while FraudWhistler evaluates
prediction stability in speaker-recognition systems~\cite{272238,294508}. Purification methods instead reconstruct the received
signal before inference. AudioPure uses diffusion-based reconstruction, and
TDAP introduces audio-language guidance to preserve task-relevant content
during purification~\cite{wu2023defendingadversarialaudiodiffusion,
chen24_interspeech}. Their use with Audio LLMs requires preserving acoustic
information needed by downstream tasks. Routine codec processing provides
limited protection, as codec-robust attacks can remain effective after
lossy compression~\cite{roh2026codecrobustattacksaudiollms}.

\paragraph{Physical-channel and sensor protection}

Physical-channel defenses detect artifacts introduced when malicious
signals are converted into microphone input. EarArray uses attenuation
differences across microphones to detect ultrasonic commands~\cite{DBLP:conf/ndss/ZhangJLQX21}. NormDetect models the time--frequency
distribution of benign audio, whereas MicGuard identifies residual carrier
and spectral artifacts associated with several out-of-band injection
mechanisms~\cite{285423,299545}. Their effectiveness depends on the sensing
configuration and the physical characteristics of the attack channel.

\paragraph{Acoustic privacy transformation}

Input processing can also reduce identity-bearing information before audio
is used by later components. Stream-Voice-Anon performs real-time speaker
anonymization while preserving linguistic utility~\cite{kuzmin2026streamvoiceanonenhancingutilityrealtime}. SecureSpeech
extends this objective by transforming both speaker identity and sensitive
textual information~\cite{soh2025securespeech}. These methods place privacy
protection close to the point where sensitive acoustic information first
enters the system.

Most defenses in this area have been evaluated on speech recognition,
speaker verification, or speech privacy tasks. Their effectiveness in Audio
LLM-based agents, where non-linguistic and paralinguistic information may
also be required by downstream processing, remains less established.

\subsection{Provenance and Authority Defenses}
\label{subsec:defense_provenance}

For provenance and authority, the central objective is to preserve the link
between perceived speech, its source, and the permissions granted within the
current interaction. In the notation of Eq.~\eqref{eq:provenance_authority},
protection must prevent errors in $\hat{s}$ or $\hat{d}$ from being
converted into $\hat{a}=1$ for an unauthorized request.

\paragraph{Selective hearing and intended-address detection}

Shared acoustic environments require the system to distinguish speech that
belongs to the active interaction from speech that is merely audible.
Bystander Privacy Fine-Tuning (BPFT) trains Audio LLMs to suppress
bystander-related responses while preserving comprehension of the intended
speaker~\cite{zhan-etal-2026-protecting}. Device-directed speech detection provides a complementary voice-assistant
mechanism. Adaptive knowledge distillation can distinguish user queries
addressed to the device from background speech and side conversations while
retaining a lightweight deployment target~\cite{chi2025adaptiveddsd}. These mechanisms reduce
unintended admission of ambient speech, but they do not by themselves
authenticate the speaker or determine whether an admitted request is
authorized to trigger an external action.

\paragraph{Instruction provenance and trusted-context separation}

Instruction provenance controls which perceived content may guide agent
behavior after it enters context. StruQ separates trusted instructions from untrusted data and trains the
model to preserve that distinction~\cite{307694}. At the agent level, Task
Shield checks whether each instruction and tool call serves the
user-specified objective, thereby constraining indirect instructions that
deviate from the trusted task~\cite{jia-etal-2025-task}. Applying these mechanisms to Audio LLM-based
agents requires an additional binding step. Source and
intended-address information from the acoustic interaction must remain
associated with the content as it is converted into context and later
execution requests.

\paragraph{Voice authenticity and impersonation protection}

Voice-identity defenses protect the credentials used to associate speech
with a user. Existing approaches include proactive resistance to voice
cloning, provenance marking for synthesized or adapted speech, and physical
liveness verification~\cite{307846,307838,11430623}. These mechanisms reduce the likelihood that a
reproduced or generated voice is accepted under another user's identity.

Current defenses address selective hearing, device-directed speech,
instruction provenance, and voice authenticity through separate mechanisms.
Integrated protection that binds acoustic source, intended address, user
authority, and downstream action permission remains limited.

\subsection{State and Reasoning Defenses}
\label{subsec:defense_context_reasoning}

State and reasoning safeguards focus on acoustic information after it enters
the agent, including how it is interpreted, incorporated into context, and
retained for later use.

\paragraph{Audio-native safety filtering and representation control}

Audio-native guardrails inspect information that may be missed by
transcript-only moderation. AudioGuard combines waveform-level detection
with policy-grounded content screening, while SALMONN-Guard jointly inspects
speech, non-speech audio, and text to detect compositional harms~\cite{kang2026audioguardcomprehensiveaudiosafety,
yang2026speechaudiocompositionalattacksmultimodal}. Other defenses intervene
directly in internal representations. SPIRIT performs post-hoc activation
patching against speech jailbreaks, and SARSteer derives refusal steering
directions while ablating components associated with benign speech to reduce
over-refusal~\cite{djanibekov-etal-2025-spirit,
lin2026sarsteersafeguardinglargeaudiolanguage}. These approaches operate at inference time, although guard models add
latency and representation-level methods require access to internal model
states.

\paragraph{Safety hardening during model adaptation}

Safety behavior can also be strengthened before deployment by adapting the
target model against harmful speech inputs. Existing methods use adversarial
training or safety-oriented fine-tuning to improve robustness while
controlling benign over-refusal~\cite{alexos25_interspeech,yang-etal-2025-reshaping}. Their effectiveness
depends on how well the adaptation data represents the acoustic and
behavioral variations encountered during use.

\paragraph{Persistent-state and memory protection}

Persistent interaction requires controls over memory admission, provenance,
retention, retrieval, and reuse. Proactive Privacy Amnesia identifies and replaces privacy-sensitive
memories to reduce later disclosure~\cite{ICLR2025_8c234d9c}. Agent-specific defenses
address memory integrity more directly. A-MemGuard combines consensus-based
validation with a dual-memory structure that records corrective lessons after
detected failures~\cite{wei2025amemguard}, while CognitiveGuard combines
perceptual sanitization before storage with retrieval-time consistency checks
for multimodal long-term memory~\cite{qian-2026-visual}. SENTINEL targets a
different failure mode involving forged reasoning traces. Its reasoning guard analyzes
candidate memory entries for structural evidence of fabricated rationale
before those entries are reused by later agent decisions~\cite{karamchandani2026forgedreasoning}.

Memory protection also requires preserving the authority attached to stored
information. PPMF keeps provenance outside lossy memory consolidation and
gates tool calls according to both action risk and the authority of
action-relevant memories~\cite{xu2026memoryprovenancelaundering}. This design
connects persistent-state integrity with later execution authorization by ensuring that
memory consolidation does not promote a low-trust observation into
high-authority user history. Current memory defenses remain largely non-audio and depend on memory
architectures, provenance metadata, and trusted enforcement points that are
not yet standardized across Audio LLM-based agents.

Immediate safety filtering and representation-level control are now studied
directly in Audio and Speech LLMs, while persistent-memory defenses increasingly
address memory integrity, reasoning history, and provenance preservation.
Shared-device state and authority carried from stored acoustic observations
into later actions remain largely unevaluated.

\subsection{Tool and Action Execution Defenses}
\label{subsec:defense_execution}

\begin{figure}[t]
    \centering
    \includegraphics[width=\columnwidth]{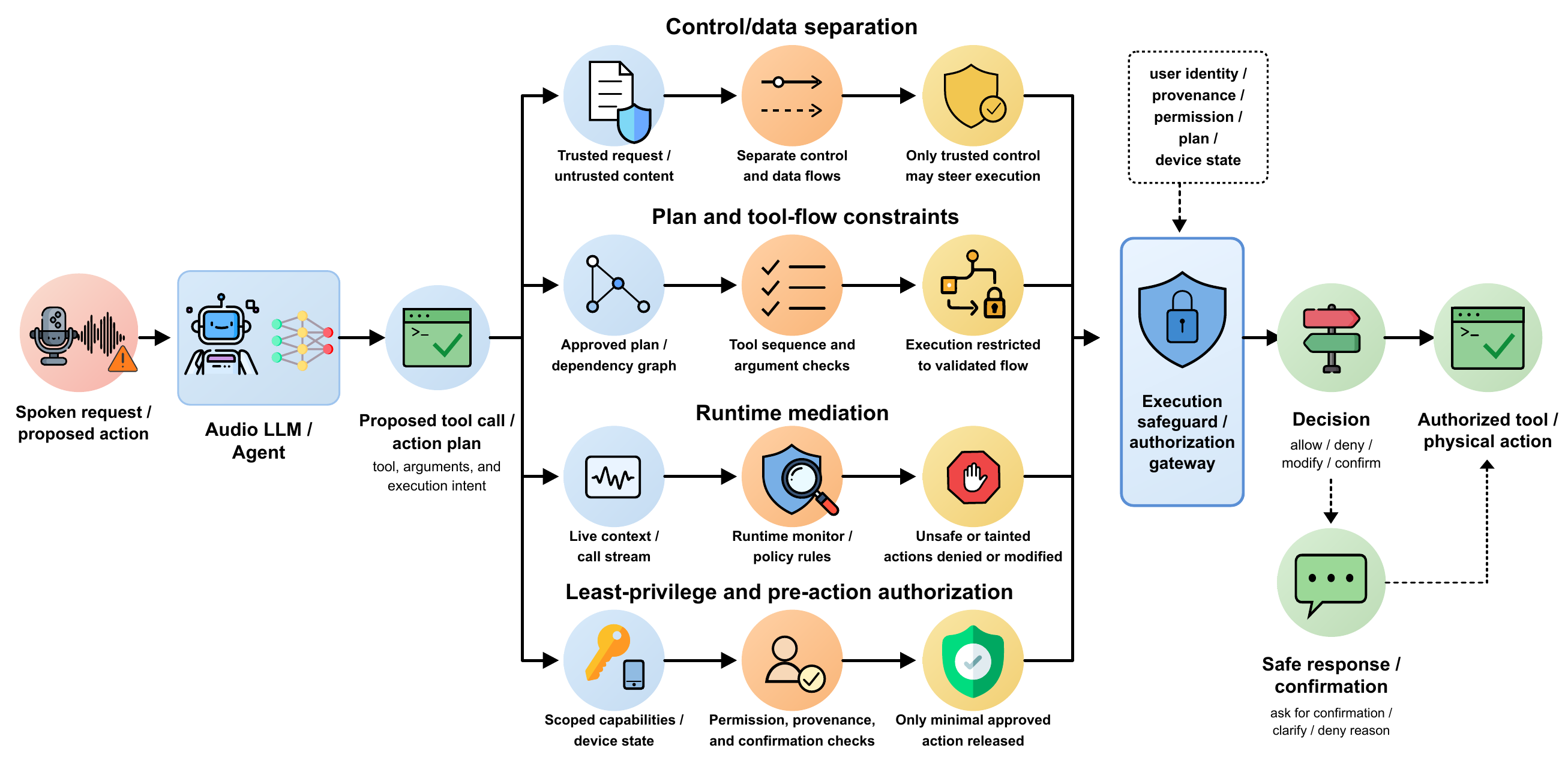}
    \caption{Overview of Tool and Action Execution Defenses.}
    \label{fig:defense_tool_execution}
\end{figure}

At the execution boundary, safeguards constrain which tool calls or external
actions may proceed. Representative enforcement points span planning,
runtime mediation, and execution authorization, as summarized in
Fig.~\ref{fig:defense_tool_execution}.

\paragraph{Control and tool-flow constraints}

Structural defenses restrict how untrusted information can redefine an
agent's execution plan. CaMeL derives control and data flows from the trusted
user request and applies capability constraints to tool calls~\cite{debenedetti2025defeatingpromptinjectionsdesign}. IPIGuard plans a
Tool Dependency Graph before interacting with untrusted content and restricts
later execution to that structure~\cite{an-etal-2025-ipiguard}. Task Shield
takes a complementary goal-alignment approach by checking whether each
instruction and tool call contributes to the user-specified task~\cite{jia-etal-2025-task}. Related work formalizes constraints over tool
sequences and information flows~\cite{doshi2026verifiablysafetooluse}.
Together, these mechanisms reduce the ability of untrusted tool outputs to
silently redefine the task being executed.

\paragraph{Runtime authorization and mediation}

Execution-time mediation checks proposed operations against trusted policy
before their effects are committed. AgentSpec provides a specification
language for enforcing runtime constraints as agent state and permissions
evolve~\cite{wang2025agentspeccustomizableruntimeenforcement}. VIGIL
addresses tool-stream injection with a verify-before-commit design in which the
agent may reason speculatively over dynamic tool feedback, but execution is
conditioned on an intent-grounded verification step~\cite{lin-etal-2026-vigil}. These mechanisms separate reasoning flexibility
from execution authority. PPMF provides a complementary memory-aware
authorization mechanism by conditioning high-risk tool use on the preserved
authority of action-relevant memories~\cite{xu2026memoryprovenancelaundering}. Their effectiveness nevertheless
depends on the quality and coverage of trusted policies, provenance metadata,
or intent representations used by the mediator.

\paragraph{Least privilege and action constraints}

Capability restriction bounds the authority available to an agent even when
its proposed action is unsafe. MiniScope reconstructs permission
relationships among tool calls and grants only the capabilities required for
the current task~\cite{zhu2025miniscopeprivilegeframeworkauthorizing}.
SEAgent formulates agent-tool security as privilege control and combines
Mandatory Access Control (MAC) with Attribute-Based Access Control (ABAC) policies over an information
flow graph~\cite{ji2026seagent}. At the physical-action boundary, RoboGuard
applies formal safety constraints before robot actions are executed~\cite{ravichandran2026safetyguardrailsllmenabledrobots}.

Structural tool-flow constraints, runtime mediation, and least-privilege
enforcement have received substantial attention in agent-security studies. Existing
execution controls rarely consume acoustic provenance, intended-address
information, or speaker-specific authority as first-class authorization
inputs.

\subsection{Edge--Cloud Runtime Defenses}
\label{subsec:defense_runtime}

When execution is distributed, protection must extend beyond the model to
the runtime connecting agent components. The main concerns are leakage from
collaborative inference, compromise of local state, and failures in
communication or resource control.

\paragraph{Privacy-preserving collaborative inference}

Split inference requires intermediate representations to cross the
local--remote boundary. PriPert adapts perturbation strength to the
reconstruction vulnerability of different layers, while DEL combines
embedding projection with differentially private communication~\cite{fan2026doesserverseeunderstanding,
gu2026differentiallyprivatecommunicationefficient}. These approaches reduce
information recoverable from transmitted states. Their application to Audio
LLMs must also account for acoustic attributes retained in intermediate
representations.

\paragraph{Secure local execution}

Trusted execution can isolate security-sensitive computation on devices
that otherwise expose model artifacts and runtime state. DarkneTZ places
selected neural-network layers inside a Trusted Execution Environment (TEE), while
ShadowNet protects sensitive operations and securely outsources expensive
computation to untrusted accelerators~\cite{Mo_2020,10179382}. AgenTEE
extends confidential execution to the agent level by placing the agent
runtime, inference engine, and third-party applications in independently
attested Virtual Machines (VMs) connected through explicit,
verifiable channels~\cite{abdollahi2026agentee}. However, continuous
acoustic buffers and state specific to Audio LLMs have not yet been studied
under comparable deployment conditions.

\paragraph{Resource and availability control}

Runtime safeguards can bound computation consumed by adversarial or
pathological requests. PD$^{3}$F combines resource-aware scheduling with
adaptive termination to control excessive LLM generation~\cite{zhang-etal-2025-pd3f}. Edge deployments of Audio LLMs require analogous limits
for continuously arriving audio and long-running agent interactions, where
resource pressure can affect the availability of the local service.

\paragraph{Communication integrity and coordination}

Hybrid execution requires local and remote components to exchange state
under consistent provenance and policy. AegisMCP monitors tool-mediated edge-agent activity through a streaming
heterogeneous graph~\cite{zhan2025aegismcponlinegraphintrusion}. AgenTEE
provides a complementary mechanism on a single edge device by requiring
mutually attested components and verifiable communication channels~\cite{abdollahi2026agentee}. Comparable protection for hybrid Audio
LLM-based agents must additionally preserve source and authority associations
when state is synchronized, routed, or recovered through fallback paths.

Current runtime defenses cover split-inference privacy, resource control,
confidential execution, and communication monitoring. Audio-specific
evaluation remains limited, especially for continuous sensing, state
synchronization, and policy-preserving local--cloud fallback.

\subsection{Lifecycle Transformation Defenses}
\label{subsec:defense_lifecycle}

\begin{figure}[t]
    \centering
    \includegraphics[width=\columnwidth]{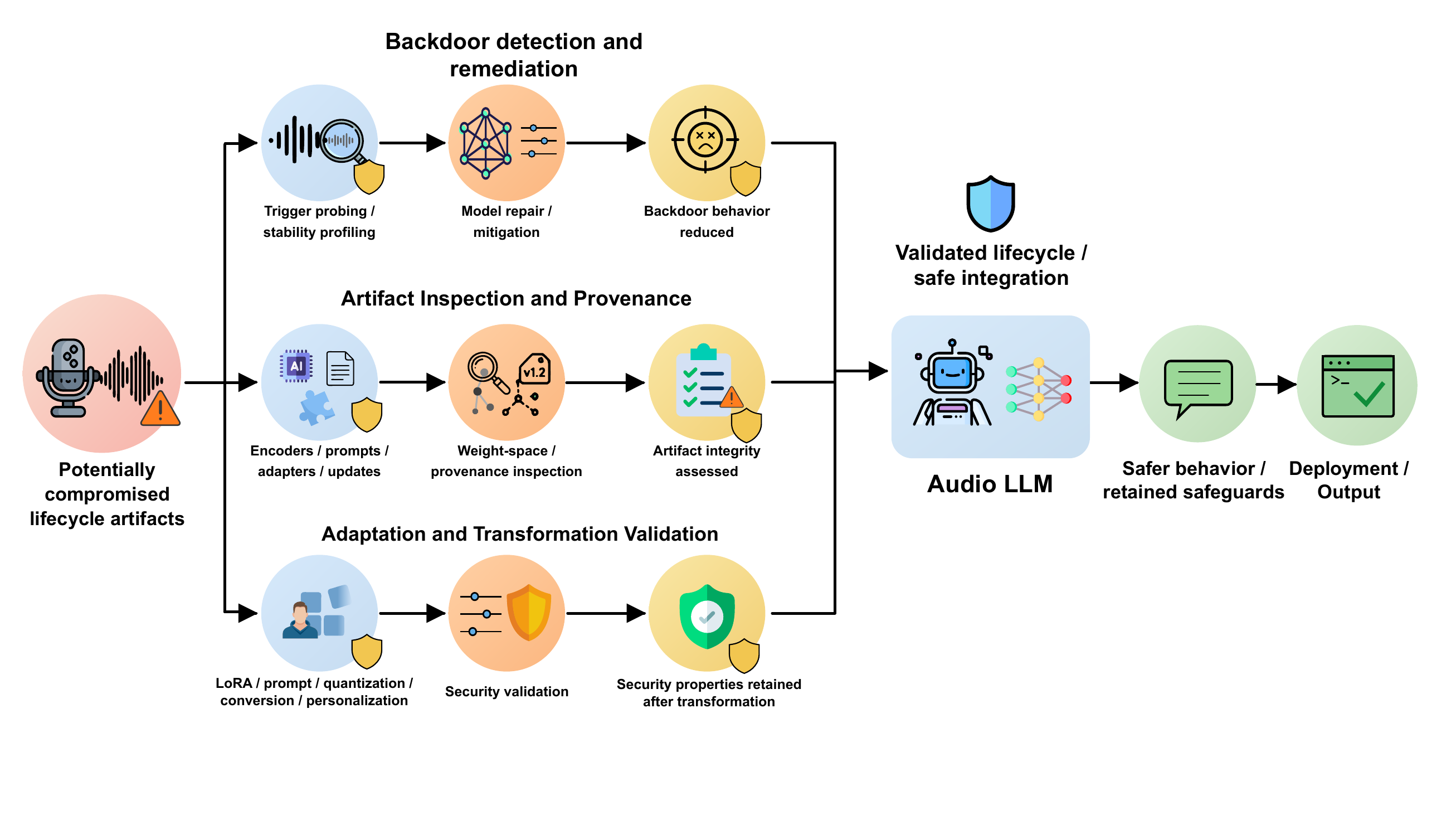}
    \caption{Overview of Lifecycle Transformation Defenses.}
    \label{fig:defense_lifecycle}
\end{figure}

Lifecycle defenses address the model artifacts produced through training,
adaptation, integration, and deployment transformation. Their purpose is to
detect malicious modifications and preserve security properties as the model
is converted into its deployed form. Relative to
Eq.~\eqref{eq:lifecycle_attack}, these safeguards either constrain
attacker-controlled changes $\Delta$ or validate the resulting artifact
$\theta_d$ before deployment. The corresponding safeguards span artifact
inspection, remediation, and post-transformation validation, as summarized
in Fig.~\ref{fig:defense_lifecycle}.

\paragraph{Backdoor detection and remediation}

Backdoor defenses identify trigger-dependent behavior or repair compromised
models before deployment. STEP detects audio backdoors through black-box
stability profiling, while GN-FT suppresses compromised behavior by
regularizing abnormal gradient patterns~\cite{wang2026stepdetectingaudiobackdoor,
zhou2025gradientnormbasedfinetuningbackdoor}. Component-level analysis of
pipelines for Speech LLMs further shows that backdoor persistence depends on
which module is compromised~\cite{fortier2026backdoorslivecomponentlevelanalysis}. Validation should therefore assess the assembled model rather than isolated
components alone.

\paragraph{Artifact inspection and provenance}

Externally obtained components and lightweight adaptations can be examined
before integration. PEFTGuard detects backdoored parameter-efficient
adapters from their weight patterns~\cite{Sun_2025}, while model-provenance mechanisms record relationships
among upstream models and downstream derivatives~\cite{wang2024mitigatingdownstreammodelrisks}. These mechanisms provide complementary checks on artifact integrity and
origin before the deployed configuration is formed.

\paragraph{Adaptation and transformation validation}

Model customization can alter previously established safety behavior. Safe
LoRA constrains parameter-efficient updates toward safety-aligned
directions, Targeted Vaccine protects selected safety-critical layers, and
Safe Delta adjusts post-fine-tuning parameter changes~\cite{hsu2025safelorasilverlining,
liu2025targetedvaccinesafetyalignment,pmlr-v267-lu25g}.
For quantized deployments, EFRAP modifies rounding to suppress
quantization-conditioned backdoors, while QuantGuard controls rounding
behavior to disrupt backdoor activation~\cite{Li_2024_CVPR,zheng2026breakingroundingtrapsecuring}. Security testing should therefore be
applied to the artifact that will actually be deployed.

Current studies on Audio and Speech LLMs emphasize backdoor detection and component-level analysis. Fine-tuning safeguards, adapter-screening methods, and quantization-aware defenses address later lifecycle stages, but few studies validate the complete transformed configurations of Audio LLM-based agents after integration.

\section{Open Issues}
\label{sec:openissues}
\begin{table*}[!t]
\centering
\caption{Representative evaluation resources relevant to security of edge-deployed Audio LLM-based agents, organized by evaluation problem.}
\label{tab:edge_evaluation}

\footnotesize
\renewcommand{\arraystretch}{0.82}
\setlength{\tabcolsep}{2.35pt}
\setlength{\aboverulesep}{0.25ex}
\setlength{\belowrulesep}{0.25ex}

\resizebox{\textwidth}{!}{
\begin{tabular}{
p{3.1cm}
c
p{5.8cm}
c c c c c c c
}
\toprule

\multirow{2}{*}{\textbf{Benchmark}} &
\multirow{2}{*}{\textbf{Year}} &
\multirow{2}{*}{\textbf{Reported Measures}} &
\multicolumn{7}{c}{\textbf{Evaluation Coverage}} \\

\cmidrule(lr){4-10}

&&&
\textbf{Sec} &
\textbf{Ac} &
\textbf{Util} &
\textbf{Priv} &
\textbf{Act} &
\textbf{Res} &
\textbf{Edge} \\

\midrule

\rowcolor{gray!15}
\multicolumn{10}{l}{
\textit{Audio Jailbreak and Adversarial Safety}
} \\

JALMBench
\cite{peng2026jalmbenchbenchmarkingjailbreakvulnerabilities}
&
2026
&
Attack success rate and attack efficiency
&
E & P & P & -- & -- & -- & -- \\

\rowcolor{RowGray}
AJailBench
\cite{song-etal-2026-audio}
&
2026
&
Attack success, transferability, relevance, and perceptual similarity
&
E & P & E & -- & -- & -- & -- \\

Jailbreak-AudioBench
\cite{cheng2025jailbreakaudiobench}
&
2025
&
Attack success under speed, pitch, emotion, and accent variations
&
E & P & P & -- & -- & -- & -- \\

\rowcolor{RowGray}
AudioJailbreak
\cite{chen2026audiojailbreakjailbreakattacksendtoend}
&
2026
&
Attack success, stealthiness, transferability, and over-the-air success
&
E & E & P & -- & -- & P & P \\

Cost-aware attack--defense study
\cite{feng2026audiojailbreakslargeaudiolanguage}
&
2026
&
Attack success, benign refusal, defense effectiveness, and latency
&
E & P & E & -- & -- & E & P \\

\midrule

\rowcolor{gray!15}
\multicolumn{10}{l}{
\textit{Comprehensive Safety and Trustworthiness}
} \\

AudioTrust
\cite{li2026audiotrustbenchmarkingmultifacetedtrustworthiness}
&
2026
&
Safety, privacy, authentication, robustness, and fairness measures
&
E & P & P & E & P & -- & -- \\

\rowcolor{RowGray}
Omni-SafetyBench
\cite{pan2025omnisafetybenchbenchmarksafetyevaluation}
&
2025
&
Cross-modal safety, refusal behavior, and multimodal risk evaluation
&
E & P & P & P & P & -- & -- \\

\midrule

\rowcolor{gray!15}
\multicolumn{10}{l}{
\textit{Privacy, Identity, and Multi-User Interaction}
} \\

SH-Bench
\cite{zhan-etal-2026-protecting}
&
2026
&
Selective efficacy, task accuracy, and bystander information leakage
&
P & P & E & E & -- & -- & -- \\

\rowcolor{RowGray}
VoxPrivacy
\cite{wang2026voxprivacybenchmarkevaluatinginteractional}
&
2026
&
Interactional privacy and speaker-conditioned information-flow control
&
P & P & E & E & -- & -- & -- \\

VoicePrivacy
\cite{tomashenko2024voiceprivacy}
&
2024
&
Identity privacy, speech utility, and attacker-knowledge settings
&
P & P & E & E & -- & -- & -- \\

\midrule

\rowcolor{gray!15}
\multicolumn{10}{l}{
\textit{Acoustic Robustness and Real-World Speech}
} \\

RSA-Bench
\cite{zhang-etal-2026-rsa}
&
2026
&
WER, task accuracy, and LLM-based reasoning evaluation
&
P & E & E & -- & -- & -- & -- \\

\rowcolor{RowGray}
WildSpeech-Bench
\cite{zhang2025wildspeechbenchbenchmarkingendtoendspeechllms}
&
2025
&
LLM-based evaluation, speech quality, and real-world variability
&
P & E & E & -- & -- & -- & -- \\

\midrule

\rowcolor{gray!15}
\multicolumn{10}{l}{
\textit{Long-Context, Persistent State, and Deployment Efficiency}
} \\

ChronosAudio
\cite{luo2026chronosaudiocomprehensivelongaudiobenchmark}
&
2026
&
Long-audio accuracy, temporal localization, and degradation measures
&
-- & P & E & -- & -- & P & -- \\

\rowcolor{RowGray}
AudioMarathon
\cite{he2025audiomarathoncomprehensivebenchmarklongcontext}
&
2025
&
Long-context accuracy, latency, and memory consumption
&
-- & P & E & -- & -- & E & -- \\

\rowcolor{RowGray}
PersistBench
\cite{pulipaka2026persistbenchlongtermmemoriesforgotten}
&
2026
&
Cross-domain memory leakage and memory-induced sycophancy
&
P & -- & P & E & -- & -- & -- \\

PerMemSafe
\cite{an-etal-2026-permemsafe}
&
2026
&
Implicit personalized safety across long-horizon self-evolving agents
&
E & -- & E & P & P & -- & -- \\

URO-Bench
\cite{yan-etal-2025-uro}
&
2025
&
LLM-based quality, UTMOS, WER/CER, and first-packet latency
&
P & P & E & -- & -- & E & -- \\

\rowcolor{RowGray}
VocalBench
\cite{liu2026vocalbenchbenchmarkingvocalconversational}
&
2026
&
Accuracy, speech quality, WER, refusal rate, and preserve rate
&
P & P & E & -- & -- & P & -- \\

\midrule

\rowcolor{gray!15}
\multicolumn{10}{l}{
\textit{Voice Agents, Tool Use, and Adaptive Red Teaming}
} \\

Aegis
\cite{li2026aegisgovernanceintegritysecurity}
&
2026
&
Pipeline-oriented red teaming, privacy leakage, and unsafe agent behavior
&
E & P & P & E & E & P & P \\

\rowcolor{RowGray}
Audio2Tool
\cite{pahwa2026audio2toolspeakcallact}
&
2026
&
Speech-based tool-call accuracy and argument generation
&
-- & P & E & -- & P & -- & -- \\

AgentDojo
\cite{NEURIPS2024_97091a51}
&
2024
&
Prompt-injection success, utility, and defense effectiveness
&
E & -- & E & P & E & -- & -- \\

\rowcolor{RowGray}
GAP
\cite{cartagena2026mindgaptextsafety}
&
2026
&
Text--tool safety divergence and forbidden tool-call behavior
&
E & -- & P & P & E & -- & -- \\

ACIArena
\cite{an-etal-2026-aciarena}
&
2026
&
Cascading injection across external inputs, agent profiles, and inter-agent messages
&
E & -- & P & P & E & -- & -- \\

\rowcolor{RowGray}
AutoDojo
\cite{ma2026autodojoadaptiveblackboxattacks}
&
2026
&
Adaptive attack success and defense degradation
&
E & -- & P & P & E & -- & -- \\

\bottomrule
\end{tabular}
}

\vspace{1pt}
\parbox{\textwidth}{
\footnotesize
\textit{Note:}
E = Explicit Evaluation; P = Partial/Indirect Evaluation;
-- = Not Substantially Examined. Sec = Security Effectiveness; Ac = Acoustic/Physical Validity;
Util = Benign Utility; Priv = Privacy/Identity;
Act = Tool/Action Consequences; Res = Resource/Runtime Evaluation;
Edge = Final On-Device, Embedded, Wearable, or Comparable Deployment.
}
\end{table*}

Table~\ref{tab:edge_evaluation} summarizes how current evaluation resources
cover different parts of the deployed-agent workflow. Acoustic robustness,
jailbreak behavior, privacy, persistent state, tool use, and runtime behavior
are typically evaluated by separate benchmarks. Few resources jointly test
how failures propagate across these components in deployed Audio LLM-based agents.

\subsection{Evaluation Gaps}

\noindent\textbf{1) Configuration-Complete Deployment Evaluation.}

Security results obtained on a source model do not automatically characterize
the configuration that executes at the edge. Tiny-Align illustrates
on-device adaptation~\cite{qin2025tinyalignbridgingautomaticspeech}, while
quantization studies demonstrate that precision conversion can change security
behavior~\cite{chen2025roundingguidedbackdoorinjectiondeep,
ma2024quantizationbackdoors}. Codec processing can also alter attack
behavior~\cite{roh2026codecrobustattacksaudiollms}, and hybrid systems such
as ASTA distribute execution across local and remote components~\cite{torkamani2025adaptiveedgecloudinferencespeechtoaction}. Physical evaluation additionally requires explicit reporting of
playback, propagation, sensing, and preprocessing conditions. Evaluation
should therefore target the final deployed configuration and report security
together with benign utility, latency, memory, communication, and energy.
For hybrid agents, evaluations should separately characterize local, remote, offline, and fallback execution modes because each may exhibit different routing, policy, latency, resource, and security behaviors.

\noindent\textbf{2) Cross-Boundary and Adaptive Agent Evaluation.}

Static, single-stage benchmarks do not capture attacks whose influence
propagates across context, memory, tools, or cooperating components.
AgentDojo~\cite{NEURIPS2024_97091a51} and PIArena~\cite{geng-etal-2026-piarena} show that prompt-injection defenses can change
substantially under adaptive attacks and heterogeneous tasks, while ACIArena~\cite{an-etal-2026-aciarena} evaluates malicious instructions that cascade
through external inputs, agent profiles, and inter-agent messages. For Audio
LLM-based agents, evaluation should therefore track complete attack
trajectories across stages. An acoustic observation may first enter context,
later be stored, cross a runtime boundary, and only then influence an external
operation. Benchmarks should record where influence is introduced, where it is
amplified or filtered, and which downstream effects are reached.

\noindent\textbf{3) Stateful and Multi-Session Evaluation.}

Persistent state introduces a temporal dimension that one-turn safety tests
cannot measure. PersistBench identifies cross-domain leakage and
memory-induced sycophancy from long-term memories~\cite{pulipaka2026persistbenchlongtermmemoriesforgotten}, and PerMemSafe
shows that personalized safety risks can emerge gradually as self-evolving
agents accumulate user-specific memory~\cite{an-etal-2026-permemsafe}.
Evaluation of Audio LLM-based agents should additionally test who caused a
memory write,
which user or session owns the resulting state, whether provenance survives
consolidation, and how retained state affects later users and actions.
Relevant metrics therefore extend beyond immediate attack success to include
unauthorized retention, cross-user retrieval, delayed activation, provenance
loss, and recovery after deletion or correction.

\noindent\textbf{4) Consequence-Aware and Reproducible System Evaluation.}

Model refusal is not an adequate proxy for agent safety. GAP demonstrates
cases in which text refuses a harmful request while a tool call still
performs the prohibited operation~\cite{cartagena2026mindgaptextsafety}.
Evaluation should therefore separate perceived input, accepted instruction,
state update, proposed action, authorization decision, executed action, and
observable consequence. The tested attacker model and configuration should
also be reported explicitly, including model and safeguard versions, memory
state, tool permissions, routing mode, and deployment artifact. This
separation supports reproducible analysis of where unsafe influence enters,
propagates, and reaches execution.

The preceding gaps concern how deployed agents should be evaluated. We next
turn to security properties and mechanisms that current systems themselves
still lack.

\subsection{Open Security Problems}

\noindent\textbf{1) Instruction Provenance and Authority.}

The acoustic channel exposes an agent to intended users, bystanders, media,
replay or synthesized speech, and other environmental sources. SH-Bench~\cite{zhan-etal-2026-protecting} studies selective hearing in shared
speaker settings, while device-directed speech detection~\cite{chi2025adaptiveddsd} addresses whether speech is intended for the
device. Neither question alone determines whether an observation is
authorized to affect the current task. This distinction becomes more
consequential in smart-home control, where permitted actions can depend on
current device state, environmental conditions, and timing. SimuHome models
state inquiry, implicit intent inference, explicit device control, and workflow
scheduling in a Matter-grounded smart-home environment~\cite{seo2026simuhome}.
Complementary smart-home security evidence from HomeGuardian, although not an
LLM-agent system, further shows that temporal context, infrequent behavior,
noisy user activity, and cold-start conditions complicate behavioral security
monitoring~\cite{zou2026homeguardian}. The unresolved system-level requirement
is therefore to preserve the association among source, intended address,
session, task, context, and authority as content passes into reasoning and
execution. Provenance must persist across transcription, embedding, and other representation changes so that downstream authorization can still bind decisions to the original source, context, and authority.

\noindent\textbf{2) Persistent Multi-User State and Memory Provenance.}

Personalization converts transient acoustic interaction into reusable state.
CarMem demonstrates long-term preference extraction, maintenance, and retrieval
for a multi-session in-car LLM voice assistant~\cite{kirmayr-etal-2025-carmem},
whereas AgentPoison~\cite{chen2024agentpoison} and MINJA~\cite{dong2025minja}
show how persistent agent memory can be poisoned through later retrieval and
reuse. SENTINEL protects reasoning-history integrity~\cite{karamchandani2026forgedreasoning}, and PPMF preserves platform
provenance so that low-trust observations are not consolidated into
high-authority history~\cite{xu2026memoryprovenancelaundering}. In shared
vehicles or household devices, however, a stored preference or instruction may
outlive the session and speaker that created it and later affect another user or
action. Audio LLM-based agents therefore require memory admission that
preserves speaker and session attribution, isolation of state across users, and
deletion or correction mechanisms that remove security-relevant influence from
subsequent retrieval and execution. These properties should remain enforceable
even when memories are summarized, merged, or personalized over long periods.

\noindent\textbf{3) Deployment Consistency and Safe Fallback.}

Hybrid systems using Audio LLMs, such as ASTA~\cite{torkamani2025adaptiveedgecloudinferencespeechtoaction} and CoFi-Agent~\cite{zhang2026bridgingperceptiongaplightweight}, distribute functions
across local and remote execution paths. Adaptation and quantization can
further change the deployed artifact~\cite{qin2025tinyalignbridgingautomaticspeech,
chen2025roundingguidedbackdoorinjectiondeep}. These transitions are especially
important in device-control settings, where connectivity or computation may
degrade while the physical environment and device state continue to change.
Assurance should therefore cover both the deployed configuration and its
permitted transitions. A fallback path may reduce functionality, but it should
not expand authority, discard provenance, weaken privacy controls, bypass
confirmation, or act on stale execution context. Maintaining this property
requires synchronized policy and state information together with explicit trust
semantics across local and cloud paths.

\noindent\textbf{4) Action-Level Containment, Verification, and Recovery.}

Action-capable agents require controls beyond output moderation. Audio2Tool
shows that speech-driven tool use already spans smart-car, smart-home, and
wearable domains and becomes substantially harder under compositional and
acoustic complexity~\cite{pahwa2026audio2toolspeakcallact}. At the physical end
of the execution spectrum, RoboGuard demonstrates through simulation and
real-world robot experiments that unsafe LLM-generated plans can translate into
harmful physical behavior unless constraints are enforced before execution~\cite{ravichandran2026safetyguardrailsllmenabledrobots}. Tool and actuator
interfaces should therefore enforce bounded capabilities, independent
authorization, and observable commit points at which unsafe operations can
still be stopped. Execution evidence should also remain distinct from
model-generated history. Proof-of-Execution Memory records completed
safety-relevant operations in a tamper-evident ledger written by the trusted
action layer, preventing forged memory from claiming that an operation already
occurred~\cite{rahman2026proofexecutionmemory}. For Audio LLM-based agents,
comparable mechanisms could support confirmation, rollback, compensation, and
post-incident attribution for digital and physical actions. Evaluation should
measure not only whether an unsafe action is proposed, but whether its effects
can be prevented, interrupted, reversed, or contained when upstream safeguards
fail.

\noindent\textbf{5) Security Beyond Speech-Centric Agents.}

Most existing security evaluations of Audio and Speech LLMs remain
speech-centric, although audio-language capabilities are already extending to
non-speech domains. NatureLM-audio targets animal vocalizations and other
bioacoustic tasks~\cite{robinson2025naturelm}, while a domain-adapted Audio LLM
has been evaluated for acoustic fault diagnosis and semantic description of
coal-mine equipment~\cite{cui2026industrial_audio_llm}. These studies show that
non-speech audio-language processing is already relevant to bioacoustic and
industrial settings, but they do not yet constitute a mature body of
end-to-end, action-capable agent-security evaluations. In such settings, provenance may be determined by sensor placement, event source, temporal context, equipment identity, and domain-specific semantics. Downstream effects may include diagnosis, alerting, maintenance recommendations, and control decisions with operational consequences. Future work should therefore test how the six S2A
analysis areas transfer to non-speech auditory tasks and develop provenance,
authority, and evaluation mechanisms that do not assume spoken interaction.

\section{Conclusion}
\label{sec:conclusion}
This survey examined the security of edge-deployed Audio LLM-based agents through the Acoustic Sensing-to-Action (S2A) system model. Across different application profiles, the same system stages recur, while their security importance changes with state, authority, environmental context, and the consequences of execution. The analysis therefore identifies the deployed agent configuration as the relevant unit of security assessment.

The reviewed literature covers this system path unevenly. Acoustic
manipulation, model-level safety, and privacy have received substantial
attention, whereas persistent memory, execution authorization, distributed
runtime, and lifecycle assurance remain less developed. Security evaluation
should therefore account for the state, permissions, runtime path, and
transformations through which acoustic information can produce digital or
physical effects. Trustworthy deployment requires these properties to remain enforceable across agent operation and permitted configuration changes.


\bibliographystyle{IEEEtran}
\bibliography{references}

@String(CVPR   = {CVPR})

@String(NeurIPS= {NeurIPS})

@String(ICLR   = {ICLR})

@String(AAAI   = {AAAI})

@String(IJCAI  = {IJCAI})

@String(ICASSP = {ICASSP})

@IEEEtranBSTCTL{IEEEexample:BSTcontrol,
  CTLuse_forced_etal       = "yes",
  CTLmax_names_forced_etal = "6",
  CTLnames_show_etal       = "1"
}

@misc{zhang2023speechgptempoweringlargelanguage,
  author = {Dong Zhang and Shimin Li and Xin Zhang and Jun Zhan and Pengyu Wang and Yaqian Zhou and Xipeng Qiu},
  title  = {{SpeechGPT}: Empowering Large Language Models with Intrinsic Cross-Modal Conversational Abilities},
  year   = {2023},
  note   = {arXiv:2305.11000}
}

@misc{defossez2024moshispeechtextfoundationmodel,
  author = {Alexandre Défossez and Laurent Mazaré and Manu Orsini and Amélie Royer and Patrick Pérez and Hervé Jégou and Edouard Grave and Neil Zeghidour},
  title  = {Moshi: a speech-text foundation model for real-time dialogue},
  year   = {2024},
  note   = {arXiv:2410.00037}
}

@misc{li2023securityprivacyproblemsvoice,
  author = {Jingjin Li and Chao Chen and Lei Pan and Mostafa Rahimi Azghadi and Hossein Ghodosi and Jun Zhang},
  title  = {Security and Privacy Problems in Voice Assistant Applications: A Survey},
  year   = {2023},
  note   = {arXiv:2304.09486}
}

@inproceedings{Zhang_2017,
  author    = {Zhang, Guoming and Yan, Chen and Ji, Xiaoyu and Zhang, Tianchen and Zhang, Taimin and Xu, Wenyuan},
  title     = {{DolphinAttack}: Inaudible Voice Commands},
  booktitle = {Proc. ACM SIGSAC Conf. Comput. Commun. Security (CCS)},
  month     = oct,
  year      = {2017},
  pages     = {103--117},
  doi       = {10.1145/3133956.3134052}
}

@inproceedings{255276,
  author    = {Takeshi Sugawara and Benjamin Cyr and Sara Rampazzi and Daniel Genkin and Kevin Fu},
  title     = {Light Commands: {Laser-Based} Audio Injection Attacks on {Voice-Controllable} Systems},
  booktitle = {Proc. 29th USENIX Security Symp. (USENIX Security 20)},
  month     = aug,
  year      = {2020},
  pages     = {2631--2648}
}

@misc{zheng2024reviewedgellms,
  author = {Zheng, Yue and Chen, Yuhao and Qian, Bin and Shi, Xiufang and Shu, Yuanchao and Chen, Jiming},
  title  = {A Review on Edge Large Language Models: Design, Execution, and Applications},
  year   = {2024},
  note   = {arXiv:2410.11845}
}

@misc{yang2025holisticlalm,
  author = {Yang, Chih-Kai and Ho, Neo S. and Lee, Hung-yi},
  title  = {Towards Holistic Evaluation of Large Audio-Language Models: A Comprehensive Survey},
  year   = {2025},
  note   = {arXiv:2505.15957}
}

@misc{luo2026trustworthylalmsurvey,
  author = {Luo, Kaiwen and Zhou, Zhenhong and Wang, Leo and Lin, Liang and Xiao, Yang and Shao, Tianyu and Zhang, Yuanhe and Li, Yuxuan and Yu, Miao and Lyu, Kailin and Zhang, Jiaming and Liu, Dongrui and Sun, Li and Wu, Yueming and Li, Kai and Dang, Ting and Das, Rohan Kumar and Li, Xinfeng and Liang, Siyuan and Wang, Qiufeng and Ma, Xingjun and Chen, Jing and Wang, Kun and Dong, Junhao and Zou, Deqing and Cheng, Yu and Hu, Xia and Zeng, Zhigang and Su, Sen and Liu, Yang and Jiang, Yu-Gang and Yu, Philip S. and Ong, Yew-Soon},
  title  = {A Survey of Large Audio Language Models: Generalization, Trustworthiness, and Outlook},
  year   = {2026},
  note   = {arXiv:2605.20266}
}

@misc{wang2024blspbootstrappinglanguagespeechpretraining,
  author = {Chen Wang and Minpeng Liao and Zhongqiang Huang and Jinliang Lu and Junhong Wu and Yuchen Liu and Chengqing Zong and Jiajun Zhang},
  title  = {{BLSP}: Bootstrapping Language-Speech Pre-training via Behavior Alignment of Continuation Writing},
  year   = {2024},
  note   = {arXiv:2309.00916}
}

@misc{shu2023llasmlargelanguagespeech,
  author = {Yu Shu and Siwei Dong and Guangyao Chen and Wenhao Huang and Ruihua Zhang and Daochen Shi and Qiqi Xiang and Yemin Shi},
  title  = {{LLaSM}: Large Language and Speech Model},
  year   = {2023},
  note   = {arXiv:2308.15930}
}

@misc{rubenstein2023audiopalmlargelanguagemodel,
  author = {Paul K. Rubenstein and Chulayuth Asawaroengchai and Duc Dung Nguyen and Ankur Bapna and Zalán Borsos and Félix de Chaumont Quitry and Peter Chen and Dalia El Badawy and Wei Han and Eugene Kharitonov and Hannah Muckenhirn and Dirk Padfield and James Qin and Danny Rozenberg and Tara Sainath and Johan Schalkwyk and Matt Sharifi and Michelle Tadmor Ramanovich and Marco Tagliasacchi and Alexandru Tudor and Mihajlo Velimirović and Damien Vincent and Jiahui Yu and Yongqiang Wang and Vicky Zayats and Neil Zeghidour and Yu Zhang and Zhishuai Zhang and Lukas Zilka and Christian Frank},
  title  = {{AudioPaLM}: A Large Language Model That Can Speak and Listen},
  year   = {2023},
  note   = {arXiv:2306.12925}
}

@misc{tang2024salmonngenerichearingabilities,
  author = {Changli Tang and Wenyi Yu and Guangzhi Sun and Xianzhao Chen and Tian Tan and Wei Li and Lu Lu and Zejun Ma and Chao Zhang},
  title  = {{SALMONN}: Towards Generic Hearing Abilities for Large Language Models},
  year   = {2024},
  note   = {arXiv:2310.13289}
}

@misc{chu2023qwenaudioadvancinguniversalaudio,
  author = {Yunfei Chu and Jin Xu and Xiaohuan Zhou and Qian Yang and Shiliang Zhang and Zhijie Yan and Chang Zhou and Jingren Zhou},
  title  = {{Qwen-Audio}: Advancing Universal Audio Understanding via Unified Large-Scale Audio-Language Models},
  year   = {2023},
  note   = {arXiv:2311.07919}
}

@inproceedings{choi2025desamodeviceelderfriendlysmart,
  author    = {Youngwon Choi and Donghyuk Jung and Hwayeon Kim},
  title     = {{DESAMO}: A Device for Elder-Friendly Smart Homes Powered by Embedded {LLM} with Audio Modality},
  booktitle = {Adjunct Proceedings of the 38th Annual ACM Symposium on User Interface Software and Technology},
  pages     = {87:1--87:3},
  year      = {2025},
  publisher = {ACM},
  doi       = {10.1145/3746058.3758344}
}

@misc{torkamani2025adaptiveedgecloudinferencespeechtoaction,
  author = {Mohammad Jalili Torkamani and Israt Zarin},
  title  = {Adaptive Edge-Cloud Inference for Speech-to-Action Systems Using {ASR} and Large Language Models},
  year   = {2025},
  note   = {arXiv:2512.12769}
}

@misc{zhang2026bridgingperceptiongaplightweight,
  author = {Hengfan Zhang and Yueqian Lin and Hai Helen Li and Yiran Chen},
  title  = {Bridging the Perception Gap: A Lightweight Coarse-to-Fine Architecture for Edge Audio Systems},
  year   = {2026},
  note   = {arXiv:2601.15676}
}

@inproceedings{qin2025tinyalignbridgingautomaticspeech,
  author    = {Ruiyang Qin and Dancheng Liu and Gelei Xu and Amir Nassereldine and Zheyu Yan and Chenhui Xu and Yuting Hu and Shaocong Wang and X. Sharon Hu and Jinjun Xiong and Yiyu Shi},
  title     = {{Tiny-Align}: Bridging Automatic Speech Recognition and Large Language Model on Edge},
  booktitle = {2025 IEEE/ACM International Conference on Computer Aided Design (ICCAD)},
  pages     = {1--9},
  year      = {2025},
  publisher = {IEEE},
  doi       = {10.1109/ICCAD66269.2025.11240668}
}

@misc{soundhound_onchip,
  author       = {{SoundHound AI}},
  title        = {{SoundHound} to Offer On-Chip Voice {AI} with {NVIDIA} That Delivers In-Vehicle Generative {AI} Responses With No Connectivity Required},
  howpublished = {SoundHound AI},
  month        = mar,
  year         = {2024},
  note         = {{Accessed:} Jul. 29, 2026. [Online]. Available: \url{https://www.soundhound.com/newsroom/press-releases/soundhound-to-offer-on-chip-voice-ai-with-nvidia-that-delivers-in-vehicle-generative-ai-responses-with-no-connectivity-required/}}
}

@misc{gemma3n,
  author       = {Gonzalez, Lucas and Shivanna, Rakesh},
  title        = {Announcing {Gemma 3n} Preview: Powerful, Efficient, Mobile-First {AI}},
  howpublished = {Google Developers Blog},
  month        = may,
  year         = {2025},
  note         = {{Accessed:} Jul. 29, 2026. [Online]. Available: \url{https://developers.googleblog.com/en/introducing-gemma-3n/}}
}

@misc{lfm25_audio,
  author       = {{Liquid AI}},
  title        = {Introducing {LFM2.5}: The Next Generation of On-Device {AI}},
  howpublished = {Liquid AI Blog},
  month        = jan,
  year         = {2026},
  note         = {{Accessed:} Jul. 29, 2026. [Online]. Available: \url{https://www.liquid.ai/blog/introducing-lfm2-5-the-next-generation-of-on-device-ai}}
}

@inproceedings{299709,
  author    = {Guoming Zhang and Xiaohui Ma and Huiting Zhang and Zhijie Xiang and Xiaoyu Ji and Yanni Yang and Xiuzhen Cheng and Pengfei Hu},
  title     = {{LaserAdv}: Laser Adversarial Attacks on Speech Recognition Systems},
  booktitle = {Proc. 33rd USENIX Security Symp. (USENIX Security 24)},
  month     = aug,
  year      = {2024},
  pages     = {3945--3961}
}

@misc{carlini2018audioadversarialexamplestargeted,
  author = {Nicholas Carlini and David Wagner},
  title  = {Audio Adversarial Examples: Targeted Attacks on Speech-to-Text},
  year   = {2018},
  note   = {arXiv:1801.01944}
}

@inproceedings{pmlr-v97-qin19a,
  author    = {Qin, Yao and Carlini, Nicholas and Cottrell, Garrison and Goodfellow, Ian and Raffel, Colin},
  title     = {Imperceptible, Robust, and Targeted Adversarial Examples for Automatic Speech Recognition},
  booktitle = {Proc. 36th Int. Conf. Mach. Learn. (ICML)},
  month     = {09--15 Jun},
  year      = {2019},
  pages     = {5231--5240}
}

@inproceedings{kang2025advwave,
  author    = {Mintong Kang and Chejian Xu and Bo Li},
  title     = {{AdvWave}: Stealthy Adversarial Jailbreak Attack against Large Audio-Language Models},
  booktitle = {Proc. 13th Int. Conf. Learn. Representations (ICLR)},
  year      = {2025},
  note      = {[Online]. Available: \url{https://openreview.net/forum?id=0BujOfTqab}}
}

@inproceedings{ma-etal-2025-universal,
  author    = {Ma, Rao  and
      Qian, Mengjie  and
      Raina, Vyas  and
      Gales, Mark  and
      Knill, Kate},
  title     = {Universal Acoustic Adversarial Attacks for Flexible Control of Speech-{LLM}s},
  booktitle = {Findings Assoc. Comput. Linguistics: EMNLP 2025},
  month     = nov,
  year      = {2025},
  pages     = {18248--18262},
  doi       = {10.18653/v1/2025.findings-emnlp.990}
}

@misc{ziv2025breakingaudiolargelanguage,
  author = {Roee Ziv and Raz Lapid and Moshe Sipper},
  title  = {Breaking Audio Large Language Models by Attacking Only the Encoder: A Universal Targeted Latent-Space Audio Attack},
  year   = {2025},
  note   = {arXiv:2512.23881}
}

@misc{dingeto2026goodsoundsadversarialjailbreaking,
  author = {Hiskias Dingeto and Taeyoun Kwon and Dasol Choi and Bodam Kim and DongGeon Lee and Haon Park and JaeHoon Lee and Jongho Shin},
  title  = {When Good Sounds Go Adversarial: Jailbreaking Audio-Language Models with Benign Inputs},
  year   = {2026},
  note   = {arXiv:2508.03365}
}

@misc{wang2026grmutilityawarejailbreakattacks,
  author = {Yunqiang Wang and Hengyuan Na and Di Wu and Miao Hu and Guocong Quan},
  title  = {{GRM}: Utility-Aware Jailbreak Attacks on Audio {LLMs} via Gradient-Ratio Masking},
  year   = {2026},
  note   = {arXiv:2604.09222}
}

@misc{fang2026sparsetokenssufficejailbreaking,
  author = {Zheng Fang and Xiaosen Wang and Shenyi Zhang and Shaokang Wang and Zhijin Ge},
  title  = {Sparse Tokens Suffice: Jailbreaking Audio Language Models via Token-Aware Gradient Optimization},
  year   = {2026},
  note   = {arXiv:2605.04700}
}

@misc{roh2026codecrobustattacksaudiollms,
  author = {Jaechul Roh and Jean-Philippe Monteuuis and Jonathan Petit and Amir Houmansadr},
  title  = {Codec-Robust Attacks on Audio {LLMs}},
  year   = {2026},
  note   = {arXiv:2605.20519}
}

@inproceedings{197215,
  author    = {Nicholas Carlini and Pratyush Mishra and Tavish Vaidya and Yuankai Zhang and Micah Sherr and Clay Shields and David Wagner and Wenchao Zhou},
  title     = {Hidden Voice Commands},
  booktitle = {Proc. 25th USENIX Security Symp. (USENIX Security 16)},
  month     = aug,
  year      = {2016},
  pages     = {513--530}
}

@misc{yuan2018commandersongsystematicapproachpractical,
  author = {Xuejing Yuan and Yuxuan Chen and Yue Zhao and Yunhui Long and Xiaokang Liu and Kai Chen and Shengzhi Zhang and Heqing Huang and Xiaofeng Wang and Carl A. Gunter},
  title  = {{CommanderSong}: A Systematic Approach for Practical Adversarial Voice Recognition},
  year   = {2018},
  note   = {arXiv:1801.08535}
}

@misc{yang2026speechaudiocompositionalattacksmultimodal,
  author = {Yudong Yang and Xuezhen Zhang and Zhifeng Han and Siyin Wang and Jimin Zhuang and Zengrui Jin and Jing Shao and Guangzhi Sun and Chao Zhang},
  title  = {Speech-Audio Compositional Attacks on Multimodal {LLMs} and Their Mitigation with {SALMONN-Guard}},
  year   = {2026},
  note   = {arXiv:2511.10222}
}

@inproceedings{yan2020surfingattack,
  author    = {Yan, Qiben and Liu, Kehai and Zhou, Qin and Guo, Hanqing and Zhang, Ning},
  title     = {{SurfingAttack}: Interactive Hidden Attack on Voice Assistants Using Ultrasonic Guided Waves},
  booktitle = {Proc. Netw. Distrib. Syst. Security Symp. (NDSS)},
  month     = feb,
  year      = {2020},
  doi       = {10.14722/ndss.2020.24068}
}

@inproceedings{287266,
  author    = {Qi Xia and Qian Chen and Shouhuai Xu},
  title     = {{Near-Ultrasound} Inaudible Trojan (Nuit): Exploiting Your Speaker to Attack Your Microphone},
  booktitle = {Proc. 32nd USENIX Security Symp. (USENIX Security 23)},
  month     = aug,
  year      = {2023},
  pages     = {4589--4606}
}

@misc{ling2026sirenswhisper,
  author = {Ling, Zijian and Hu, Pingyi and Gao, Xiuyong and
                   Ma, Xiaojing and Zhou, Man and Feng, Jun and
                   Lu, Songfeng and Zhang, Dongmei and Zhu, Bin Benjamin},
  title  = {{Sirens' Whisper}: Inaudible Near-Ultrasonic Jailbreaks
                   of Speech-Driven {LLM}s},
  year   = {2026},
  note   = {arXiv:2603.13847}
}

@inproceedings{Yakura_2019,
  author    = {Yakura, Hiromu and Sakuma, Jun},
  title     = {Robust Audio Adversarial Example for a Physical Attack},
  booktitle = {Proc. Int. Joint Conf. Artif. Intell. (IJCAI)},
  month     = aug,
  year      = {2019},
  pages     = {5334--5341},
  doi       = {10.24963/ijcai.2019/741}
}

@inproceedings{chen2020metamorph,
  author    = {Chen, Tao and Shangguan, Longfei and Li, Zhenjiang
               and Jamieson, Kyle},
  title     = {{Metamorph}: Injecting Inaudible Commands into
               Over-the-Air Voice Controlled Systems},
  booktitle = {Proc. Netw. Distrib. Syst. Security Symp. (NDSS)},
  month     = feb,
  year      = {2020},
  doi       = {10.14722/ndss.2020.23055}
}

@article{10368060,
  author  = {Chen, Meng and Lu, Li and Yu, Jiadi and Ba, Zhongjie and Lin, Feng and Ren, Kui},
  title   = {{AdvReverb}: Rethinking the Stealthiness of Audio Adversarial Examples to Human Perception},
  journal = {IEEE Trans. Inf. Forensics Security},
  volume  = {19},
  pages   = {1948--1962},
  year    = {2024},
  doi     = {10.1109/TIFS.2023.3345639}
}

@misc{chen2026hijackinglargeaudiolanguagemodels,
  author = {Meng Chen and Kun Wang and Li Lu and Jiaheng Zhang and Tianwei Zhang},
  title  = {Hijacking Large Audio-Language Models via Context-Agnostic and Imperceptible Auditory Prompt Injection},
  year   = {2026},
  note   = {arXiv:2604.14604}
}

@inproceedings{sadasivan-etal-2026-attackers,
  author    = {Sadasivan, Vinu Sankar  and
      Feizi, Soheil  and
      Mathews, Rajiv  and
      Wang, Lun},
  title     = {Attacker{'}s Noise Can Manipulate Your Audio-based {LLM} in the Real World},
  booktitle = {Proc. 19th Conf. Eur. Chapter Assoc. Comput. Linguistics (EACL)},
  month     = mar,
  year      = {2026},
  pages     = {1430--1440},
  doi       = {10.18653/v1/2026.eacl-long.66}
}

@inproceedings{song-etal-2026-audio,
  author    = {Song, Zirui  and
      Jiang, Qian  and
      Cui, Mingxuan  and
      Li, Mingzhe  and
      Gao, Lang  and
      Zhang, Zeyu  and
      Xu, Zixiang  and
      Wang, Yanbo  and
      Ouyang, Guangxian  and
      Chen, Zhenhao  and
      Chen, Xiuying},
  title     = {Audio Jailbreak: An Open Comprehensive Benchmark for Jailbreaking Large Audio-Language Models},
  booktitle = {Proc. 64th Annu. Meeting Assoc. Comput. Linguistics (ACL)},
  month     = jul,
  year      = {2026},
  pages     = {27294--27308},
  doi       = {10.18653/v1/2026.acl-long.1259}
}

@misc{shen2024voicejailbreakattacksgpt4o,
  author = {Xinyue Shen and Yixin Wu and Michael Backes and Yang Zhang},
  title  = {Voice Jailbreak Attacks Against {GPT-4o}},
  year   = {2024},
  note   = {arXiv:2405.19103}
}

@inproceedings{yu-etal-2026-now,
  author    = {Yu, Ye  and
      Jin, Haibo  and
      Yu, Yaoning  and
      Zhuang, Jun  and
      Wang, Haohan},
  title     = {Now You Hear Me: Audio Narrative Attacks Against Large Audio{--}Language Models},
  booktitle = {Proc. 19th Conf. Eur. Chapter Assoc. Comput. Linguistics (EACL)},
  month     = mar,
  year      = {2026},
  pages     = {5925--5939},
  doi       = {10.18653/v1/2026.eacl-long.278}
}

@inproceedings{roh2025multilingual,
  author    = {Jaechul Roh and Virat Shejwalkar and Amir Houmansadr},
  title     = {Multilingual and  Multi-Accent Jailbreaking of Audio {LLM}s},
  booktitle = {Proc. 2nd Conf. Language Modeling (COLM)},
  year      = {2025},
  note      = {[Online]. Available: \url{https://openreview.net/forum?id=yGa8CYT8kS}}
}

@article{li2026stylebreak,
  author  = {Li, Hongyi and Zhou, Chengxuan and Wang, Chu and
             Liang, Sicheng and Chen, Yanting and Xie, Qinlin and
             Ye, Jiawei and Wu, Jie},
  title   = {{StyleBreak}: Revealing Alignment Vulnerabilities in
             Large Audio-Language Models via Style-Aware Audio Jailbreak},
  journal = {Proc. AAAI Conf. Artif. Intell.},
  volume  = {40},
  number  = {44},
  pages   = {37591--37599},
  month   = mar,
  year    = {2026},
  doi     = {10.1609/aaai.v40i44.41093}
}

@misc{bagdasaryan2023abusingimagessoundsindirect,
  author = {Eugene Bagdasaryan and Tsung-Yin Hsieh and Ben Nassi and Vitaly Shmatikov},
  title  = {Abusing Images and Sounds for Indirect Instruction Injection in Multi-Modal {LLMs}},
  year   = {2023},
  note   = {arXiv:2307.10490}
}

@inproceedings{hou-etal-2025-evaluating,
  author    = {Hou, Guanyu  and
      He, Jiaming  and
      Zhou, Yinhang  and
      Guo, Ji  and
      Qiao, Yitong  and
      Zhang, Rui  and
      Jiang, Wenbo},
  title     = {Evaluating Robustness of Large Audio Language Models to Audio Injection: An Empirical Study},
  booktitle = {Proc. Conf. Empirical Methods Natural Language Process. (EMNLP)},
  month     = nov,
  year      = {2025},
  pages     = {25660--25676},
  doi       = {10.18653/v1/2025.emnlp-main.1303}
}

@inproceedings{geng-etal-2025-mathsf,
  author    = {Geng, Jiahui  and
      Tran, Thy Thy  and
      Nakov, Preslav  and
      Gurevych, Iryna},
  title     = {{C}on {I}nstruction: Universal Jailbreaking of Multimodal Large Language Models via Non-Textual Modalities},
  booktitle = {Proc. 63rd Annu. Meeting Assoc. Comput. Linguistics (ACL)},
  month     = jul,
  year      = {2025},
  pages     = {2917--2933},
  doi       = {10.18653/v1/2025.acl-long.146}
}

@misc{ma2025audiojailbreakattacksexposing,
  author = {Binhao Ma and Hanqing Guo and Zhengping Jay Luo and Rui Duan},
  title  = {Audio Jailbreak Attacks: Exposing Vulnerabilities in {SpeechGPT} in a White-Box Framework},
  year   = {2025},
  note   = {arXiv:2505.18864}
}

@misc{krishnan2026optimizingmultimodaljailbreaksspoken,
  author = {Aravind Krishnan and Karolina Stańczak and Dietrich Klakow},
  title  = {On Optimizing Multimodal Jailbreaks for Spoken Language Models},
  year   = {2026},
  note   = {arXiv:2603.19127}
}

@misc{wang2026acousticinterferencenewparadigm,
  author = {Yanyun Wang and Yu Huang and Zi Liang and Xixin Wu and Li Liu},
  title  = {Acoustic Interference: A New Paradigm Weaponizing Acoustic Latent Semantic for Universal Jailbreak against Large Audio Language Models},
  year   = {2026},
  note   = {arXiv:2605.18168}
}

@inproceedings{294504,
  author    = {Meng Chen and Xiangyu Xu and Li Lu and Zhongjie Ba and Feng Lin and Kui Ren},
  title     = {Devil in the Room: Triggering Audio Backdoors in the Physical World},
  booktitle = {Proc. 33rd USENIX Security Symp. (USENIX Security 24)},
  month     = aug,
  year      = {2024},
  pages     = {7285--7302}
}

@inproceedings{10096332,
  author    = {Koffas, Stefanos and Pajola, Luca and Picek, Stjepan and Conti, Mauro},
  title     = {Going in Style: Audio Backdoors Through Stylistic Transformations},
  booktitle = {Proc. IEEE Int. Conf. Acoust., Speech Signal Process. (ICASSP)},
  year      = {2023},
  pages     = {1--5},
  doi       = {10.1109/ICASSP49357.2023.10096332}
}

@inproceedings{schoof2024emobackdoorattacksspeaker,
  author    = {Schoof, Coen and Koffas, Stefanos and Conti, Mauro and Picek, Stjepan},
  title     = {{EmoBack}: Backdoor Attacks Against Speaker Identification Using Emotional Prosody},
  booktitle = {Proc. Workshop Artif. Intell. Security (AISec)},
  year      = {2024},
  pages     = {137--148},
  doi       = {10.1145/3689932.3694773}
}

@misc{lin2025hiddennoiseunveilingbackdoors,
  author = {Liang Lin and Miao Yu and Kaiwen Luo and Yibo Zhang and Lilan Peng and Dexian Wang and Xuehai Tang and Yuanhe Zhang and Xikang Yang and Zhenhong Zhou and Kun Wang and Yang Liu},
  title  = {Hidden in the Noise: Unveiling Backdoors in Audio {LLMs} Alignment through Latent Acoustic Pattern Triggers},
  year   = {2025},
  note   = {arXiv:2508.02175}
}

@misc{cai2022vsvcbackdoorattackkeyword,
  author = {Hanbo Cai and Pengcheng Zhang and Hai Dong and Yan Xiao and Shunhui Ji},
  title  = {{VSVC}: Backdoor attack against Keyword Spotting based on Voiceprint Selection and Voice Conversion},
  year   = {2022},
  note   = {arXiv:2212.10103}
}

@misc{fortier2026backdoorslivecomponentlevelanalysis,
  author = {Alexandrine Fortier and Thomas Thebaud and Jesús Villalba and Najim Dehak and Patrick Cardinal and Peter West},
  title  = {Where Do Backdoors Live? A Component-Level Analysis of Backdoor Propagation in Speech Language Models},
  year   = {2026},
  note   = {arXiv:2510.01157}
}

@inproceedings{291116,
  author    = {Cheng{\textquoteright}an Wei and Yeonjoon Lee and Kai Chen and Guozhu Meng and Peizhuo Lv},
  title     = {Aliasing Backdoor Attacks on Pre-trained Models},
  booktitle = {Proc. 32nd USENIX Security Symp. (USENIX Security 23)},
  month     = aug,
  year      = {2023},
  pages     = {2707--2724}
}

@inproceedings{hanif-etal-2025-trojanwave,
  author    = {Hanif, Asif  and
      Agro, Maha Tufail  and
      Shamshad, Fahad  and
      Nandakumar, Karthik},
  title     = {{T}rojan{W}ave: Exploiting Prompt Learning for Stealthy Backdoor Attacks on Large Audio-Language Models},
  booktitle = {Proc. Conf. Empirical Methods Natural Language Process. (EMNLP)},
  month     = nov,
  year      = {2025},
  pages     = {18617--18633},
  doi       = {10.18653/v1/2025.emnlp-main.940}
}

@inproceedings{liu-etal-2025-loratk,
  author    = {Liu, Hongyi  and
      Zhong, Shaochen  and
      Sun, Xintong  and
      Tian, Minghao  and
      Hariri, Mohsen  and
      Liu, Zirui  and
      Tang, Ruixiang  and
      Jiang, Zhimeng  and
      Yuan, Jiayi  and
      Chuang, Yu-Neng  and
      Li, Li  and
      Choi, Soo-Hyun  and
      Chen, Rui  and
      Chaudhary, Vipin  and
      Hu, Xia},
  title     = {{L}o{RATK}: {L}o{RA} Once, Backdoor Everywhere in the Share-and-Play Ecosystem},
  booktitle = {Findings Assoc. Comput. Linguistics: EMNLP 2025},
  month     = nov,
  year      = {2025},
  pages     = {23009--23047},
  doi       = {10.18653/v1/2025.findings-emnlp.1253}
}

@inproceedings{285505,
  author    = {Xinben Gao and Lan Zhang},
  title     = {{PCAT}: Functionality and Data Stealing from Split Learning by {Pseudo-Client} Attack},
  booktitle = {Proc. 32nd USENIX Security Symp. (USENIX Security 23)},
  month     = aug,
  year      = {2023},
  pages     = {5271--5288}
}

@inproceedings{309752,
  author    = {Tian Dong and Yan Meng and Shaofeng Li and Guoxing Chen and Zhen Liu and Haojin Zhu},
  title     = {Depth Gives a False Sense of Privacy: {LLM} Internal States Inversion},
  booktitle = {Proc. 34th USENIX Security Symp. (USENIX Security 25)},
  month     = aug,
  year      = {2025},
  pages     = {1629--1648}
}

@misc{fan2026doesserverseeunderstanding,
  author = {Mingyuan Fan and Yu Liu and Fuyi Wang and Cen Chen},
  title  = {What Does the Server See? Understanding Privacy Leakage from Large Language Models in Split Inference},
  year   = {2026},
  note   = {arXiv:2605.23158}
}

@inproceedings{298278,
  author    = {Tushar Nayan and Qiming Guo and Mohammed Al Duniawi and Marcus Botacin and Selcuk Uluagac and Ruimin Sun},
  title     = {{SoK}: All You Need to Know About {On-Device} {ML} Model Extraction - The Gap Between Research and Practice},
  booktitle = {Proc. 33rd USENIX Security Symp. (USENIX Security 24)},
  month     = aug,
  year      = {2024},
  pages     = {5233--5250}
}

@inproceedings{309838,
  author    = {Zibo Gao and Junjie Hu and Feng Guo and Yixin Zhang and Yinglong Han and Siyuan Liu and Haiyang Li and Zhiqiang Lv},
  title     = {I Know What You Said: Unveiling Hardware Cache {Side-Channels} in Local Large Language Model Inference},
  booktitle = {Proc. 34th USENIX Security Symp. (USENIX Security 25)},
  month     = aug,
  year      = {2025},
  pages     = {1649--1668}
}

@inproceedings{xiao2026modelspy,
  author    = {Xiao, Rui and Feng, Sibo and Ramesh, Soundarya
               and Han, Jun and Han, Jinsong},
  title     = {Peering Inside the Black-Box: Long-Range and Scalable
               Model Architecture Snooping via {GPU} Electromagnetic
               Side-Channel},
  booktitle = {Proc. Netw. Distrib. Syst. Security Symp. (NDSS)},
  month     = feb,
  year      = {2026},
  doi       = {10.14722/ndss.2026.230141}
}

@inproceedings{294635,
  author    = {Shaofeng Li and Xinyu Wang and Minhui Xue and Haojin Zhu and Zhi Zhang and Yansong Gao and Wen Wu and Xuemin (Sherman) Shen},
  title     = {Yes, {One-Bit-Flip} Matters! Universal {DNN} Model Inference Depletion with Runtime Code Fault Injection},
  booktitle = {Proc. 33rd USENIX Security Symp. (USENIX Security 24)},
  month     = aug,
  year      = {2024},
  pages     = {1315--1330}
}

@inproceedings{255272,
  author    = {Fan Yao and Adnan Siraj Rakin and Deliang Fan},
  title     = {{DeepHammer}: Depleting the Intelligence of Deep Neural Networks through Targeted Chain of Bit Flips},
  booktitle = {Proc. 29th USENIX Security Symp. (USENIX Security 20)},
  month     = aug,
  year      = {2020},
  pages     = {1463--1480}
}

@inproceedings{haque23_interspeech,
  author    = {Mirazul Haque and Rutvij Shah and Simin Chen and Berrak Sisman and Cong Liu and Wei Yang},
  title     = {{SlothSpeech}: Denial-of-service attack against speech recognition models},
  booktitle = {Proc. Interspeech},
  year      = {2023},
  pages     = {1274--1278},
  doi       = {10.21437/Interspeech.2023-1118}
}

@misc{shumailov2021spongeexamplesenergylatencyattacks,
  author = {Ilia Shumailov and Yiren Zhao and Daniel Bates and Nicolas Papernot and Robert Mullins and Ross Anderson},
  title  = {Sponge Examples: Energy-Latency Attacks on Neural Networks},
  year   = {2021},
  note   = {arXiv:2006.03463}
}

@inproceedings{zhang-etal-2025-crabs,
  author    = {Zhang, Yuanhe  and
      Zhou, Zhenhong  and
      Zhang, Wei  and
      Wang, Xinyue  and
      Jia, Xiaojun  and
      Liu, Yang  and
      Su, Sen},
  title     = {Crabs: Consuming Resource via Auto-generation for {LLM}-{D}o{S} Attack under Black-box Settings},
  booktitle = {Findings Assoc. Comput. Linguistics: ACL 2025},
  month     = jul,
  year      = {2025},
  pages     = {11128--11150},
  doi       = {10.18653/v1/2025.findings-acl.580}
}

@inproceedings{chen-etal-2026-naturalsloth,
  author    = {Chen, Yiming  and
      Li, Zexin  and
      Yue, Xianghu  and
      Tan, Robby T.  and
      Li, Haizhou},
  title     = {{N}atural{S}loth: Revisiting Denial-of-Service Attacks on Large Language Models},
  booktitle = {Proc. 64th Annu. Meeting Assoc. Comput. Linguistics (ACL)},
  month     = jul,
  year      = {2026},
  pages     = {19685--19702},
  doi       = {10.18653/v1/2026.acl-long.901}
}

@misc{chen2025roundingguidedbackdoorinjectiondeep,
  author = {Xiangxiang Chen and Peixin Zhang and Jun Sun and Wenhai Wang and Jingyi Wang},
  title  = {Rounding-Guided Backdoor Injection in Deep Learning Model Quantization},
  year   = {2025},
  note   = {arXiv:2510.09647}
}

@article{ma2024quantizationbackdoors,
  author  = {Ma, Hua and Qiu, Huming and Gao, Yansong and
             Zhang, Zhi and Abuadbba, Alsharif and Xue, Minhui and
             Fu, Anmin and Zhang, Jiliang and Al-Sarawi, Said F.
             and Abbott, Derek},
  title   = {Quantization Backdoors to Deep Learning Commercial Frameworks},
  journal = {IEEE Trans. Dependable Secure Comput.},
  volume  = {21},
  number  = {3},
  pages   = {1155--1172},
  month   = may,
  year    = {2024},
  doi     = {10.1109/TDSC.2023.3271956}
}

@misc{zhan2026systemslevelattacksurfaceedge,
  author = {Zhonghao Zhan and Krinos Li and Yefan Zhang and Hamed Haddadi},
  title  = {Systems-Level Attack Surface of Edge Agent Deployments on {IoT}},
  year   = {2026},
  note   = {arXiv:2602.22525}
}

@misc{lei2019insecurityhomedigitalvoice,
  author = {Xinyu Lei and Guan-Hua Tu and Alex X. Liu and Kamran Ali and Chi-Yu Li and Tian Xie},
  title  = {The Insecurity of Home Digital Voice Assistants -- {Amazon Alexa} as a Case Study},
  year   = {2019},
  note   = {arXiv:1712.03327}
}

@inproceedings{shi2026toolhijacker,
  author    = {Shi, Jiawen and Yuan, Zenghui and Tie, Guiyao and
               Zhou, Pan and Gong, Neil Zhenqiang and Sun, Lichao},
  title     = {Prompt Injection Attack to Tool Selection in {LLM} Agents},
  booktitle = {Proc. Netw. Distrib. Syst. Security Symp. (NDSS)},
  month     = feb,
  year      = {2026},
  doi       = {10.14722/ndss.2026.230675}
}

@inproceedings{zhan-etal-2024-injecagent,
  author    = {Zhan, Qiusi  and
      Liang, Zhixiang  and
      Ying, Zifan  and
      Kang, Daniel},
  title     = {{I}njec{A}gent: Benchmarking Indirect Prompt Injections in Tool-Integrated Large Language Model Agents},
  booktitle = {Findings Assoc. Comput. Linguistics: ACL 2024},
  month     = aug,
  year      = {2024},
  pages     = {10471--10506},
  doi       = {10.18653/v1/2024.findings-acl.624}
}

@inproceedings{ye-etal-2024-toolsword,
  author    = {Ye, Junjie  and
      Li, Sixian  and
      Li, Guanyu  and
      Huang, Caishuang  and
      Gao, Songyang  and
      Wu, Yilong  and
      Zhang, Qi  and
      Gui, Tao  and
      Huang, Xuanjing},
  title     = {{T}ool{S}word: Unveiling Safety Issues of Large Language Models in Tool Learning Across Three Stages},
  booktitle = {Proc. 62nd Annu. Meeting Assoc. Comput. Linguistics (ACL)},
  month     = aug,
  year      = {2024},
  pages     = {2181--2211},
  doi       = {10.18653/v1/2024.acl-long.119}
}

@inproceedings{yuan-etal-2024-r-judge,
  author    = {Yuan, Tongxin  and
      He, Zhiwei  and
      Dong, Lingzhong  and
      Wang, Yiming  and
      Zhao, Ruijie  and
      Xia, Tian  and
      Xu, Lizhen  and
      Zhou, Binglin  and
      Li, Fangqi  and
      Zhang, Zhuosheng  and
      Wang, Rui  and
      Liu, Gongshen},
  title     = {{R}-Judge: Benchmarking Safety Risk Awareness for {LLM} Agents},
  booktitle = {Findings Assoc. Comput. Linguistics: EMNLP 2024},
  month     = nov,
  year      = {2024},
  pages     = {1467--1490},
  doi       = {10.18653/v1/2024.findings-emnlp.79}
}

@misc{cartagena2026mindgaptextsafety,
  author = {Arnold Cartagena and Ariane Teixeira},
  title  = {Mind the {GAP}: Text Safety Does Not Transfer to Tool-Call Safety in {LLM} Agents},
  year   = {2026},
  note   = {arXiv:2602.16943}
}

@inproceedings{NEURIPS2024_97091a51,
  author    = {Debenedetti, Edoardo and Zhang, Jie and Balunovic, Mislav and Beurer-Kellner, Luca and Fischer, Marc and Tram\`{e}r, Florian},
  title     = {{AgentDojo}: A Dynamic Environment to Evaluate Prompt Injection Attacks and Defenses for {LLM} Agents},
  booktitle = {Proc. Int. Conf. Neural Inf. Process. Syst. (NeurIPS)},
  year      = {2024},
  pages     = {82895--82920},
  doi       = {10.52202/079017-2636}
}

@misc{li2025lesdissonancescrosstoolharvesting,
  author = {Zichuan Li and Jian Cui and Xiaojing Liao and Luyi Xing},
  title  = {Les Dissonances: Cross-Tool Harvesting and Polluting in Pool-of-Tools Empowered {LLM} Agents},
  year   = {2025},
  note   = {arXiv:2504.03111}
}

@misc{li2026aegisgovernanceintegritysecurity,
  author = {Xiang Li and Pin-Yu Chen and Wenqi Wei},
  title  = {Aegis: Towards Governance, Integrity, and Security of {AI} Voice Agents},
  year   = {2026},
  note   = {arXiv:2602.07379}
}

@misc{wang2026voxprivacybenchmarkevaluatinginteractional,
  author = {Yuxiang Wang and Hongyu Liu and Dekun Chen and Xueyao Zhang and Zhizheng Wu},
  title  = {{VoxPrivacy}: A Benchmark for Evaluating Interactional Privacy of Speech Language Models},
  year   = {2026},
  note   = {arXiv:2601.19956}
}

@inproceedings{wang-etal-2026-hearsay,
  author    = {Wang, Jin  and
      Luo, Kaiwen  and
      Lin, Liang  and
      Wang, Weiliu  and
      Chen, Yitian  and
      Aloqaily, Moayad  and
      Tang, Xuehai  and
      Zhou, Zhenhong  and
      Wang, Kun  and
      Sun, Li  and
      Wen, Qingsong},
  title     = {{H}ear{S}ay Benchmark: Do Audio {LLM}s Leak What They Hear?},
  booktitle = {Findings Assoc. Comput. Linguistics: ACL 2026},
  month     = jul,
  year      = {2026},
  pages     = {19312--19331},
  doi       = {10.18653/v1/2026.findings-acl.964}
}

@inproceedings{he-etal-2026-audioprivacy,
  author    = {He, Jiabei  and
      Zhang, Yanzhe  and
      Zhou, Jiaming  and
      Wang, Hui  and
      Sun, Haoqin  and
      Qin, Yong},
  title     = {{A}udio{P}rivacy: Parallel Audio Dataset for Speaker Profiling with Diverse Audio Types and Rich Attributes},
  booktitle = {Findings Assoc. Comput. Linguistics: ACL 2026},
  month     = jul,
  year      = {2026},
  pages     = {5735--5748},
  doi       = {10.18653/v1/2026.findings-acl.283}
}

@misc{kuzmin2026privacypreservingendtoendfullduplexspeech,
  author = {Nikita Kuzmin and Tao Zhong and Jiajun Deng and Yingke Zhu and Tristan Tsoi and Tianxiang Cao and Simon Lui and Kong Aik Lee and Eng Siong Chng},
  title  = {Privacy-Preserving End-to-End Full-Duplex Speech Dialogue Models},
  year   = {2026},
  note   = {arXiv:2603.08179}
}

@misc{dong2026membershipinferenceattackslarge,
  author = {Jia-Kai Dong and Yu-Xiang Lin and Hung-Yi Lee},
  title  = {Membership Inference Attacks against Large Audio Language Models},
  year   = {2026},
  note   = {arXiv:2603.28378}
}

@inproceedings{zhan-etal-2026-protecting,
  author    = {Zhan, Xiao  and
      Sun, Guangzhi  and
      Such, Jose  and
      Woodland, Phil},
  title     = {Protecting Bystander Privacy via Selective Hearing in Audio {LLM}s},
  booktitle = {Proc. 64th Annu. Meeting Assoc. Comput. Linguistics (ACL)},
  month     = jul,
  year      = {2026},
  pages     = {15180--15192},
  doi       = {10.18653/v1/2026.acl-long.693}
}

@inproceedings{kirmayr-etal-2025-carmem,
  author    = {Kirmayr, Johannes and Stappen, Lukas and Schneider, Phillip and Matthes, Florian and Andre, Elisabeth},
  title     = {{C}ar{M}em: Enhancing Long-Term Memory in {LLM} Voice Assistants through Category-Bounding},
  booktitle = {Proceedings of the 31st International Conference on Computational Linguistics: Industry Track},
  month     = jan,
  year      = {2025},
  address   = {Abu Dhabi, UAE},
  publisher = {Association for Computational Linguistics},
  pages     = {343--357},
  url       = {https://aclanthology.org/2025.coling-industry.29/}
}

@misc{pulipaka2026persistbenchlongtermmemoriesforgotten,
  author = {Sidharth Pulipaka and Oliver Chen and Manas Sharma and Taaha S Bajwa and Vyas Raina and Ivaxi Sheth},
  title  = {{PersistBench}: When Should Long-Term Memories Be Forgotten by {LLMs}?},
  year   = {2026},
  note   = {arXiv:2602.01146}
}

@inproceedings{yamagishi2021asvspoof,
  author    = {Yamagishi, Junichi and Wang, Xin and
               Todisco, Massimiliano and Sahidullah, Md and
               Patino, Jose and Nautsch, Andreas and Liu, Xuechen and
               Lee, Kong Aik and Kinnunen, Tomi and Evans, Nicholas and
               Delgado, H{\'e}ctor},
  title     = {{ASVspoof 2021}: Accelerating Progress in Spoofed and
               Deepfake Speech Detection},
  booktitle = {Proc. Autom. Speaker Verification Spoofing Countermeasures Challenge (ASVspoof)},
  year      = {2021},
  pages     = {47--54},
  doi       = {10.21437/ASVSPOOF.2021-8}
}

@inproceedings{307846,
  author    = {Zhisheng Zhang and Derui Wang and Qianyi Yang and Pengyang Huang and Junhan Pu and Yuxin Cao and Kai Ye and Jie Hao and Yixian Yang},
  title     = {{SafeSpeech}: Robust and Universal Voice Protection Against Malicious Speech Synthesis},
  booktitle = {Proc. 34th USENIX Security Symp. (USENIX Security 25)},
  month     = aug,
  year      = {2025},
  pages     = {4581--4600}
}

@inproceedings{307838,
  author    = {Wei Zong and Yang-Wai Chow and Willy Susilo and Joonsang Baek and Seyit Camtepe},
  title     = {{AudioMarkNet}: Audio Watermarking for Deepfake Speech Detection},
  booktitle = {Proc. 34th USENIX Security Symp. (USENIX Security 25)},
  month     = aug,
  year      = {2025},
  pages     = {4663--4682}
}

@inproceedings{272238,
  author    = {Shehzeen Hussain and Paarth Neekhara and Shlomo Dubnov and Julian McAuley and Farinaz Koushanfar},
  title     = {{WaveGuard}: Understanding and Mitigating Audio Adversarial Examples},
  booktitle = {Proc. 30th USENIX Security Symp. (USENIX Security 21)},
  month     = aug,
  year      = {2021},
  pages     = {2273--2290}
}

@inproceedings{294508,
  author    = {Kun Wang and Xiangyu Xu and Li Lu and Zhongjie Ba and Feng Lin and Kui Ren},
  title     = {{FraudWhistler}: A Resilient, Robust and Plug-and-play Adversarial Example Detection Method for Speaker Recognition},
  booktitle = {Proc. 33rd USENIX Security Symp. (USENIX Security 24)},
  month     = aug,
  year      = {2024},
  pages     = {7303--7320}
}

@misc{wu2023defendingadversarialaudiodiffusion,
  author = {Shutong Wu and Jiongxiao Wang and Wei Ping and Weili Nie and Chaowei Xiao},
  title  = {Defending against Adversarial Audio via Diffusion Model},
  year   = {2023},
  note   = {arXiv:2303.01507}
}

@inproceedings{chen24_interspeech,
  author    = {Sizhou Chen and Yibo Bai and Jiadi Yao and Xiao-Lei Zhang and Xuelong Li},
  title     = {Textual-driven adversarial purification for speaker verification},
  booktitle = {Proc. Interspeech},
  year      = {2024},
  pages     = {527--531},
  doi       = {10.21437/Interspeech.2024-21}
}

@inproceedings{DBLP:conf/ndss/ZhangJLQX21,
  author    = {Guoming Zhang and
                  Xiaoyu Ji and
                  Xinfeng Li and
                  Gang Qu and
                  Wenyuan Xu},
  title     = {{EarArray}: Defending against {DolphinAttack} via Acoustic Attenuation},
  booktitle = {Proc. Netw. Distrib. Syst. Security Symp. (NDSS)},
  year      = {2021},
  doi       = {10.14722/ndss.2021.24551}
}

@inproceedings{285423,
  author    = {Xinfeng Li and Xiaoyu Ji and Chen Yan and Chaohao Li and Yichen Li and Zhenning Zhang and Wenyuan Xu},
  title     = {Learning Normality is Enough: A Software-based Mitigation against Inaudible Voice Attacks},
  booktitle = {Proc. 32nd USENIX Security Symp. (USENIX Security 23)},
  month     = aug,
  year      = {2023},
  pages     = {2455--2472}
}

@inproceedings{299545,
  author    = {Tiantian Liu and Feng Lin and Zhongjie Ba and Li Lu and Zhan Qin and Kui Ren},
  title     = {{MicGuard}: A Comprehensive Detection System against Out-of-band Injection Attacks for Different Level Microphone-based Devices},
  booktitle = {Proc. 33rd USENIX Security Symp. (USENIX Security 24)},
  month     = aug,
  year      = {2024},
  pages     = {3963--3978}
}

@misc{kang2026audioguardcomprehensiveaudiosafety,
  author = {Mintong Kang and Chen Fang and Bo Li},
  title  = {{AudioGuard}: Toward Comprehensive Audio Safety Protection Across Diverse Threat Models},
  year   = {2026},
  note   = {arXiv:2604.08867}
}

@inproceedings{alexos25_interspeech,
  author    = {Antonios Alexos and Raghuveer Peri and Sai Muralidhar Jayanthi and Metehan Cekic and Srikanth Vishnubhotla and Kyu J. Han and Srikanth Ronanki},
  title     = {Defending speech-enabled {LLM}s against adversarial jailbreak threats},
  booktitle = {Proc. Interspeech},
  year      = {2025},
  pages     = {2048--2052},
  doi       = {10.21437/Interspeech.2025-1921}
}

@inproceedings{yang-etal-2025-reshaping,
  author    = {Yang, Hao  and
      Qu, Lizhen  and
      Shareghi, Ehsan  and
      Haffari, Gholamreza},
  title     = {Reshaping Representation Space to Balance the Safety and Over-rejection in Large Audio Language Models},
  booktitle = {Proc. Conf. Empirical Methods Natural Language Process. (EMNLP)},
  month     = nov,
  year      = {2025},
  pages     = {10067--10079},
  doi       = {10.18653/v1/2025.emnlp-main.510}
}

@inproceedings{djanibekov-etal-2025-spirit,
  author    = {Djanibekov, Amirbek  and
      Mukhituly, Nurdaulet  and
      Inui, Kentaro  and
      Aldarmaki, Hanan  and
      Lukas, Nils},
  title     = {{SPIRIT}: Patching Speech Language Models against Jailbreak Attacks},
  booktitle = {Proc. Conf. Empirical Methods Natural Language Process. (EMNLP)},
  month     = nov,
  year      = {2025},
  pages     = {14503--14520},
  doi       = {10.18653/v1/2025.emnlp-main.734}
}

@misc{lin2026sarsteersafeguardinglargeaudiolanguage,
  author = {Weilin Lin and Jianze Li and Hui Xiong and Li Liu},
  title  = {{SARSteer}: Safeguarding Large Audio-Language Models via Safe-Ablated Refusal Steering},
  year   = {2026},
  note   = {arXiv:2510.17633}
}

@inproceedings{307694,
  author    = {Sizhe Chen and Julien Piet and Chawin Sitawarin and David Wagner},
  title     = {{StruQ}: Defending Against Prompt Injection with Structured Queries},
  booktitle = {Proc. 34th USENIX Security Symp. (USENIX Security 25)},
  month     = aug,
  year      = {2025},
  pages     = {2383--2400}
}

@misc{feng2026audiojailbreakslargeaudiolanguage,
  author = {Bo-Han Feng and Yu-Hsuan Li Liang and Chien-Feng Liu and You-Hsuan Chang and Yun-Nung Chen},
  title  = {Audio Jailbreaks in Large Audio-Language Models: Taxonomy, Attack-Defense Analysis, and Cost-Aware Evaluation},
  year   = {2026},
  note   = {arXiv:2605.30031}
}

@misc{wang2026stepdetectingaudiobackdoor,
  author = {Kun Wang and Meng Chen and Junhao Wang and Yuli Wu and Li Lu and Chong Zhang and Peng Cheng and Jiaheng Zhang and Kui Ren},
  title  = {{STEP}: Detecting Audio Backdoor Attacks via Stability-based Trigger Exposure Profiling},
  year   = {2026},
  note   = {arXiv:2603.18103}
}

@misc{zhou2025gradientnormbasedfinetuningbackdoor,
  author = {Nanjun Zhou and Weilin Lin and Li Liu},
  title  = {Gradient Norm-based Fine-Tuning for Backdoor Defense in Automatic Speech Recognition},
  year   = {2025},
  note   = {arXiv:2502.01152}
}

@inproceedings{Sun_2025,
  author    = {Sun, Zhen and Cong, Tianshuo and Liu, Yule and Lin, Chenhao and He, Xinlei and Chen, Rongmao and Han, Xingshuo and Huang, Xinyi},
  title     = {{PEFTGuard}: Detecting Backdoor Attacks Against Parameter-Efficient Fine-Tuning},
  booktitle = {Proc. IEEE Symp. Security Privacy (SP)},
  month     = may,
  year      = {2025},
  pages     = {1713--1731},
  doi       = {10.1109/sp61157.2025.00161}
}

@misc{wang2024mitigatingdownstreammodelrisks,
  author = {Keyu Wang and Abdullah Norozi Iranzad and Scott Schaffter and Meg Risdal and Doina Precup and Jonathan Lebensold},
  title  = {Mitigating Downstream Model Risks via Model Provenance},
  year   = {2024},
  note   = {arXiv:2410.02230}
}

@misc{hsu2025safelorasilverlining,
  author = {Chia-Yi Hsu and Yu-Lin Tsai and Chih-Hsun Lin and Pin-Yu Chen and Chia-Mu Yu and Chun-Ying Huang},
  title  = {Safe {LoRA}: the Silver Lining of Reducing Safety Risks when Fine-tuning Large Language Models},
  year   = {2025},
  note   = {arXiv:2405.16833}
}

@misc{liu2025targetedvaccinesafetyalignment,
  author = {Guozhi Liu and Weiwei Lin and Tiansheng Huang and Ruichao Mo and Qi Mu and Li Shen},
  title  = {Targeted Vaccine: Safety Alignment for Large Language Models against Harmful Fine-Tuning via Layer-wise Perturbation},
  year   = {2025},
  note   = {arXiv:2410.09760}
}

@inproceedings{pmlr-v267-lu25g,
  author    = {Lu, Ning and Liu, Shengcai and Wu, Jiahao and Chen, Weiyu and Zhang, Zhirui and Ong, Yew-Soon and Wang, Qi and Tang, Ke},
  title     = {Safe Delta: Consistently Preserving Safety when Fine-Tuning {LLM}s on Diverse Datasets},
  booktitle = {Proc. 42nd Int. Conf. Mach. Learn. (ICML)},
  month     = {13--19 Jul},
  year      = {2025},
  pages     = {40537--40559}
}

@misc{gu2026differentiallyprivatecommunicationefficient,
  author = {Yujie Gu and Richeng Jin and Xiaoyu Ji and Yier Jin and Wenyuan Xu},
  title  = {Differentially Private and Communication Efficient Large Language Model Split Inference via Stochastic Quantization and Soft Prompt},
  year   = {2026},
  note   = {arXiv:2602.11513}
}

@inproceedings{Mo_2020,
  author    = {Mo, Fan and Shamsabadi, Ali Shahin and Katevas, Kleomenis and Demetriou, Soteris and Leontiadis, Ilias and Cavallaro, Andrea and Haddadi, Hamed},
  title     = {{DarkneTZ}: towards model privacy at the edge using trusted execution environments},
  booktitle = {Proc. Int. Conf. Mobile Syst., Appl., Services (MobiSys)},
  month     = jun,
  year      = {2020},
  pages     = {161--174},
  doi       = {10.1145/3386901.3388946}
}

@inproceedings{10179382,
  author    = {Sun, Zhichuang and Sun, Ruimin and Liu, Changming and Chowdhury, Amrita Roy and Lu, Long and Jha, Somesh},
  title     = {{ShadowNet}: A Secure and Efficient On-device Model Inference System for Convolutional Neural Networks},
  booktitle = {Proc. IEEE Symp. Security Privacy (SP)},
  year      = {2023},
  pages     = {1596--1612},
  doi       = {10.1109/SP46215.2023.10179382}
}

@inproceedings{zhang-etal-2025-pd3f,
  author    = {Zhang, Yuanhe  and
      Wang, Xinyue  and
      Gao, Haoran  and
      Zhou, Zhenhong  and
      Meng, Fanyu  and
      Zhang, Yuyao  and
      Su, Sen},
  title     = {$PD^3F$: A Pluggable and Dynamic {D}o{S}-Defense Framework against resource consumption attacks targeting Large Language Models},
  booktitle = {Findings Assoc. Comput. Linguistics: EMNLP 2025},
  month     = nov,
  year      = {2025},
  pages     = {3641--3671},
  doi       = {10.18653/v1/2025.findings-emnlp.195}
}

@inproceedings{Li_2024_CVPR,
  author    = {Li, Boheng and Cai, Yishuo and Li, Haowei and Xue, Feng and Li, Zhifeng and Li, Yiming},
  title     = {Nearest is Not Dearest: Towards Practical Defense against Quantization-conditioned Backdoor Attacks},
  booktitle = {Proc. IEEE/CVF Conf. Comput. Vis. Pattern Recognit. (CVPR)},
  month     = jun,
  year      = {2024},
  pages     = {24523--24533}
}

@misc{zheng2026breakingroundingtrapsecuring,
  author = {Aoying Zheng and Anqi Du and Zizhuang Deng and Yuxuan Chen},
  title  = {Breaking the Rounding Trap: Securing {LLMs} against Quantization-Conditioned Backdoors},
  year   = {2026},
  note   = {arXiv:2606.29239}
}

@misc{zhan2025aegismcponlinegraphintrusion,
  author = {Zhonghao Zhan and Amir Al Sadi and Krinos Li and Hamed Haddadi},
  title  = {{AegisMCP}: Online Graph Intrusion Detection for Tool-Augmented {LLMs} on Edge Devices},
  year   = {2025},
  note   = {arXiv:2510.19462}
}

@misc{debenedetti2025defeatingpromptinjectionsdesign,
  author = {Edoardo Debenedetti and Ilia Shumailov and Tianqi Fan and Jamie Hayes and Nicholas Carlini and Daniel Fabian and Christoph Kern and Chongyang Shi and Andreas Terzis and Florian Tramèr},
  title  = {Defeating Prompt Injections by Design},
  year   = {2025},
  note   = {arXiv:2503.18813}
}

@inproceedings{an-etal-2025-ipiguard,
  author    = {An, Hengyu  and
      Zhang, Jinghuai  and
      Du, Tianyu  and
      Zhou, Chunyi  and
      Li, Qingming  and
      Lin, Tao  and
      Ji, Shouling},
  title     = {{IPIG}uard: A Novel Tool Dependency Graph-Based Defense Against Indirect Prompt Injection in {LLM} Agents},
  booktitle = {Proc. Conf. Empirical Methods Natural Language Process. (EMNLP)},
  month     = nov,
  year      = {2025},
  pages     = {1023--1039},
  doi       = {10.18653/v1/2025.emnlp-main.53}
}

@misc{doshi2026verifiablysafetooluse,
  author = {Aarya Doshi and Yining Hong and Congying Xu and Eunsuk Kang and Alexandros Kapravelos and Christian Kästner},
  title  = {Towards Verifiably Safe Tool Use for {LLM} Agents},
  year   = {2026},
  note   = {arXiv:2601.08012}
}

@misc{wang2025agentspeccustomizableruntimeenforcement,
  author = {Haoyu Wang and Christopher M. Poskitt and Jun Sun},
  title  = {{AgentSpec}: Customizable Runtime Enforcement for Safe and Reliable {LLM} Agents},
  year   = {2025},
  note   = {arXiv:2503.18666}
}

@misc{zhu2025miniscopeprivilegeframeworkauthorizing,
  author = {Jinhao Zhu and Kevin Tseng and Gil Vernik and Xiao Huang and Shishir G. Patil and Vivian Fang and Raluca Ada Popa},
  title  = {{MiniScope}: A Least Privilege Framework for Authorizing Tool Calling Agents},
  year   = {2025},
  note   = {arXiv:2512.11147}
}

@article{ravichandran2026safetyguardrailsllmenabledrobots,
  author  = {Zachary Ravichandran and Alexander Robey and Vijay Kumar and George J. Pappas and Hamed Hassani},
  title   = {Safety Guardrails for {LLM}-Enabled Robots},
  journal = {IEEE Robotics and Automation Letters},
  volume  = {11},
  number  = {4},
  pages   = {4649--4656},
  month   = apr,
  year    = {2026},
  doi     = {10.1109/LRA.2026.3667488}
}

@misc{ma2026autodojoadaptiveblackboxattacks,
  author = {Xinhang Ma and Taoran Li and Chaowei Xiao and Zhiyuan Yu and Ning Zhang and Yevgeniy Vorobeychik},
  title  = {{AutoDojo}: Adaptive Black-Box Attacks Reveal the Limits of {IPI} Defenses and Task-Specification Effects in {LLM} Agents},
  year   = {2026},
  note   = {arXiv:2606.15057}
}

@misc{tomashenko2024voiceprivacy,
  author = {Natalia Tomashenko and Xiaoxiao Miao and Pierre Champion and Sarina Meyer and Xin Wang and Emmanuel Vincent and Michele Panariello and Nicholas Evans and Junichi Yamagishi and Massimiliano Todisco},
  title  = {The {VoicePrivacy} 2024 Challenge Evaluation Plan},
  year   = {2024},
  note   = {arXiv:2404.02677}
}

@misc{kuzmin2026streamvoiceanonenhancingutilityrealtime,
  author = {Nikita Kuzmin and Songting Liu and Kong Aik Lee and Eng Siong Chng},
  title  = {Stream-Voice-Anon: Enhancing Utility of Real-Time Speaker Anonymization via Neural Audio Codec and Language Models},
  year   = {2026},
  note   = {arXiv:2601.13948}
}

@inproceedings{soh2025securespeech,
  author    = {Soh, Belinda Hui Hui and Miao, Xiaoxiao and Wang, Xin},
  title     = {{SecureSpeech}: Prompt-Based Speaker and Content Protection},
  booktitle = {Proc. IEEE Int. Joint Conf. Biometrics (IJCB)},
  year      = {2025},
  pages     = {1--8},
  doi       = {10.1109/IJCB65343.2025.11410690}
}

@article{11430623,
  author  = {Sun, Xiping and Chen, Jing and Wu, Cong and He, Kun and Xu, Haozhe and Feng, Yebo and Du, Ruiying and Chen, Xianhao},
  title   = {{MagLive}: Robust Voice Liveness Detection on Smartphones Using Magnetic Pattern Changes},
  journal = {IEEE Trans. Inf. Forensics Security},
  volume  = {21},
  pages   = {2978--2991},
  year    = {2026},
  doi     = {10.1109/TIFS.2026.3673129}
}

@inproceedings{ICLR2025_8c234d9c,
  author    = {Kuo, Martin and Zhang, Jingyang and Zhang, Jianyi and Tang, Minxue and DiValentin, Louis and Ding, Aolin and Sun, Jingwei and Chen, William and Hass, Amin and Chen, Tianlong and Chen, Yiran  and Li, Hai},
  title     = {Proactive Privacy Amnesia for Large Language Models: Safeguarding {PII} with Negligible Impact on Model Utility},
  booktitle = {Proc. Int. Conf. Learn. Representations (ICLR)},
  year      = {2025},
  pages     = {55698--55719}
}

@misc{peng2026jalmbenchbenchmarkingjailbreakvulnerabilities,
  author = {Zifan Peng and Yule Liu and Zhen Sun and Mingchen Li and Zeren Luo and Jingyi Zheng and Wenhan Dong and Xinlei He and Xuechao Wang and Yingjie Xue and Shengmin Xu and Xinyi Huang},
  title  = {{JALMBench}: Benchmarking Jailbreak Vulnerabilities in Audio Language Models},
  year   = {2026},
  note   = {arXiv:2505.17568}
}

@inproceedings{cheng2025jailbreakaudiobench,
  author    = {Cheng, Hao and Xiao, Erjia and Shao, Jing and Wang, Yichi and Yang, Le and Shen, Chao and Torr, Philip and Gu, Jindong and Xu, Renjing},
  title     = {{Jailbreak-AudioBench}: In-Depth Evaluation and Analysis of Jailbreak Threats for Large Audio Language Models},
  booktitle = {Proc. Int. Conf. Neural Inf. Process. Syst. (NeurIPS)},
  volume    = {38},
  year      = {2025},
  doi       = {10.52202/085713-0361}
}

@misc{chen2026audiojailbreakjailbreakattacksendtoend,
  author = {Guangke Chen and Fu Song and Zhe Zhao and Xiaojun Jia and Yang Liu and Yanchen Qiao and Weizhe Zhang and Weiping Tu and Yuhong Yang and Bo Du},
  title  = {{AudioJailbreak}: Jailbreak Attacks against End-to-End Large Audio-Language Models},
  year   = {2026},
  note   = {arXiv:2505.14103}
}

@misc{li2026audiotrustbenchmarkingmultifacetedtrustworthiness,
  author = {Kai Li and Can Shen and Yile Liu and Jirui Han and Kelong Zheng and Xuechao Zou and Lionel Z. Wang and Shun Zhang and Xingjian Du and Hanjun Luo and Yingbin Jin and Xinxin Xing and Ziyang Ma and Yue Liu and Yifan Zhang and Junfeng Fang and Kun Wang and Yibo Yan and Gelei Deng and Haoyang Li and Yiming Li and Xiaobin Zhuang and Tianlong Chen and Qingsong Wen and Tianwei Zhang and Yang Liu and Haibo Hu and Zhizheng Wu and Xiaolin Hu and Eng-Siong Chng and Wenyuan Xu and XiaoFeng Wang and Wei Dong and Xinfeng Li},
  title  = {{AudioTrust}: Benchmarking the Multifaceted Trustworthiness of Audio Large Language Models},
  year   = {2026},
  note   = {arXiv:2505.16211}
}

@misc{pan2025omnisafetybenchbenchmarksafetyevaluation,
  author = {Leyi Pan and Zheyu Fu and Yunpeng Zhai and Shuchang Tao and Sheng Guan and Shiyu Huang and Lingzhe Zhang and Zhaoyang Liu and Bolin Ding and Felix Henry and Aiwei Liu and Lijie Wen},
  title  = {{Omni-SafetyBench}: A Benchmark for Safety Evaluation of Audio-Visual Large Language Models},
  year   = {2025},
  note   = {arXiv:2508.07173}
}

@inproceedings{zhang-etal-2026-rsa,
  author    = {Zhang, Yibo  and
      Luo, Kaiwen  and
      Lin, Liang  and
      Yan, Shilinlu  and
      Wang, Jin  and
      Guo, Yaoqi  and
      Chen, Yitian  and
      Qin, Yalan  and
      Zhou, Zhenhong  and
      Wang, Kun  and
      Sun, Li},
  title     = {{RSA}-Bench: Benchmarking Audio Large Models in Real-World Acoustic Scenarios},
  booktitle = {Findings Assoc. Comput. Linguistics: ACL 2026},
  month     = jul,
  year      = {2026},
  pages     = {18350--18374},
  doi       = {10.18653/v1/2026.findings-acl.913}
}

@misc{zhang2025wildspeechbenchbenchmarkingendtoendspeechllms,
  author = {Linhao Zhang and Jian Zhang and Bokai Lei and Chuhan Wu and Aiwei Liu and Wei Jia and Xiao Zhou},
  title  = {{WildSpeech-Bench}: Benchmarking End-to-End {SpeechLLMs} in the Wild},
  year   = {2025},
  note   = {arXiv:2506.21875}
}

@misc{luo2026chronosaudiocomprehensivelongaudiobenchmark,
  author = {Kaiwen Luo and Liang Lin and Yibo Zhang and Moayad Aloqaily and Jialiang Tao and Dexian Wang and Zhenhong Zhou and Junwei Zhang and Kun Wang and Li Sun and Qingsong Wen},
  title  = {{ChronosAudio}: A Comprehensive Long-Audio Benchmark for Evaluating Audio-Large Language Models},
  year   = {2026},
  note   = {arXiv:2601.04876}
}

@misc{he2025audiomarathoncomprehensivebenchmarklongcontext,
  author = {Peize He and Zichen Wen and Yubo Wang and Yuxuan Wang and Xiaoqian Liu and Jiajie Huang and Zehui Lei and Zhuangcheng Gu and Xiangqi Jin and Jiabing Yang and Kai Li and Zhifei Liu and Weijia Li and Cunxiang Wang and Conghui He and Linfeng Zhang},
  title  = {{AudioMarathon}: A Comprehensive Benchmark for Long-Context Audio Understanding and Efficiency in Audio {LLMs}},
  year   = {2025},
  note   = {arXiv:2510.07293}
}

@inproceedings{yan-etal-2025-uro,
  author    = {Yan, Ruiqi  and
      Li, Xiquan  and
      Chen, Wenxi  and
      Niu, Zhikang  and
      Yang, Chen  and
      Ma, Ziyang  and
      Yu, Kai  and
      Chen, Xie},
  title     = {{URO}-Bench: Towards Comprehensive Evaluation for End-to-End Spoken Dialogue Models},
  booktitle = {Findings Assoc. Comput. Linguistics: EMNLP 2025},
  month     = nov,
  year      = {2025},
  pages     = {17211--17242},
  doi       = {10.18653/v1/2025.findings-emnlp.933}
}

@misc{liu2026vocalbenchbenchmarkingvocalconversational,
  author = {Heyang Liu and Yuhao Wang and Ziyang Cheng and Hongcheng Liu and Yiqi Li and Yixuan Hou and Ronghua Wu and Qunshan Gu and Yanfeng Wang and Yu Wang},
  title  = {{VocalBench}: Benchmarking the Vocal Conversational Abilities for Speech Interaction Models},
  year   = {2026},
  note   = {arXiv:2505.15727}
}

@inproceedings{pahwa2026audio2toolspeakcallact,
  author    = {Ramit Pahwa and Apoorva Beedu and Parivesh Priye and Rutu Gandhi and Saloni Takawale and Aruna Baijal and Zengli Yang},
  title     = {{Audio2Tool}: Speak, Call, Act -- A Dataset for Benchmarking Speech Tool Use},
  booktitle = {Proc. Interspeech 2026},
  year      = {2026},
  pages     = {3453--3457},
  doi       = {10.21437/Interspeech.2026-2857}
}

@misc{rahimov2026audiollmagentsurvey,
  title        = {Safety and Robustness of Audio-{LLM} Agents: A Comprehensive Survey of Risks, Defenses, and Trustworthiness Evaluation},
  author       = {Rahimov, Akmal and Bekbayev, Nurlan and Nurakun Kyzy, Zhumagul},
  year         = {2026},
  month        = jul,
  note         = {Preprint},
  doi          = {10.13140/RG.2.2.36106.02242},
  url          = {https://www.researchgate.net/publication/409023785_Safety_and_Robustness_of_Audio-LLM_Agents_A_Comprehensive_Survey_of_Risks_Defenses_and_Trustworthiness_Evaluation}
}

@article{xu2026llmagentsecurityduality,
  title        = {{LLM} Agents Security Duality: A Comprehensive Survey of Self-Security and Empowered Cybersecurity},
  author       = {Xu, Yiwei and Zhuang, Yong and Liu, Xuanming and Zhang, Tian and Xiao, Bowen and Xu, Xiaoyang and Jiang, Delong and Wang, Juan and Hu, Hongxin},
  journal      = {Artificial Intelligence Review},
  volume       = {59},
  pages        = {174},
  year         = {2026},
  doi          = {10.1007/s10462-026-11563-0}
}

@misc{ling2026securellmagents,
  title        = {Toward Secure {LLM} Agents: Threat Surfaces, Attacks, Defenses, and Evaluation},
  author       = {Ling, Yuchen and Yu, Shengcheng and Chen, Zhenyu and Fang, Chunrong},
  year         = {2026},
  eprint       = {2606.10749},
  archivePrefix= {arXiv},
  primaryClass = {cs.CR},
  url          = {https://arxiv.org/abs/2606.10749}
}

@misc{lin2026agentmemorysecuritysurvey,
  title        = {A Survey on Long-Term Memory Security in {LLM} Agents: Attacks, Defenses, and Governance Across the Memory Lifecycle},
  author       = {Lin, Zehao and Hao, Xixuan and Fu, Renyu and Cui, Shaobo and Chen, Kai and Li, Chunyu and Li, Zhiyu and Xiong, Feiyu},
  year         = {2026},
  eprint       = {2604.16548},
  archivePrefix= {arXiv},
  primaryClass = {cs.CR},
  url          = {https://arxiv.org/abs/2604.16548}
}

@article{dib2026agenticedgeintelligencesurvey,
  title        = {Communication-Efficient Agentic Edge Intelligence: A Survey of Federated Multi-Agent Information Fusion for {LLM}-Driven Edge {AI}},
  author       = {Dib, Omar},
  journal      = {Information Fusion},
  volume       = {136},
  pages        = {104577},
  year         = {2026},
  doi          = {10.1016/j.inffus.2026.104577}
}

@inproceedings{abdollahi2026agentee,
  title     = {{AgenTEE}: Confidential {LLM} Agent Execution on Edge Devices},
  author    = {Abdollahi, Sina and Maheri, Mohammad M. and Forough, Javad and Al Sadi, Amir and Millar, Josh and Kotz, David and Kogias, Marios and Haddadi, Hamed},
  booktitle = {Proceedings of the Sixth European Workshop on Machine Learning and Systems (EuroMLSys)},
  month     = apr,
  year      = {2026},
  pages     = {473--480},
  publisher = {ACM},
  doi       = {10.1145/3805621.3807660}
}

@inproceedings{chi2025adaptiveddsd,
  title        = {Adaptive Knowledge Distillation for Device-Directed Speech Detection},
  author       = {Chi, Hyung-gun and Pesce, Florian and Chang, Wonil and Rudovic, Oggi and Argueta, Arturo and Braun, Stefan and Garg, Vineet and Abdelaziz, Ahmed Hussen},
  booktitle    = {Interspeech 2025},
  pages        = {5788--5792},
  year         = {2025},
  doi          = {10.21437/Interspeech.2025-30}
}

@inproceedings{chen2024agentpoison,
  title        = {{AgentPoison}: Red-Teaming {LLM} Agents via Poisoning Memory or Knowledge Bases},
  author       = {Chen, Zhaorun and Xiang, Zhen and Xiao, Chaowei and Song, Dawn and Li, Bo},
  booktitle    = {Advances in Neural Information Processing Systems},
  volume       = {37},
  year         = {2024},
  url          = {https://proceedings.neurips.cc/paper_files/paper/2024/hash/eb113910e9c3f6242541c1652e30dfd6-Abstract-Conference.html}
}

@inproceedings{dong2025minja,
  title        = {Memory Injection Attacks on {LLM} Agents via Query-Only Interaction},
  author       = {Dong, Shen and Xu, Shaochen and He, Pengfei and Li, Yige and Tang, Jiliang and Liu, Tianming and Liu, Hui and Xiang, Zhen J.},
  booktitle    = {Advances in Neural Information Processing Systems},
  volume       = {38},
  year         = {2025},
  doi          = {10.52202/085713-1554},
  url          = {https://proceedings.neurips.cc/paper_files/paper/2025/hash/42a97bbd9844d2bf68596730af80bcdf-Abstract-Conference.html}
}

@inproceedings{wang2025agentmemoryprivacy,
  title        = {Unveiling Privacy Risks in {LLM} Agent Memory},
  author       = {Wang, Bo and He, Weiyi and Zeng, Shenglai and Xiang, Zhen and Xing, Yue and Tang, Jiliang and He, Pengfei},
  booktitle    = {Proceedings of the 63rd Annual Meeting of the Association for Computational Linguistics (Volume 1: Long Papers)},
  pages        = {25241--25260},
  year         = {2025},
  doi          = {10.18653/v1/2025.acl-long.1227},
  url          = {https://aclanthology.org/2025.acl-long.1227/}
}

@misc{wei2025amemguard,
  title        = {{A-MemGuard}: A Proactive Defense Framework for {LLM}-Based Agent Memory},
  author       = {Wei, Qianshan and Yang, Tengchao and Wang, Yaochen and Li, Xinfeng and Li, Lijun and Yin, Zhenfei and Zhan, Yi and Holz, Thorsten and Lin, Zhiqiang and Wang, XiaoFeng},
  year         = {2025},
  eprint       = {2510.02373},
  archivePrefix= {arXiv},
  primaryClass = {cs.CR},
  url          = {https://arxiv.org/abs/2510.02373}
}

@inproceedings{wagner2025selma,
  author    = {Dominik Wagner and Alexander W. Churchill and Siddharth Sigtia and Erik Marchi},
  title     = {{SELMA}: A Speech-Enabled Language Model for Virtual Assistant Interactions},
  booktitle = {ICASSP 2025 - 2025 IEEE International Conference on Acoustics, Speech and Signal Processing (ICASSP)},
  pages     = {1--5},
  year      = {2025},
  publisher = {IEEE},
  doi       = {10.1109/ICASSP49660.2025.10890139}
}

@misc{dash2026untrustedinputtrustedmemory,
  author        = {Pritam Dash and Tongyu Ge and Aditi Jain and Tanmay Shah and Zhiwei Shang},
  title         = {From Untrusted Input to Trusted Memory: A Systematic Study of Memory Poisoning Attacks in {LLM} Agents},
  year          = {2026},
  note          = {arXiv:2606.04329},
  eprint        = {2606.04329},
  archivePrefix = {arXiv},
  url           = {https://arxiv.org/abs/2606.04329}
}

@misc{pulipaka2026hiddenmemory,
  author        = {Sidharth Pulipaka and Stanislau Hlebik and Leonidas Raghav and Sahar Abdelnabi and Vyas Raina and Ivaxi Sheth and Mario Fritz},
  title         = {Hidden in Memory: Sleeper Memory Poisoning in {LLM} Agents},
  year          = {2026},
  note          = {arXiv:2605.15338},
  eprint        = {2605.15338},
  archivePrefix = {arXiv},
  url           = {https://arxiv.org/abs/2605.15338}
}

@misc{tian2026injecmem,
  author        = {Hanling Tian and Gengyu Zhang and Zeyang Sha and Jingying Wang and Yuhang Liu and Zhehao Huang and Kun Yang and Xiaolin Huang},
  title         = {{InjecMEM}: Memory Injection Attack on {LLM} Agent Memory Systems},
  year          = {2026},
  note          = {arXiv:2608.23471},
  eprint        = {2608.23471},
  archivePrefix = {arXiv},
  url           = {https://arxiv.org/abs/2608.23471}
}

@inproceedings{qian-2026-visual,
  author    = {Jiachen Qian},
  title     = {Visual Inception: Compromising Long-term Planning in Agentic Recommenders via Multimodal Memory Poisoning},
  booktitle = {Proceedings of the 64th Annual Meeting of the Association for Computational Linguistics (Volume 1: Long Papers)},
  year      = {2026},
  month     = jul,
  pages     = {20846--20862},
  address   = {San Diego, California, United States},
  publisher = {Association for Computational Linguistics},
  doi       = {10.18653/v1/2026.acl-long.954},
  url       = {https://aclanthology.org/2026.acl-long.954/}
}

@inproceedings{feng-etal-2026-backdooragent,
  author    = {Yunhao Feng and Yige Li and Yutao Wu and Yingshui Tan and Yanming Guo and Yifan Ding and Kun Zhai and Xingjun Ma and Yu-Gang Jiang},
  title     = {{BackdoorAgent}: A Unified Framework for Backdoor Attacks on {LLM}-based Agents},
  booktitle = {Findings of the Association for Computational Linguistics: ACL 2026},
  year      = {2026},
  month     = jul,
  pages     = {16115--16127},
  address   = {San Diego, California, United States},
  publisher = {Association for Computational Linguistics},
  doi       = {10.18653/v1/2026.findings-acl.791},
  url       = {https://aclanthology.org/2026.findings-acl.791/}
}

@inproceedings{zhang-pei-2026-llm,
  author    = {Wuyang Zhang and Shichao Pei},
  title     = {Your {LLM} Agent Can Leak Your Data: Data Exfiltration via Backdoored Tool Use},
  booktitle = {Findings of the Association for Computational Linguistics: ACL 2026},
  year      = {2026},
  month     = jul,
  pages     = {25105--25129},
  address   = {San Diego, California, United States},
  publisher = {Association for Computational Linguistics},
  doi       = {10.18653/v1/2026.findings-acl.1257},
  url       = {https://aclanthology.org/2026.findings-acl.1257/}
}

@inproceedings{yan-etal-2026-evo,
  author    = {Bingyu Yan and Xiaoming Zhang and JinYu Hou and Chaozhuo Li and Ziyi Zhou and Yiming Hei and Litian Zhang},
  title     = {{Evo-Attacker}: Memory-Augmented Reinforcement Learning for Long-Horizon Tool Attacks on {LLM}-{MAS}},
  booktitle = {Proceedings of the 64th Annual Meeting of the Association for Computational Linguistics (Volume 1: Long Papers)},
  year      = {2026},
  month     = jul,
  pages     = {7286--7300},
  address   = {San Diego, California, United States},
  publisher = {Association for Computational Linguistics},
  doi       = {10.18653/v1/2026.acl-long.330},
  url       = {https://aclanthology.org/2026.acl-long.330/}
}

@inproceedings{lin-etal-2026-vigil,
  author    = {Junda Lin and Zhaomeng Zhou and Zhi Zheng and Shuochen Liu and Tong Xu and Yong Chen and Enhong Chen},
  title     = {{VIGIL}: Defending {LLM} Agents Against Tool-Stream Injection via Verify-Before-Commit},
  booktitle = {Proceedings of the 64th Annual Meeting of the Association for Computational Linguistics (Volume 1: Long Papers)},
  year      = {2026},
  month     = jul,
  pages     = {9764--9785},
  address   = {San Diego, California, United States},
  publisher = {Association for Computational Linguistics},
  doi       = {10.18653/v1/2026.acl-long.443},
  url       = {https://aclanthology.org/2026.acl-long.443/}
}

@inproceedings{jia-etal-2025-task,
  author    = {Feiran Jia and Tong Wu and Xin Qin and Anna Squicciarini},
  title     = {The Task Shield: Enforcing Task Alignment to Defend Against Indirect Prompt Injection in {LLM} Agents},
  booktitle = {Proceedings of the 63rd Annual Meeting of the Association for Computational Linguistics (Volume 1: Long Papers)},
  year      = {2025},
  month     = jul,
  pages     = {29680--29697},
  address   = {Vienna, Austria},
  publisher = {Association for Computational Linguistics},
  doi       = {10.18653/v1/2025.acl-long.1435},
  url       = {https://aclanthology.org/2025.acl-long.1435/}
}

@misc{ji2026seagent,
  author        = {Zimo Ji and Daoyuan Wu and Wenyuan Jiang and Pingchuan Ma and Zongjie Li and Yudong Gao and Shuai Wang and Yingjiu Li},
  title         = {Taming Various Privilege Escalation in {LLM}-Based Agent Systems: A Mandatory Access Control Framework},
  year          = {2026},
  note          = {arXiv:2601.11893},
  eprint        = {2601.11893},
  archivePrefix = {arXiv},
  url           = {https://arxiv.org/abs/2601.11893}
}

@inproceedings{an-etal-2026-aciarena,
  author    = {Hengyu An and Minxi Li and Jinghuai Zhang and Naen Xu and Chunyi Zhou and Changjiang Li and Xiaogang Xu and Tianyu Du and Shouling Ji},
  title     = {{ACIArena}: Toward Unified Evaluation for Agent Cascading Injection},
  booktitle = {Proceedings of the 64th Annual Meeting of the Association for Computational Linguistics (Volume 1: Long Papers)},
  year      = {2026},
  month     = jul,
  pages     = {10049--10066},
  address   = {San Diego, California, United States},
  publisher = {Association for Computational Linguistics},
  doi       = {10.18653/v1/2026.acl-long.457},
  url       = {https://aclanthology.org/2026.acl-long.457/}
}

@misc{karamchandani2026forgedreasoning,
  author        = {Neeraj Karamchandani and Piyush Nagasubramaniam and Sencun Zhu and Dinghao Wu},
  title         = {Your Agent's Memories Are Not Its Own: Forged Reasoning Attacks on {LLM} Agent Memory and Defenses},
  year          = {2026},
  note          = {arXiv:2607.05029},
  eprint        = {2607.05029},
  archivePrefix = {arXiv},
  url           = {https://arxiv.org/abs/2607.05029}
}

@misc{xu2026memoryprovenancelaundering,
  author        = {Jinghan Xu and Yiyong Xiao and Wanru Shao and Hankai Liu and Xinjin Li},
  title         = {Memory Provenance Laundering in {LLM} Agents: A Non-Amplification Firewall for Persistent Memory},
  year          = {2026},
  note          = {arXiv:2607.29167},
  eprint        = {2607.29167},
  archivePrefix = {arXiv},
  url           = {https://arxiv.org/abs/2607.29167}
}

@inproceedings{an-etal-2026-permemsafe,
  author    = {Hengyu An and Minxi Li and Naen Xu and Chunyi Zhou and Xiaogang Xu and Tianyu Du and Jinbao Li and Shouling Ji},
  title     = {{PerMemSafe}: Benchmarking Implicit Personalized Safety of Long Horizon Self-Evolving Agents},
  booktitle = {Findings of the Association for Computational Linguistics: ACL 2026},
  year      = {2026},
  month     = jul,
  pages     = {6415--6433},
  address   = {San Diego, California, United States},
  publisher = {Association for Computational Linguistics},
  doi       = {10.18653/v1/2026.findings-acl.320},
  url       = {https://aclanthology.org/2026.findings-acl.320/}
}

@inproceedings{geng-etal-2026-piarena,
  author    = {Runpeng Geng and Chenlong Yin and Yanting Wang and Ying Chen and Jinyuan Jia},
  title     = {{PIArena}: A Platform for Prompt Injection Evaluation},
  booktitle = {Proceedings of the 64th Annual Meeting of the Association for Computational Linguistics (Volume 1: Long Papers)},
  year      = {2026},
  month     = jul,
  pages     = {33170--33192},
  address   = {San Diego, California, United States},
  publisher = {Association for Computational Linguistics},
  doi       = {10.18653/v1/2026.acl-long.1533},
  url       = {https://aclanthology.org/2026.acl-long.1533/}
}

@misc{rahman2026proofexecutionmemory,
  author        = {Md Habibur Rahman and Jaeho Kim},
  title         = {Proof-of-Execution Memory: Defending {LLM} Agents Against Forged-Reasoning Attacks by Verifying What Actually Happened},
  year          = {2026},
  note          = {arXiv:2608.16032},
  eprint        = {2608.16032},
  archivePrefix = {arXiv},
  url           = {https://arxiv.org/abs/2608.16032}
}

@inproceedings{seo2026simuhome,
  title     = {SimuHome: A Temporal- and Environment-Aware Benchmark for Smart Home {LLM} Agents},
  author    = {Seo, Gyuhyeon and Yang, Jungwoo and Pyo, Junseong and Kim, Nalim and Lee, Jonggeun and Jo, Yohan},
  booktitle = {The Fourteenth International Conference on Learning Representations},
  year      = {2026},
  url       = {https://openreview.net/forum?id=LCS1WsGvha}
}

@article{zou2026homeguardian,
  title   = {HomeGuardian: Detecting Behavioral Anomalies Through Implicit Profiling in Smart Home Systems},
  author  = {Zou, Qiang and Xu, Z. and Xiao, J. and Li, Q. and Zhao, D. and Li, R. and Tang, W. and Li, K. and Zuo, X. and Jiang, Y. and Lyu, Michael R.},
  journal = {IEEE Transactions on Mobile Computing},
  year    = {2026},
  doi     = {10.1109/TMC.2026.3730188}
}

@inproceedings{robinson2025naturelm,
  title     = {NatureLM-audio: An Audio-Language Foundation Model for Bioacoustics},
  author    = {Robinson, David and Miron, Marius and Hagiwara, Masato and Pietquin, Olivier},
  booktitle = {International Conference on Learning Representations},
  year      = {2025},
  url       = {https://openreview.net/forum?id=hJVdwBpWjt}
}

@article{cui2026industrial_audio_llm,
  title   = {Domain-Adaptive Audio Large Language Model for Acoustic Fault Diagnosis and Semantic Description of Coal Mine Equipment},
  author  = {Cui, Daming and Zhang, Xin and Ma, Qiang},
  journal = {Algorithms},
  volume  = {19},
  number  = {9},
  pages   = {778},
  year    = {2026},
  doi     = {10.3390/a19090778}
}
\end{document}